\documentclass[11pt]{article}
\usepackage{amssymb,amsmath}
\usepackage{bm} 
\usepackage{booktabs} 
\usepackage{array}
\usepackage{latexsym}
\usepackage{graphicx}
\usepackage{color}
\usepackage{datetime}
\usepackage[nosort]{cite}
\usepackage{verbatim}
\usepackage{enumerate}
\usepackage{chngpage} 
\usepackage{mathrsfs}
\usepackage{euscript}
\usepackage{psfrag}

\usepackage{blindtext}%
\usepackage[theorems,breakable]{tcolorbox}%

\usepackage{multirow,graphicx}
\usepackage{makecell}
\usepackage{float}
\usepackage{tikz}
\usetikzlibrary{decorations.text}
\usepackage{enumitem}
\setlist{itemsep=0pt}
\usetikzlibrary{math}
\usepackage{setspace}

\usepackage{mciteplus}

\usepackage[colorlinks=true,      linkcolor=blue,      urlcolor=blue,      
            filecolor=blue,      citecolor=blue,       pdfstartview=FitH,     
						pdfpagemode=UseNone,      bookmarksopen=true]{hyperref}  
\usepackage[all]{hypcap}     

\usepackage{tikz}
\usepackage{pgfplots}
\usetikzlibrary{decorations.text,intersections, pgfplots.fillbetween}
\usetikzlibrary{math}
 \usetikzlibrary{snakes,3d,shapes.geometric,shadows.blur}
\usepackage{tikz-3dplot}

\usepackage{blindtext}%
\usepackage[theorems,breakable]{tcolorbox}%

\newtcbtheorem{mytheo}{}%
{colback=gray!5,colframe=gray!35!black,fonttitle=\bfseries}{th}

\newtcbtheorem{lem}{}{%
        colback=green!5,%
        colframe=green!35!black,%
        fonttitle=\bfseries,title %
    }{lem}

\definecolor{amaranthred}{rgb}{0.83,0.13,0.18}
\definecolor{amazon}{rgb}{0.23,0.48,0.34}
\definecolor{bdazzledblue}{rgb}{0.18,0.35,0.58}
\definecolor{absolutezero}{rgb}{0.0,0.28,0.73}
\definecolor{bitterlemon}{rgb}{0.79,0.88,0.05}
\definecolor{byzantine}{rgb}{0.74,0.2,0.64}
\definecolor{turquoise}{rgb}{0.19, 0.84, 0.78}
\definecolor{burgundy}{rgb}{0.5, 0.0, 0.13}

\newcommand{\comm}[1]{} 

\def\({\left(}
\def\){\right)}
\def\[{\left[}
\def\]{\right]}

\def\One{{\hbox{ 1\kern-.8mm l}}}

\def\barray{\begin{array}}
\def\earray{\end{array}}
\def\be{\begin{equation}}
\def\ee{\end{equation}}
\def\bea{\begin{eqnarray}}
\def\eea{\end{eqnarray}}
\def\bal{\begin{align}}
\def\eal{\end{align}}
\def\nn{\nonumber}

\def\-{\,-\,}
\def\={\,=\,}
\def\+{\,+\,}
\def\equi{\,\equiv\,}

\numberwithin{equation}{section} 

\definecolor{cardinal}{rgb}{0.6,0,0}
\definecolor{darkgreen}{rgb}{0,0.4,0}
\definecolor{golden}{rgb}{0.92, 0.7, 0}
\definecolor{midnight}{rgb}{0, 0, 0.5}
\definecolor{darkblue}{rgb}{0, 0, 0.7}
\definecolor{purple}{rgb}{0.5, 0, 0.5}
\definecolor{gray}{rgb}{0.25, 0.25, 0.25}

\definecolor{amaranthred}{rgb}{0.83,0.13,0.18}
\definecolor{amazon}{rgb}{0.23,0.48,0.34}
\definecolor{bdazzledblue}{rgb}{0.18,0.35,0.58}
\definecolor{absolutezero}{rgb}{0.0,0.28,0.73}
\definecolor{bitterlemon}{rgb}{0.79,0.88,0.05}
\definecolor{byzantine}{rgb}{0.74,0.2,0.64}
\definecolor{turquoise}{rgb}{0.19, 0.84, 0.78}
\definecolor{burgundy}{rgb}{0.5, 0.0, 0.13}

\def\IR{\mathbb{R}}

\def\cA{{\cal A}}

\def\cH{{\cal H}}
\def\cJ{{\cal J}}

\def\cM{{\cal M}}

\def\cO{{\cal O}}

\def\cV{{\cal V}}

\usetikzlibrary{positioning}

\tikzset{
 diffuse color/.initial = black,                       
}

\tikzset{
 linear opacity/.initial=0.5,                          
 linear stroke/.style = {                              
   preaction={                                         
     draw=\pgfkeysvalueof{/tikz/diffuse color},        
     line width = (2.0-#1)*\pgflinewidth,              
     opacity=\pgfkeysvalueof{/tikz/linear opacity},white}},  
 diffuse gradient/.style={                             
   draw = none,                                        
   linear opacity=#1,                                  
   linear stroke/.list={0.0,#1,...,1.0}},              
 diffuse gradient/.default=1,                          
}

\tikzset{
 non-linear stroke/.style = {                          
   preaction={                                         
     draw=\pgfkeysvalueof{/tikz/diffuse color},        
     line width = (2.0-#1)*\pgflinewidth,              
     opacity=#1,white}},                                     
 diffuse falloff/.style={                              
   draw = none,                                        
   non-linear stroke/.list={0.0,#1,...,1.0}},          
 diffuse falloff/.default=1,                           
}
 
\begin{document}

\phantom{AAA}
\vspace{-10mm}

\begin{flushright}
%
%
\end{flushright}

\vspace{2cm}

\begin{center}

{\fontsize{20}{20}\selectfont {\bf Electromagnetic Entrapment in Gravity} }

\vspace{1.5cm}

{\large{\bf {Pierre Heidmann$^{1,2}$ and Madhur Mehta$^{1}$}}}

\vspace{7mm}

\centerline{$^1$ Department of Physics,}
\centerline{$^2$ Center for Cosmology and AstroParticle Physics (CCAPP),}
\centerline{The Ohio State University,}
\centerline{191 W Woodruff Ave, Columbus, OH 43210, USA}

\vspace{7mm} 

{\footnotesize\upshape\ttfamily heidmann.5@osu.edu,~  mehta.493@osu.edu} 

\vspace{15mm}
 
\textsc{Abstract}

\end{center}

\begin{adjustwidth}{6mm}{6mm} 
 
\vspace{-2mm}
\noindent

\noindent 
We derive specific properties of electromagnetism when gravitational effects are not negligible and analyze their impact on new physics at the horizons of black holes. We show that a neutral configuration of charges in a region of high redshift, characterized by a large $g^{tt}$, produces a highly localized electromagnetic field that vanishes just beyond that region.
This phenomenon implies the existence of extensive families of spacetime structures generated by electromagnetic degrees of freedom that are as compact as black holes. We construct neutral bound states of extremal black holes in four dimensions and in five dimensions, where one direction is compact. These geometries are indistinguishable from a neutral black hole, referred to as \emph{distorted Schwarzschild}, except in an infinitesimal region near its horizon where the entrapped electromagnetic structures start to manifest. The five-dimensional solutions satisfy various criteria for describing black hole microstructure: they increase in size with the Newton constant, are as compact as the Schwarzschild black hole, and have an entropy that scales like $M^2$.

\bigskip

\end{adjustwidth}

\vspace{8mm}
 

\thispagestyle{empty}

\newpage


\baselineskip=14pt
\parskip=2.5pt
\setcounter{tocdepth}{2}

\tableofcontents


\baselineskip=15pt
\parskip=3pt
\newpage

\section{Introduction}

In many ways, nothing special happens at the horizon of a black hole --- a region of spacetime in vacuum with generally low curvature. An infalling observer should barely notice the crossing, and the associated dynamics could be expected to be entirely classical under general relativity. However, the horizon plays a key role in the thermodynamics of black holes, contributing to the ongoing theoretical conflict between quantum mechanics and general relativity. The Bekenstein-Hawking formula establishes a link between the size of the horizon and the entropy of black holes, reflecting the extensive phase space of microstates within \cite{Bekenstein:1973ur}. Additionally, the semi-classical mechanism proposed by Hawking \cite{Hawking:1975vcx}, involving thermal radiation from the horizon region, gives rise to the information loss paradox associated with black hole evaporation.

There is a growing consensus that unitarity at the horizon cannot be restored by small quantum corrections in Hawking radiation, necessitating new physics at the horizon scale \cite{Mathur:2009hf,Almheiri:2012rt}. It is expected not only to resolve the information loss paradox but also to provide a microscopic description of black hole entropy, involving a novel quantum-gravitational form of matter with an extensive phase space.

In this paper, we aim to address the question of novel horizon-scale physics from a classical perspective, assuming that if $\cO(1)$ corrections appear at a scale vastly larger than the Planck and string scales, they might impact classical dynamics. This is also motivated by recent results \cite{Mathur:2023uoe} showing that any ultra-compact classical spacetime structures developping a region of high redshift are governed by the same thermodynamic properties as a black hole. While lacking singularities or horizons to prompt a breakdown in the classical picture, these entities still require for the emergence of new gravitational degrees of freedom, describing new physics in strong-gravity environments. 

Our focus is on electromagnetism when gravity is non-negligible and electromagnetic degrees of freedom at the horizon scale. Electromagnetic degrees of freedom, associated with brane dynamics in string theory, have been previously theorized as key ingredients for describing the extensive phase space within black hole microstructures, particularly for supersymmetric solutions \cite{Strominger:1996sh,Gibbons:2013tqa,Bena:2013dka,Bena:2022rna}. An intriguing question, arising from these constructions, when they admit a classical description in supergravity, is: how can a widespread electromagnetic structure of the size of a black hole horizon resemble infinitesimally-closely a single-center solution?

In flat space, an extended charge distribution typically leads to an infinite series of electromagnetic multipole moments, significantly distinguishing it from a simple monopole. We will explore to what extent this property is altered when gravity is introduced and shed light on the physical mechanism behind this modification, which we name \emph{electromagnetic entrapment.} Electromagnetic entrapment is the characteristic of any charge configuration to generate an electromagnetic field with all multipole moments being infinitesimal, except the monopole term, when they are in a region of high redshift. This region of high redshift can be induced either by the gravitational backreaction of the charges themselves or by a fixed gravitational background, such as a black hole or a neutron star.

Electromagnetic entrapment suggests the emergence of novel degrees of freedom at the horizon scale, allowing nontrivial spacetime structures supported by electromagnetic flux to nucleate there and be indistinguishable from each other just outside the high-redshift region. We first illustrate this principle via probe derivation in curved spacetime before explicitly constructing these spacetime structures using the electrostatic Ernst formalism in four dimensions and in five dimensions with one direction being compact.

The spacetimes correspond to neutral bound states of extremal black holes held apart by struts in four dimensions or distributed along a smooth topological bubble in five dimensions. We demonstrate that they admit an entrapment limit where they are indistinguishable from a neutral black hole with a singular horizon, which we refer to as a \emph{distorted Schwarzschild.} They only differ in the near-horizon region where the entrapped electromagnetic flux begins to manifest and characterizes each bound state.

The five-dimensional geometries are particularly interesting as they involve only physical sources: extremal black holes and a vacuum bubble. We show that they meet many criteria to describe the novel gravitational phase of matter expected to emerge at the horizon scale. First, they are confined to exist at this scale and are as compact as a distorted Schwarzschild black hole. Second, like black holes, they grow in size as the Newton constant increases, which is a very unusual feature for any form of matter in gravity. Finally, they span an extensive phase space for which the entropy scales as $M^2$ like the Schwarzschild entropy.

Thus, electromagnetic entrapment could be a key feature to classically capture the new physics that should emerge at the horizon of black holes, as it allows the extraction of novel horizon-scale structures in classical theories of gravity that replace the vacuum region at the horizon and provide a description of black hole microstructures.

Section \ref{sec:probe} illustrates the concept of electromagnetic entrapment through probe derivations, demonstrating the effect of highly-redshifted spacetimes on electromagnetic fields generated by test particles. In section \ref{sec:GravAndElecPot}, we extend to neutral configurations of self-gravitating charges, revealing their ability to entrap their own electromagnetic field as they reach a critical size associated with their ``Schwarzschild radius''. Section \ref{sec:DisSchw} introduces a black hole solution in vacuum with a singular horizon, named a distorted Schwarzschild black hole. Finally, section \ref{sec:EntrapLimGR} analyzes neutral bound states of extremal black holes in four and five dimensions, and investigates their electromagnetic entrapment. Section \ref{sec:Conclusion} concludes with reflections and future outlook.

\noindent The reader interested in additional details on the static Ernst formalism in four and five dimensions can refer to Appendix \ref{app:Ernst}.  Appendix \ref{app:DisSchw} analyzes the gravitational signatures of the distorted Schwarzschild geometry, while Appendix \ref{app:TwoBH4d} provides a metric analysis of the black-hole bound states constructed in this paper

\subsection{Summary of results}

Our study begins with a review of known electrostatic and magnetostatic properties in black hole backgrounds. The electromagnetic fields generated by a test charge around a Schwarzschild black hole were studied in \cite{Copson1928ElectrostaticsIA,1968CMaPh...8..245I,1971JMP....12.1845C,PhysRevD.8.3259,Linet:1976sq,Garfinkle_2021}. Notably, as a particle approaches the horizon, the fields become nearly identical, except in the immediate vicinity of the particle, to those generated by the same charge at the center of the black hole, despite the considerable distance from the center. Consequently, the intense gravitational field naturally results in the \emph{vanishing of all electromagnetic multipole moments except the monopole term}. This result will be extended to arbitrary spacetime structures exhibiting a region of high redshift in an upcoming paper \cite{EntrapmentProbe}, a derivation that we summarize in the present paper.

\begin{figure}[t]
\begin{center}
\includegraphics[width= \textwidth]{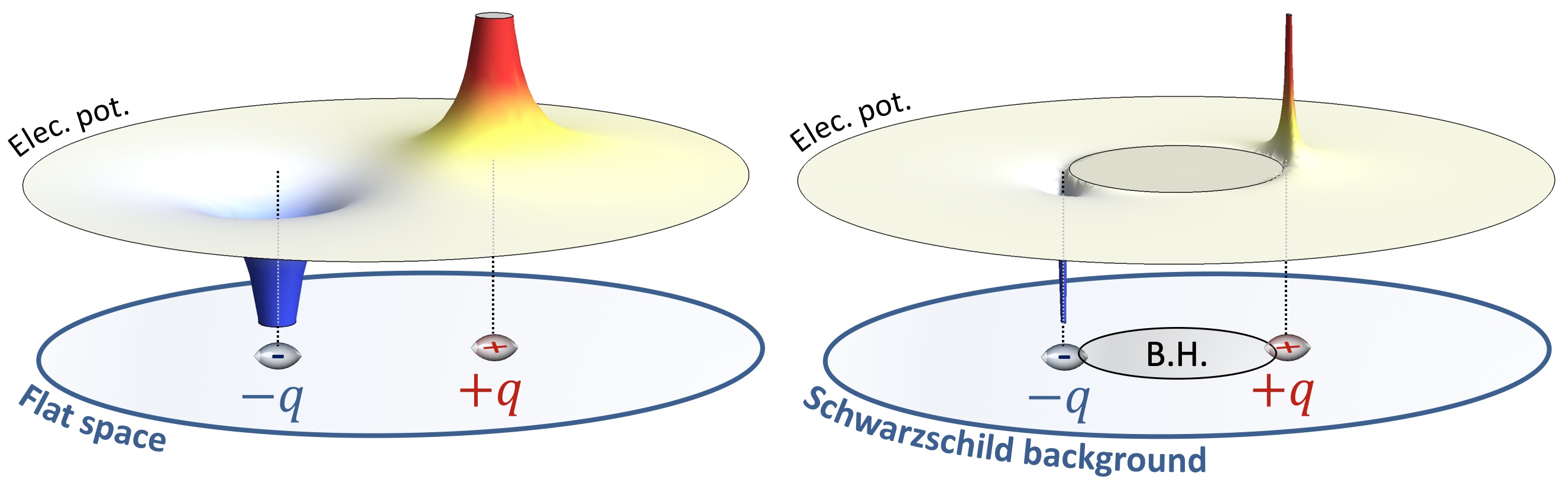}
\caption{Electric potential generated by two equal charges of opposite sign in flat space (left-hand side) and in the near-horizon region of a Schwarzschild black hole (right-hand side). The charges are separated by the same distance in both cases. For the black hole, the charges are positioned diametrically opposite and are situated at a distance of $0.2M$ from the horizon. Due to electromagnetic entrapment, the electric field becomes highly localized around the sources, causing all electric multipole moments to vanish.}
\label{fig:EMEntrapIntro}
\end{center}
\end{figure}

A noteworthy consequence of this property arises when considering not just one charge but arbitrary \emph{neutral} configurations of charges. As these configurations enter a high redshift region or a near-horizon region, all charges are virtually ``transferred'' at the center of spacetime. Thus, the electromagnetic fields become increasingly localized around the sources, vanishing elsewhere, with all electromagnetic multipoles disappearing. This suggest a significant reduction in electromagnetic interaction in this region, opening up a large phase of electromagnetic degrees of freedom without impacting the physics just outside the high-redshift region. In contrast to flat space, where a nontrivial charge configuration necessarily triggers large-scale effects in terms of dipole, quadrupole, etc., electromagnetism in a strong-gravity environment renders all charge distributions indistinguishable, with fields entirely entrapped in the high-redshift region. This phenomenon is termed \emph{electromagnetic entrapment}. In Fig.\ref{fig:EMEntrapIntro}, we have represented the effect of electromagnetic entrapment on the electric potential produced by two equal charges of opposite sign in flat space (left) and in the near-horizon region of a Schwarzschild black hole (right).

Electromagnetic entrapment can have far-reaching implications in various areas of gravitational and electromagnetic dynamics. It could affect the physics of ultra-compact gravitational objects like neutron stars and significantly modify nuclear and molecular physics in high-redshifted spacetime regions compared to flat space. This paper specifically focuses on its implications for a new black hole physics.

One significant limitation of the probe derivation is that the gravitational field is produced by a fixed background, whereas the electromagnetic field results from probe charges. To investigate the impact of electromagnetic entrapment in a new horizon-scale physics, it is essential to derive spacetime structures of \emph{self-gravitating charged objects} in which both the electromagnetic and gravitational fields originate from the same sources. For this purpose, we use the static Ernst formalism, enabling the extraction of integrable systems in four-dimensional Einstein-Maxwell theory \cite{Ernst:1967wx,Ernst:1967by}, as well as in higher-dimensional theories of gravity  \cite{Bah:2020pdz,Bah:2021owp,Heidmann:2021cms}. The solution corresponding to $N$ arbitrary charged masses on a line has been derived in \cite{NoraBreton1998}. This solution inevitably leads to bound states of Reissner-Nordstr\"om black holes held apart by struts, i.e., strings with negative tension, in four dimensions \cite{Alekseev:2007re,Alekseev:2007gt,Manko:2007hi,Manko:2008gb}. However, in five dimensions, these sources can give rise to diverse gravitational entities, such as regular bound states of black holes and Kaluza-Klein bubbles, as well as smooth horizonless topological solitons \cite{Bah:2020pdz,Bah:2021owp,Bah:2021rki,Heidmann:2021cms,Bah:2022yji,Bah:2022pdn,Heidmann:2022zyd,Bah:2023ows}.

In this paper, we focus on solutions resulting from \emph{extremal} charges for simplicity.\footnote{This does not imply a reduction to the linear Majumdar-Papapetrou (or BPS) multicenter solutions \cite{PhysRevMajumdar,Papaetrou:1947ib}, as these sources can carry charges with different signs (so BPS and anti-BPS sources).} By finding the gravitational and electromagnetic fields produced by neutral configurations, we demonstrate that all configurations possess a critical and ultra-compact size where the gravitational field intensifies, leading to an increasing redshift, and the electromagnetic fields become entrapped near the sources, vanishing elsewhere. Consequently, these solutions are indistinguishable from a vacuum solution that develops a region of infinite redshift --- a black hole.

The black hole toward which all configurations converge in their entrapment limit is not a Schwarzschild black hole, as one might expect, but a deformed version of Schwarzschild that we name a \emph{ distorted Schwarzschild black hole}. It is a solution of vacuum general relativity corresponding to a Schwarzschild black hole with a nontrivial deformation along the two-sphere. Although the solution has a horizon, it is singular due to the S$^2$ deformation. Nevertheless, it still possesses a finite horizon area identical to Schwarzschild, $16 \pi M^2$, suggesting a description in terms of a similar phase space of microstates. We also analyze its gravitational signatures, such as its photon ring structure, gravitational multipoles, and physical size, demonstrating that they closely resemble those of the Schwarzschild black hole.

\begin{figure}[t]
\begin{center}
\includegraphics[width= 0.75\textwidth]{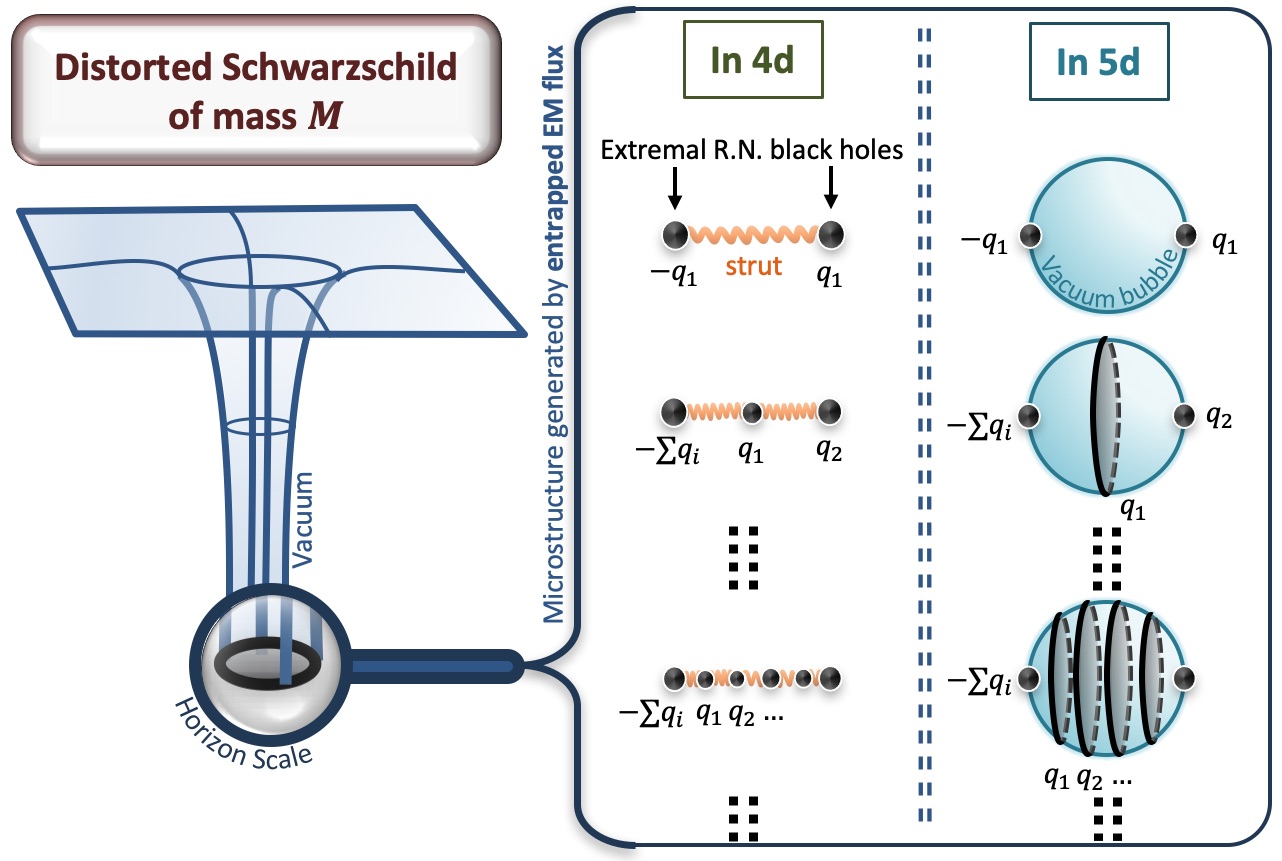}
\caption{The horizon-scale structures describing classically the horizon of a distorted Schwarzschild black hole from electromagnetic entrapment. The geometries correspond to neutral bound states of extremal black holes either separated by struts in four dimensions or distributed along a KK bubble in five dimensions. Each bound state on the right has an entropy that scales like $M^2$, akin to a Schwarzschild black hole.}
\label{fig:IntroFig}
\end{center}
\end{figure}

We analyze the spacetime structures generated by neutral bound states of extremal charges by incorporating them into Einstein-Maxwell theory in four dimensions. The solutions correspond to neutral and axially-symmetric configurations of extremal Reissner-Nordstr\"om black holes held apart by struts. Through a meticulous study of the solution consisting of two extremal black holes with opposite charges, we demonstrate that it admits an entrapment limit as their geometric distance $\ell$, approaches the value of the total mass $M$. In this regime, it is indistinguishable from a distorted Schwarzschild black hole, except in an infinitesimal region of size $\ell-M$ around its horizon, where the vacuum region of the latter is replaced by the bound state structure generated by intense and entrapped electromagnetic flux. Remarkably, the entrapment limit constrains the bound state in such a way that its entropy\footnote{The entropy of the bound state is determined by the entropies of both extremal black holes.} is half the entropy of the distorted Schwarzschild black hole computed from the Bekenstein-Hawking formula.\footnote{Note that this aligns with the Bekenstein bound \cite{PhysRevD.23.287}, where the entropy of a black hole serves as an upper limit to the entropy of any system within an equivalent area.} Furthermore, we show that the same properties apply to generic configurations of $N$ extremal black holes on a line: they ``replace'' the horizon of the distorted Schwarzschild black hole in their entrapment limit and have an entropy of the same order as half its entropy (see Fig.\ref{fig:IntroFig}).

Even if the four-dimensional solutions are relevant to illustrate the novel degrees of freedom at the horizon scale, they lack physicality due to the presence of struts. Following the mechanism in \cite{Elvang:2002br,Bah:2021owp}, we replace the strut with a Kaluza-Klein bubble in five dimensions so that the geometries correspond to physical bound states of extremal black holes on a KK bubble.

The regularity of the bubble  completely fixes the configurations in terms of their internal charges, total mass $M$ and the S$^1$ radius $R_y$. For example, the total size of the configurations $\ell$ is fixed such that $\ell-M = \cO(R_y)$. This yields significant physical implications. First, it forces the bound states to grow as the energy increases like black holes. Second, it forces them to exist only at the horizon scale as a novel form of gravitational matter describing new physics at the horizon. Indeed, macroscopic solutions, with $M \gg R_y$, exist inevitably in their entrapment limit with $\ell \sim M$. Our analysis demonstrates that all macroscopic solutions are indistinguishable from a distorted Schwarzschild black hole, except within an infinitesimal region around its horizon set by the extra-dimension radius (see Fig.\ref{fig:IntroFig}). Finally, it discretizes the phase space of solutions and each bound state has an entropy that scales as $M^2$, akin to the Schwarzschild black hole.

These constructions demonstrate the existence of a large phase space of electromagnetic degrees of freedom emerging at the horizon scale, where we have replaced the horizon of a neutral black hole with extremal black holes. Ultimately, to construct effective microstates, we need to resolve the extremal black holes in the bound states in terms of smooth, horizonless and topologically-nontrivial geometries supported by the same amount of electromagnetic flux. Similar to the microstate geometry program, we do not expect all microstates to admit such a geometrical description, but we can still build coherent (and surely atypical) states in the phase space. We conclude the discussion in this paper by demonstrating how to construct such states, called \emph{topological solitons}.

\section{Electromagnetic entrapment: a probe derivation}
\label{sec:probe}

The study of electromagnetic fields around black holes has a long history, with significant contributions from pioneer works  \cite{Copson1928ElectrostaticsIA,1968CMaPh...8..245I,1971JMP....12.1845C,PhysRevD.8.3259,Linet:1976sq,Garfinkle_2021}. These analyses demonstrate that charges near a black hole horizon induce the same field as if the charges were located at the center of the black hole, despite the considerable distance between these two regions. This result suggests different electromagnetic dynamics at the horizon scale compared to flat space.

We aim to revisit and describe these dynamics, using similar probe derivations as those found in \cite{Copson1928ElectrostaticsIA,1971JMP....12.1845C,PhysRevD.8.3259,Linet:1976sq,Garfinkle_2021}. In this section, we  provide only a concise overview of the main concepts and keep the discussion brief, as it will be the central topic of an upcoming paper \cite{EntrapmentProbe}.

Our focus will be on a key distinction that we name \emph{electromagnetic entrapment}. This phenomenon can be summarized as follows: a neutral distribution of charges in a region of high redshift will generate electromagnetic fields that are infinitesimal just outside this region and highly localized inside.

We start this section by revisiting electrostatics in flat space, illustrating that electrostatic degrees of freedom within a gravity-free environment are highly limited due to an infinite series of electric multipoles that differentiate between various configurations. Subsequently, we demonstrate how this characteristic is altered in a black hole background by reviewing the work of \cite{PhysRevD.8.3259,Linet:1976sq}. Furthermore, we offer a preview of an upcoming paper \cite{EntrapmentProbe}, demonstrating that this modification arises from a high-redshift region in the spacetime and is not solely tied to the existence of a horizon and a black hole. This section concludes by extending the discussion from electric to electromagnetic entrapment, generalizing to magnetic charge distributions.

\subsection{Electrostatics in flat space}

In flat space, a charge distribution consisting of $N$ discrete electric charges with values $q_i$, located at points $\vec{R}_i$ in space, generates an electric potential, $A$, given by:
\begin{equation}
    A(\vec{r}) \= \sum_{i=1}^N \frac{q_i}{|\vec{r}-\vec{R}_i|}.
\end{equation}
This potential can be expanded at large distances as
\begin{align}\label{multipoleexp}
     A(\vec{r}) \=  \frac{Q}{|\vec{r}|} \+ \frac{\vec{\cJ_1} \cdot \vec{r}}{|\vec{r}|^3} \+ \frac{\vec{r}\cdot \cJ_2 \cdot\vec{r}}{2|\vec{r}|^5} \+ \cdots\,,
\end{align}
where 
\begin{align}
    Q  =\sum_{i=1}^N q_i\;,\qquad \vec{\cJ_1} = \sum_{i=1}^N q_i \vec{R}_i\;,\qquad (\cJ_2)_{mn} = \sum_{i=1}^N q_i(3 R_{im}R_{in} - \delta_{mn}R_i^2)\;,
\end{align} represent the monopole, dipole, and quadrupole terms, respectively. Therefore, the configuration naturally triggers electric multipole moments that scale as:
\begin{equation}
    \cJ_l \,\sim\, q \,a^l \,.
    \label{eq:MultipolesFlat}
\end{equation}
where $a$ is the characteristic size of the configuration. These multipole moments are physical quantities that characterize how the charges are distributed in space and significantly influence the behavior of the electric field. Conversely, if one restricts these multipoles up to a certain order, it imposes strict constraints on how charges can be arranged.

Consider a scenario with a neutral configuration of $N$ charges spread out on a sphere with a radius $L$. This configuration has $3(N-1)$ internal degrees of freedom. Moreover, the $k^\text{th}$ multipole moment consists of $\frac{(k+2)(k+1)}{2}$ independent components. If the $n$ first multipole moments are forced to vanish, it ends up creating $\frac{(n+3)(n+2)(n+1)}{6}$ constraints on how these charges can be distributed. For two point charges, $q_1$ and $q_2$, positioned at coordinates $\vec{R}_1$ and $\vec{R}_2$ respectively, the configuration possesses three degrees of freedom. However, the conditions that both the total charge and the dipole moment vanish require that $q_1=-q_2$ and $\vec{R}_1=\vec{R}_2$, which leads to a vacuum configuration.

This highlights that in flat space, the number of available options for electrostatic setups significantly reduces when ensuring neutrality and demanding the electric field to vanish beyond a specific threshold, i.e. making multipoles up to a certain order vanish.

\subsection{Electrostatics in a Schwarzschild background}

\begin{figure}[t]
    \centering
    \includegraphics[width= \textwidth]{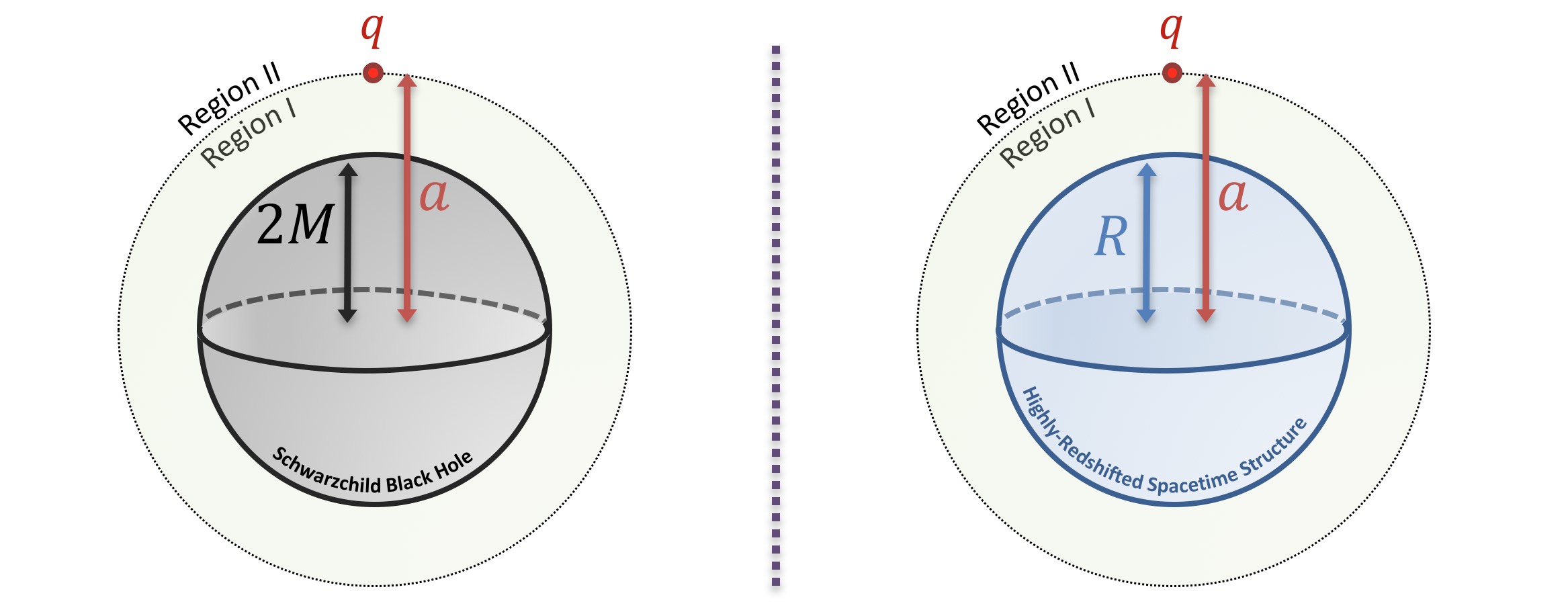}
    \caption{Illustration of a point charge $q$ situated in the vicinity of a Schwarzschild black hole (left) or a regular gravitational object with a radius $R$ (right). The charge is positioned at a distance $a$ from the object's center. The electric potential arising from the charge can be determined by considering two distinct regions, as depicted.}
    \label{fig:high red-shift}
\end{figure}

We now discuss how gravity alters electrostatic properties by revisiting known results regarding probe charges in black hole backgrounds. We consider a Schwarzschild background of mass $M$ with a static charge $q$ at a specific radius $r=a$, where $a=2M(1+\epsilon)$ and $\epsilon>0$ (see the left-hand side of Fig.\ref{fig:high red-shift}). The electric potential that satisfies Maxwell equations for such a probe configuration has a closed-form expression, provided in \cite{Copson1928ElectrostaticsIA,Linet:1976sq}. Nevertheless, we choose to follow the approach outlined in \cite{1971JMP....12.1845C,PhysRevD.8.3259}, emphasizing the multipolar structure of the electric potential.

The computation consists of expanding the electric potential in terms of Legendre polynomials and solving Maxwell's equations in curved spacetime in two regions - above and below the source (see Fig.\ref{fig:high red-shift}). After enforcing regularity at the horizon and ensuring an appropriate fall-off behavior at infinity, the electric potential takes the following form:
\begin{equation}
A(r,\theta) \= \begin{cases} 
\frac{a_0}{2M}+\sum_{l\geq 1} \frac{a_l}{r} \,u_l(r)\, P_l (\cos \theta)\,,\qquad &r < a\,,
\\
\frac{b_0}{r}+\sum_{l\geq 1} \frac{b_l}{r}\, u_l(r) \left( \int^\infty_r \frac{dr'}{2M\,u_l(r')^2}\right)\, P_l (\cos \theta)\,,\qquad &r \geq a\,,
\end{cases}
\end{equation}
where $u_l(r) = \frac{r}{2M}\,{}_2 F_1 \left(-l, 1+l,2,\frac{r}{2M}\right)$ and $(a_l,b_l)$ are constants with the following monopole term:
\begin{equation}
b_0 \=  q \,.
\end{equation}
All the higher multipole coefficients $a_l$ and $b_l$ are fixed by imposing continuity at the charge radius, $r=2M(1+\epsilon)$, and introducing a discontinuity in the derivative to enforce that the flux integral vanishes in Region I:
\begin{equation}
b_l = \epsilon \, c_l\,,\qquad l\geq 1\,, \qquad \text{where }c_l \underset{\epsilon \to 0}{=} \,\cO(1).
\end{equation}

\begin{figure}[t]
\begin{center}
\includegraphics[width= \textwidth]{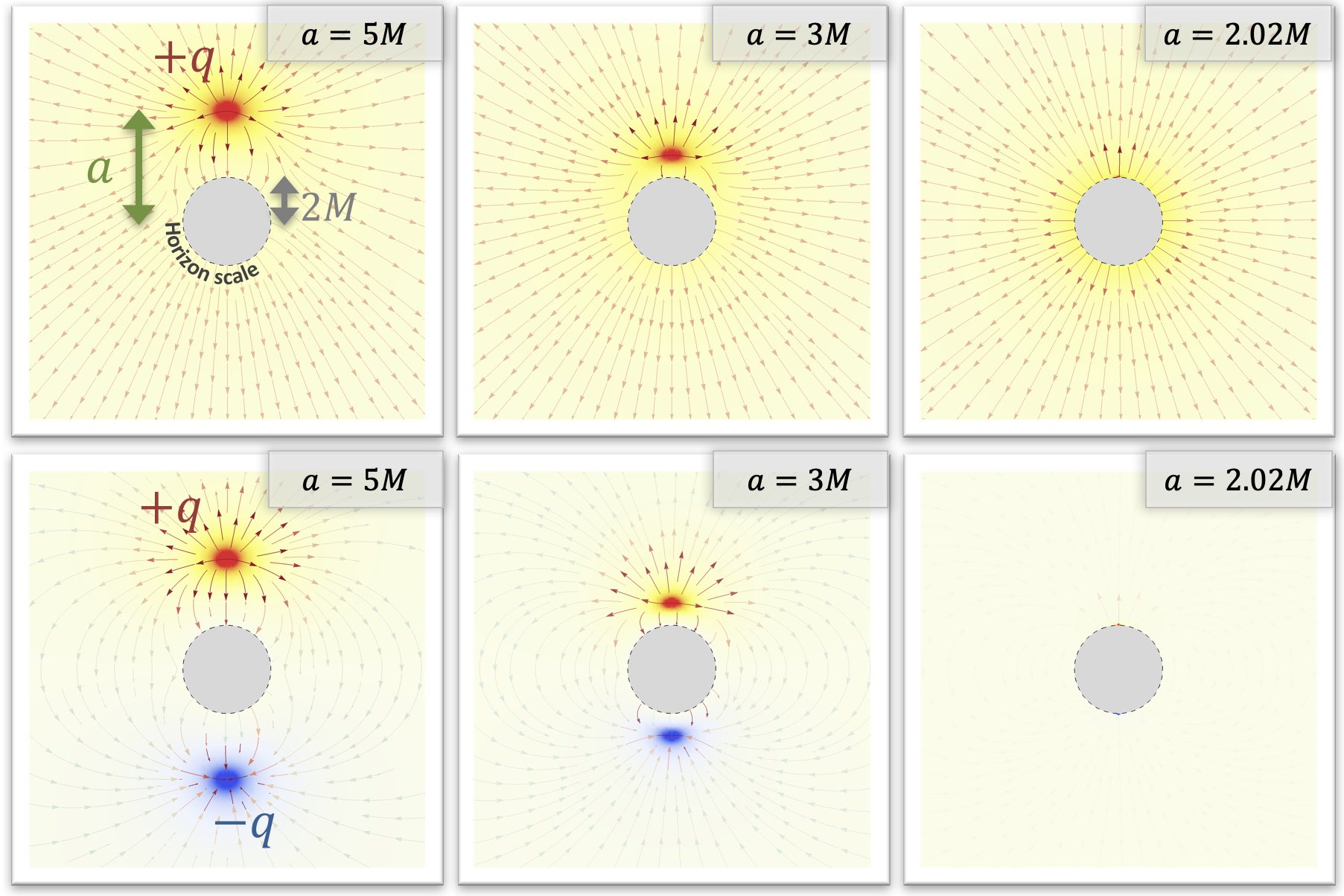}
\caption{The lines of force and electric potential produced by test charges in a Schwarzschild background and for different distances to the black hole. The first row corresponds to a single charge, and, as it approaches the horizon, the electric field becomes identical to a monopole at the center of the black hole. The second row corresponds to two test particles of opposite charges, and as the charges approach the horizon, the electric field gets entrapped in the near-horizon region with the field being predominantly zero elsewhere. For the second row, the opacity of the lines of force has been adjusted based on the intensity of the electric potential.}
\label{fig:ElecPotProbe}
\end{center}
\end{figure}

Therefore, the electric field generated by a charge positioned at a radius $r=2M(1+\epsilon)$ in a Schwarzschild background behaves as if it were emanating from a charge at the center of the black hole when $\epsilon\ll 1$. The first row in Fig.\ref{fig:ElecPotProbe} illustrates this property by representing the electric potential as well as the lines of force as the test charge approaches the horizon.

It is important to mention that, despite all $b_l$ vanishing for $l\geq 1$, the $a_l$ remain finite and scale like their flat-space expectation: $a_l = \cO (q \,a^l)$. As such, the field in the inner region keeps track of the characteristics of the charge distribution and differs from a monopole potential, but these characteristics remain completely entrapped there and are washed out outside the near-horizon region. 
\vspace{0.1cm}

As a corollary, any static distribution of $N$ charges $q_i$ near the horizon, within a radial range $2M<r<2M(1+\epsilon)$ where $\epsilon\ll 1$, produces an electric potential outside the source given by:
\begin{equation}
A \,\sim\, \frac{\sum_{i=1}^N q_i}{r} \,,\qquad r>2M(1+\epsilon)\,.
\end{equation}
The potential starts to differ in the near-horizon region, where it strongly localizes at the sources.

This property has important consequences that we can illustrate by reconsidering our simple configuration with two charges. Take two opposite charges randomly positioned on a sphere of radius $r=2M(1+\epsilon)$. All multipole moments vanish as $\cO(\epsilon)$, resulting in the electric field being zero throughout spacetime and highly localized near the sources. Such a property is achieved without imposing any constraints on the spatial degrees of freedom within the configuration. The second row of Fig.\ref{fig:ElecPotProbe} illustrates the entrapment of the electric field for such a configuration.

Therefore, from $\frac{(n+3)(n+2)(n+1)}{6}$ constraints to have the first $n$ electric multipoles vanish in flat space, we transition to a single neutrality constraint, $\sum_{i=1}^N q_i = 0$, and all multipoles become infinitesimal. The presence of the horizon triggers new properties that should translate into new physics at the horizon scale from emergent electromagnetic degrees of freedom.

\subsection{Electrostatics in a highly-redshifted background}

The primary concern arising from the previous statement is the suspicion that, even if a large phase space of charge distributions can emerge at the horizon scale, they will never remain static and will be absorbed immediately through the horizon. In this section, we present some results that will be published in a separate paper \cite{EntrapmentProbe}. We argue that this property does not depend solely on the presence of a horizon and a black hole, but rather on the existence of a region with high redshift. This will set the stage for future sections where the horizon will be replaced by regular structures that induce the same gravitational effect without the absorbing feature. 

We consider a static and spherically-symmetric background characterized by the metric:
\begin{align}
ds_4^2 = -V(r)^{-1} \,dt^2+V(r)\,dr^2+r^2\left( d\theta^2+\sin^2{\theta}\, d\phi^2\right),
\label{eq:metricProbe}
\end{align}
We aim to solve Maxwell's equations for a charge $q$ at a radius $r=a$ while keeping $V(r)$ arbitrary. We intentionally remain vague about the geometry, assuming it possesses a specific structure within a given radius $R$, with the condition that $V \to 1$ at large distances. The charge and background setup is illustrated on the right-hand side of Fig.\ref{fig:high red-shift}.

As $V$ is arbitrary, finding a closed-form solution is not feasible. However, by expanding in terms of Legendre polynomials as in the previous section, one can evaluate each harmonic using a WKB approximation. The detailed derivation will be available in \cite{EntrapmentProbe}. The multipolar expansion of the electric field also decomposes into inner and outer regions:
\begin{align}
    A(r,\theta) \sim \left\{
\begin{array}{ll}
      \frac{q}{a}+ \sum_{l\geq 1} \frac{q}{\sqrt{a r} (V(r)V(a))^{1/4}} \left(e^{\mathcal{I}_l(r) } -\kappa_l \,e^{-\mathcal{I}_l(r) }\right)\, \,P_l(\cos{\theta}), & \,\,r< a,
      \\
      \\
     \frac{q}{r} + \sum_{l\geq 1}\frac{(1-\kappa_l)\,q}{\sqrt{a r} (V(r)V(a))^{1/4}}  \,e^{- \mathcal{I}_l(r) }\,P_l(\cos{\theta}), & \,\,r\geq a, \\
\end{array} 
\right. \label{eq:EqPotWKB}
\end{align}
where we have defined the WKB integral
\begin{equation}
    \mathcal{I}_l(r) \equi \int_a^r \frac{\sqrt{l(l+1)V(r')}}{r'} dr'\,,
\end{equation}
and $\kappa_l$ is a constant bounded between $-1$ and $1$ that relies on the inner boundary condition and the specific nature of the spacetime structure (see \cite{EntrapmentProbe} for more details).

Similar to the black hole derivation, the electric potential develops a monopole charge $q$ at the center of spacetime. Additionally, through expansion at large distances, $e^{- \mathcal{I}_l(r)} \propto \left(\frac{r}{a} \right)^{-\sqrt{l(l+1)}}$, the higher-order multipole moments are given by:
\begin{equation}
    \cJ_l  \= \cO \left( \frac{q \, a^l}{V(a)^{\frac{1}{4}}}\right)\,, \qquad l\geq 1\,.
    \label{eq:MultipoleWKBApp}
\end{equation}
In contrast to flat space \eqref{eq:MultipolesFlat}, all multipole moments, excluding the monopole, are rescaled by the redshift at the location of the charge.

As a result, if the spacetime structure induces a high redshift at the charges locus so that $V(a) \sim V(R) \sim \epsilon^{-1}$, all multipole moments, except the monopole, scale as:
\begin{equation}
    \cJ_l = \cO(\epsilon^\frac{1}{4} \, q a^l ),
\end{equation}
resulting in the vanishing of all higher-order multipoles. Consequently, the electric potential outside the source, $r>a$, is then approximated by:
\begin{equation}
    A(r,\theta) \sim \frac{q}{r} \left( 1+ \cO \left(\epsilon^\frac{1}{4}\left(\frac{a}{r} \right)^l\right)\right)\,, \qquad r>a\,.
\end{equation}
Furthermore, the WKB formula \eqref{eq:EqPotWKB} shows that the electric potential deviates significantly from a monopole term in the region around the source for $|r-a |\ll \sqrt{\epsilon} \,a$. 

This computation illustrates that the dynamics of the electric field in a high-redshifted background follows the one derived for black holes despite the absence of a horizon and singularity. This result emphasizes that the crucial factor is the presence of a high redshift, ensuring that any charge approaching this region generates an electric field identical to a charge at the center of the spacetime. Deviations from this behavior only emerge in the high-redshift region near the charge. Therefore, neutral configurations of test charges will be entirely ``entrapped'' within the high-redshifted region, where $A(r,\theta)\sim 0$ above the sources and very intense around them.

\subsection{Generalization to magnetostatics}

Previous subsections have focused on electrostatic charge distributions. Through electromagnetic duality, similar results apply to magnetostatic configurations. Indeed, any electric charge distribution inducing an electric field strength, $F_e=-dA\wedge dt$, can be dualized into a magnetic charge distribution generating a magnetic field strength, $F_m=dH\wedge d\phi$, using the duality relation:
\begin{equation}
    \star F_e \= F_m \qquad \Rightarrow \qquad \begin{pmatrix}
        \partial_r H \\
        \partial_\theta H
    \end{pmatrix} 
   = \sin \theta \,
    \begin{pmatrix}
      V(r)\, \partial_\theta  A \\
       -r^2\, \partial_r A
    \end{pmatrix},
\end{equation}
where $\star$  denotes the Hodge star in the spacetime, $H$ represents the magnetic potential, and the right-hand side has been obtained from the metric \eqref{eq:metricProbe}. Hence, if the electric potential $A$ is entrapped, so is the magnetic potential $H$. This implies that it is possible to create a non-trivial charge distribution at the horizon scale, featuring not only vanishing higher multipoles of electric charges but also magnetic charges.

\section{Electromagnetic entrapment: self-gravitating charges}
\label{sec:GravAndElecPot}

In the preceding section, we have shown an intriguing interplay between gravity and electrostatic at the horizon scale and strong-gravity environments. However, a drawback  was that the gravitational effects were generated by a fixed background, whereas the electromagnetic field was induced by test charges without gravitational backreaction.

To illustrate the intricate connection between electric (magnetic) flux and gravity, it is essential to consider self-gravitating configurations of charges and investigate whether nontrivial entrapment effects emerge as the geometry approach their ``Schwarzschild radius''. However, this is a highly challenging task, as it involves solving Einstein-Maxwell field equations, which are notoriously more complex than Maxwell's equations in curved spacetimes.

Fortunately, a solution corresponding to an axisymmetric configuration of $N$ massive electric charges has been derived in \cite{NoraBreton1998} using the Ernst formalism and Sibgatullin's method \cite{Manko_1993}. While this solution is a genuinely intricate solution of Einstein-Maxwell equations, its reduction to a system of two massive charges is considerably more manageable \cite{Alekseev:2007re,Alekseev:2007gt,Manko:2007hi,Manko:2008gb,Bah:2022yji,Bah:2023ows}.

In this paper, we restrict our analysis to configurations where the sources are \emph{extremal.} This does not imply a reduction to the linear Majumdar-Papapetrou (or BPS) multicenter solution \cite{PhysRevMajumdar,Papaetrou:1947ib}, as these sources can carry charges with different signs. The complete solutions corresponding to $N$ extremal charges or two extremal charges on a line are detailed in Appendix \ref{app:NExtSources} and \ref{app:TwoExtSources} respectively.

In this section, our focus will be on the gravitational and electric potentials. The analysis of the spacetime structures are deferred to later sections. This approach allows us to emphasize the electromagnetic entrapment more effectively as the configurations become more compact and begin to produce high redshifts. We will begin by studying the two-charge configuration before discussing the generalization to $N$ charges. At the end of the section, we will see how this feature extends to magnetic fields by electromagnetic duality.

\subsection{Two extremal charges}
\label{sec:TwoExCharges}

We consider a configuration consisting of two extremal point charges of opposite electric charges $\pm q$ and total mass $M$, separated by a geometric distance $\ell$ (see Fig.\ref{fig:TwoChargeConv}). Further details regarding the complete solution can be found in Appendix \ref{app:TwoExtSources} and section \ref{sec:BSTwoBH4d}. Currently, our focus is on the electric and gravitational potentials generated by this system. The internal charges are constrained by the extremality bound (this will be more precisely derived in section \ref{sec:BSTwoBH4d}):
\begin{figure}[t]
\begin{center}
\includegraphics[width= 0.5 \textwidth]{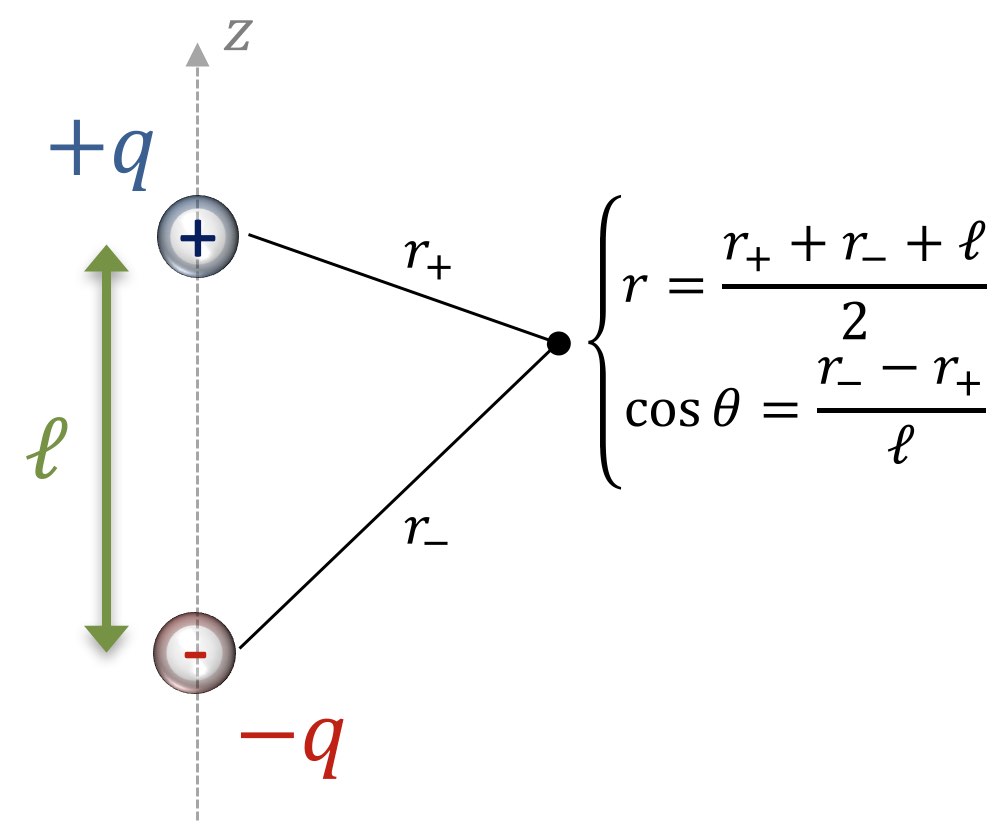}
\caption{Configuration of two extremal point sources of opposite charges and the spherical coordinates centered around it.}
\label{fig:TwoChargeConv}
\end{center}
\end{figure}
\begin{equation}
q = \frac{M}{2} \,\sqrt{\frac{\ell+M}{\ell-M}}\,.
\end{equation}
The reality of the solution necessitates that
\begin{equation}
\ell > M,
\end{equation}
which prevents the sources from being in close proximity to each other. Additionally, it is important to note that the BPS bound, $q=\frac{M}{2}$, is achieved when the charges are infinitely separated, i.e., as $\ell\to \infty$. However, at finite $\ell$, we have $q > \frac{M}{2}$ due to a portion of the individual masses being used as binding energy for the bound state.

The gravitational and electric potentials are denoted by $V$ and $A$, respectively.\footnote{We define the gravitational potential simply as the redshift $V\equiv -g^{tt}$ in four dimensions as for \eqref{eq:metricProbe}.} They are expressed as follows:
\begin{align}
V & = \left(1+\frac{2M(2r+M-\ell)}{(2r-\ell)^2-\ell^2\cos^2\theta- M^2 \sin^2 \theta}\right)^{2}, \quad A = \frac{4M\sqrt{\ell^2-M^2}\,\cos\theta}{(2r+M-\ell)^2-(\ell^2-M^2)\cos^2\theta} \,,\nonumber
\end{align}
where $(r,\theta)$ represent the spherical coordinates centered around the configuration, as illustrated in Fig.\ref{fig:TwoChargeConv}, with the condition $r\geq \ell$. The charges are located at $r=\ell$ and $\theta=0$,$\pi$.

The large $r$ expansion of the electric potential indicates that the configuration is neutral, with an electric dipole of $M \sqrt{\ell^2-M^2}$ and higher-order multipoles proportional to $\left(\ell^2-M^2 \right)^{n+\frac{1}{2}}$. \\

From the expression of the gravitational potential, it is clear that the configuration admits an ultra-compact limit where the surface at $r=\ell$ approaches an infinite redshift ($V\to \infty$) as $\ell\to M$. Consequently, in this limit, the two-charge configuration reaches its horizon scale.

Similar to the probe derivation, the electric potential tends to zero in this limit, leading to the vanishing of all electric multipoles. This phenomenon implies the entrapment of the electromagnetic flux near the sources. Specifically, for $\ell = M(1+\epsilon)$, when $r> M(1+\cO(\epsilon))$, we have:
\begin{equation}
V \sim \left( 1-\frac{\ell}{r}\right)^{-2}\,,\qquad A\sim 0\,,
\label{eq:LimGravElecPot}
\end{equation}
and the solution effectively corresponds to a vacuum solution just above the sources.

Despite the electric potential becoming of order $\epsilon$ everywhere, the electric flux does not vanish the same way. It becomes highly localized at the extremal sources. This entrapment will be better illustrated when spacetime structures are analyzed in section \ref{sec:BSTwoBH4d}. At this point, one can reasonably argue that the flux becomes highly localized as its integral close to the sources yields the charges $\pm q \sim \pm \frac{M}{\sqrt{2\epsilon}}\,\gg M$.

\begin{figure}[t]
\begin{center}
\includegraphics[width= \textwidth]{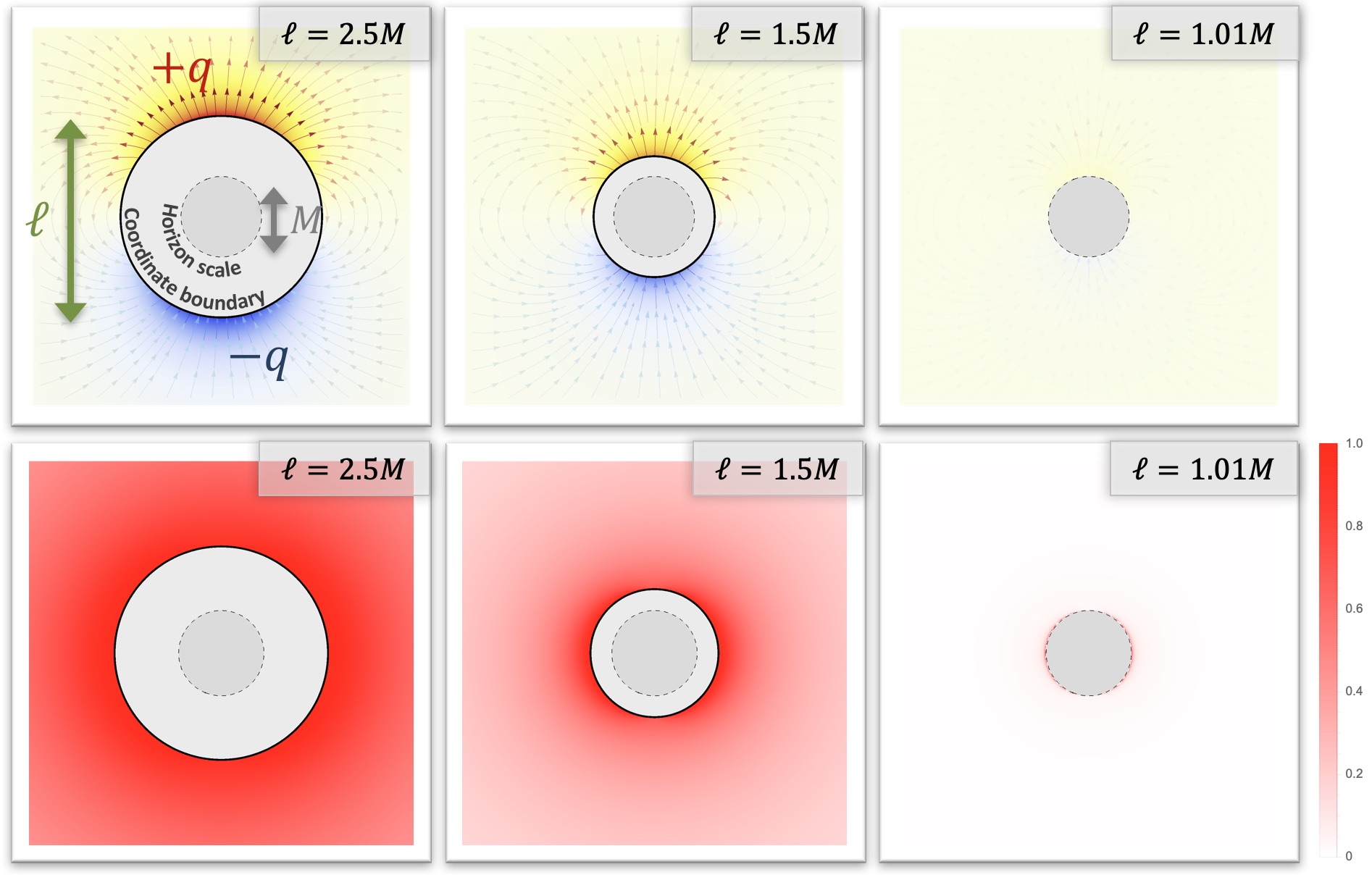}
\caption{The entrapment limit of a neutral configuration consisting of two massive charges for three different configuration sizes, $\ell$, in terms of the total mass $M$. The first row is dedicated to the electric field, depicting the lines of force and electric potential produced by the configuration. The second row illustrates the gravitational potential, indicating the deviation between the gravitational potential and the vacuum potential to which it converges, $\Delta V$ \eqref{eq:DeviationGravPot}. The $r=\ell$ circle is the boundary of the coordinate system, and the $r=M$ circle corresponds to the horizon-scale radius of the configuration. The opacity of the force lines has been adjusted according to the intensity of the electric potential. As $\ell$ decreases, the flux becomes more and more entrapped, and the local charges increase (as a function of the total mass), $q\approx 1.5M,\,2.2M,\,14.2M$.}
\label{fig:ElecFlux2charge}
\end{center}
\end{figure}

In Figure \ref{fig:ElecFlux2charge}, the gravitational and electric fields of the system are plotted to illustrate the entrapment of the flux as the configuration approaches its horizon-scale limit, $\ell \to M$.

The first row consists of the electric potential and the lines of force of the electric field. The second row represents the difference between the gravitational potential of the configuration and the vacuum potential derived from the entrapment limit \eqref{eq:LimGravElecPot}:
\begin{equation}
    \Delta V(r,\theta) \equi \left|\frac{V(r,\theta)-\left(1-\frac{\ell}{r}\right)^{-2}}{\left(1-\frac{\ell}{r}\right)^{-2}} \right|.
    \label{eq:DeviationGravPot}
\end{equation}
The figures clearly confirm the analytic arguments: as $\ell\to M$, the electric potential indeed gets entrapped in the region of high redshift and is almost zero everywhere, and the solution is indistinguishable from a vacuum solution given by the gravitational potential \eqref{eq:LimGravElecPot}.

The key insight derived from this analysis is that, even though nothing special occurs at the horizon, a distinct physics emerges at the horizon scale compared to flat space. When a neutral charge configuration approaches its ``Schwarzschild radius'', it becomes entirely invisible and indistinguishable from a vacuum solution right above where the sources are located, and becomes uniquely defined by a single parameter ($\ell$ or $M$), akin to black holes in vacuum. This implies that a substantial phase space of electromagnetic degrees of freedom should begin to manifest at the horizon scale. These degrees of freedom remain totally entrapped and highly localized in the region of high redshift without significantly altering the overall geometry. To validate the existence of this phase space, the next section will demonstrate that different charge distributions yield the same gravitational and electric potentials just outside the sources as \eqref{eq:LimGravElecPot}, while possessing distinctly different internal electromagnetic structures.

\subsection{$N$ extremal charges}
\label{sec:NExCharges}

We now consider a configuration of $N$ extremal charges on a line, for which the solution of the Einstein-Maxwell equations is detailed in Appendix \ref{app:NExtSources}. Dealing with this general solution poses two primary challenges. First, the gravitational and electric potentials, expressed in terms of matrix determinants in \eqref{eq:NExGravElecPot} and \eqref{eq:NExMatrixDet}, are intricate making direct analysis highly involved. Second, the solution is described using $2N$ parameters, but none of them correspond directly to the local charges. We have $\alpha_n$, where $n=1,\ldots,N$, corresponding to the charge positions on the symmetry axis, and $\beta_n$ are somehow related to the charges in a very indirect manner. Additionally, we introduce the overall size of the charge distribution and position the origin of the symmetry axis as follows:
\begin{equation}
    \ell \equi \alpha_N - \alpha_1\,, \qquad \sum_{n=1}^N \alpha_n =0\,.
    \label{eq:SizeConfAndOrigin}
\end{equation}

To emphasize the entrapment limit of the charge configuration, we adopt an indirect approach that avoids dealing with the intricate matrix determinants. Interested readers can find the details of the derivation in the Appendix \ref{app:NExtEntrap}. Here we summarize the main results.

First, we restrict the analysis on the symmetry axis, parametrized by a coordinate $z$, right above the sources $z\geq \alpha_N$. In this semi-infinite segment, the electric and gravitational potentials admit a much simpler form that allows analytic derivation. We showed in Appendix \ref{app:NExtEntrap} that neutral configurations of $N$ extremal charges admit an entrapment limit (at least on the symmetry axis) when the configuration reaches its ``Schwarzschild radius'', $\ell \to M$. More precisely, we found that the gravitational and electric potentials for $z> \alpha_N(1+\cO(\epsilon))$ become :
\begin{equation}
    V(\rho=0,z) \sim \left( 1-\frac{\ell}{z-\alpha_1}\right)^{-2}  \,,\qquad A(\rho=0,z) \sim 0\,. \label{eq:AxisDataApprox}
\end{equation}
where $\rho$ is the cylindrical radius measuring the distance to the symmetry axis and $\epsilon$ is an infinitesimal parameter associated to the difference $\ell-M$.

Therefore, similar to the two-charge system, the configuration of $N$ extremal charges converges toward a vacuum solution, at least on the symmetry axis, as it approaches an infinite redshift limit. Furthermore, the axis values of the vacuum solutions \eqref{eq:AxisDataApprox} correspond to the vacuum solution obtained from the entrapment of the two-charge system \eqref{eq:LimGravElecPot}. Thus, the configuration of $N$ extremal charges converges toward the exact same solution and is similarly indistinguishable (at least on the symmetry axis) from this solution, except in an infinitesimal region around the sources where the electric degrees of freedom used to produce the nontrivial charge distribution start to manifest. \\

We now aim to extend the entrapment derived on the symmetry axis to the entire space. The axis values reveal much more about the entire solution than one might initially anticipate. In fact, it has been demonstrated in \cite{Simon:1983kz,Sotiriou:2004ud} that all gravitational and electric multipole moments of an axisymmetric electrovacuum solution can be obtained by expanding these functions at large $z$. For instance, the total mass, the total charge, and the electric dipole moments are respectively (see Appendix \ref{app:NExtSources} for more details):
\begin{equation}
    M \=  -\sum_{n=1}^N \beta_n\,,\qquad Q = \sum_{n=1}^N f_n\,,\qquad \cJ = M Q + \sum_{n=1}^N f_n \beta_n\,.
    \label{eq:ConsChargeNex}
\end{equation}
where the absolute values of $f_n$ are given by
\begin{equation}
\begin{split}
    f_k^2 &\= (\beta_k-\alpha_k)^2 \,\left(\prod_{n\neq k} \frac{\beta_k-\alpha_n}{\beta_k-\beta_n}\right)^2\,.
    \label{eq:AxisDataN}
\end{split}
\end{equation}
The constraint of having a neutral configuration had simply required that:
\begin{equation}
    \sum_{n=1}^N f_n \=0.
    \label{eq:neutralityCond}
\end{equation}
As such, matching on the symmetry axis implies that the multipoles will be identical to the vacuum solution with $\epsilon$ corrections. Having the same multipole structure indicates that the solutions will match everywhere, except for a small region around the sources.

\begin{figure}[t]
\begin{center}
\includegraphics[width= \textwidth]{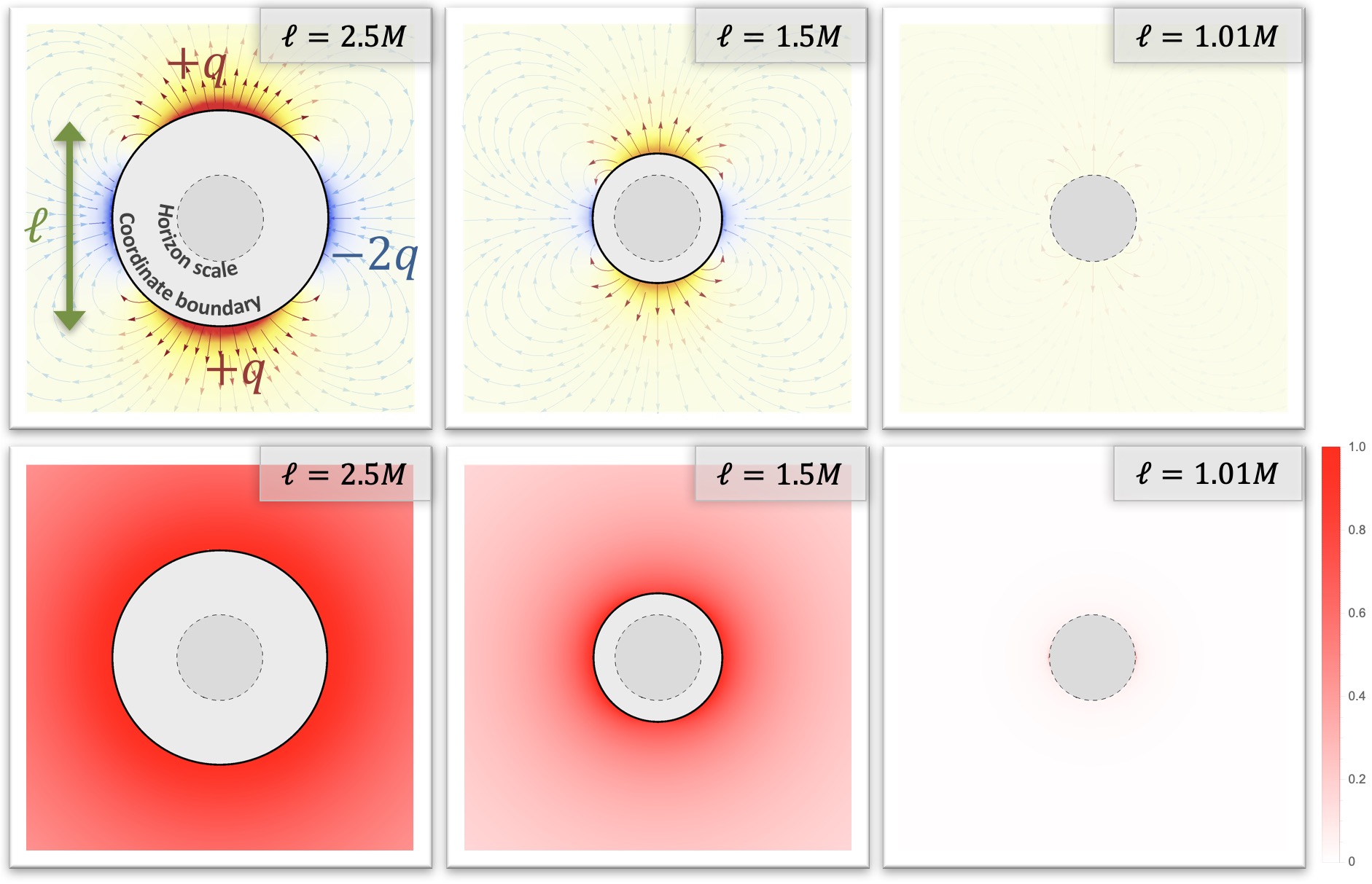}
\caption{The entrapment limit of a neutral configuration consisting of three point charges that has no electric dipole moment for three different configuration sizes, $\ell$, in terms of the ADM mass $M$. The first row is dedicated to the electric field, depicting the lines of force and electric potential produced by the configuration, with the same conventions as in Fig.\ref{fig:ElecFlux2charge}. The second row illustrates the gravitational potential, indicating the deviation between the gravitational potential and the approximate vacuum potential to which it converges, as given in \eqref{eq:DeviationGravPot}.}
\label{fig:ElecFlux3charge}
\end{center}
\end{figure}

Second, in Fig.\ref{fig:ElecFlux3charge}, we have plotted the gravitational and electric fields produced by a configuration of three extremal charges for three different configuration sizes that approach the entrapment limit $\ell\to M$. For simplicity, we have chosen the specific configurations where the electric dipole, $\cJ$ \eqref{eq:ConsChargeNex}, is strictly zero throughout the limit, which fixes the configurations in terms of their ADM mass, $M$, and their size, $\ell$, only. The fields are obtained from the generic expressions detailed in the Appendix equation \eqref{eq:NExGravElecPot}, and are plotted in the same spherical coordinates system as the previous two-charge configuration, $(r,\theta)$:
\begin{equation}
    \rho = \sqrt{r(r-\ell)}\sin\theta \,,\qquad z = \left( r- \frac{\ell}{2} \right) \cos \theta + \frac{\alpha_1+\alpha_3}{2}\,.
\end{equation}
The charges are located at the coordinate boundary $r=\ell$ for three discrete values of $\theta$, namely, $\theta=0,$ $\pi/2,$ and $\pi$. 

The first row shows the electric potential and the lines of force of the electric field. The second row corresponds to the difference between the gravitational potential and the vacuum potential derived from the entrapment limit of the two-charge configuration in \eqref{eq:DeviationGravPot}.
It is clear that as $\ell\to M$, the electric potential indeed vanishes, and the gravitational potential converges toward the vacuum one not only on the symmetry axis. This confirms that the entrapment observed on the symmetry axis extends to everywhere in spacetime.\\

In summary, we have demonstrated that arbitrary configurations of self-gravitating extremal point charges exhibit an entrapment limit, where the electric flux becomes completely localized in an infinitesimal region near the sources, accompanied by a redshift approaching infinity. Furthermore, these solutions approach the exact same vacuum solution characterized by a single parameter and a radius of infinite redshift, akin to a black hole in vacuum. This suggests that \emph{this vacuum solution possesses a microscopic description in terms of an extensive phase space of electric degrees of freedom that begin to manifest at the horizon scale}. Our results highlight the profound interplay between gravity and electromagnetism at the horizon scale, directly involving self-gravitating charged entities. 

\subsection{Magnetic configurations}

The static Ernst formalism explicitly exhibits electric-magnetic symmetry, implying that the dynamics of electric flux may also be applicable to magnetic flux. Any electric charge distribution inducing an electric field strength, $F_e=-dA\wedge dt$, can be dualized into a magnetic charge distribution generating a magnetic field strength, $F_m=dH\wedge d\phi$, using the duality relation (see Appendix \ref{app:ErnstEOM}):
\begin{equation}
    \star F_e \= F_m \qquad \Rightarrow \qquad \begin{pmatrix}
        \partial_\rho H \\
        \partial_z H
    \end{pmatrix} \= \rho \, V\, \begin{pmatrix}
      \partial_z A \\
    -\partial_\rho A
    \end{pmatrix},
    \label{eq:MagneticDualGen}
\end{equation}
where $\star$ denotes the Hodge star in the spacetime, $(\rho,z)$ are the cylindrical coordinates as previously introduced, and $V$, $A$, and $H$ represent the gravitational, electric, and magnetic potentials, respectively.

Based on this relation, it becomes apparent that the entrapment limit derived for $(A,V)$ also implies a similar entrapment limit for the magnetic dual, $(H,V)$. Hence, one can expect configurations involving magnetic charges to exhibit a limit where they reach their horizon scale and where the magnetic flux remains confined near the sources.

For instance, the solution associated with two extremal magnetic monopoles of opposite charges generates a magnetic field given by the following potential \cite{Bah:2022yji,Bah:2023ows}:
\begin{equation}
    H=\frac{2M \sqrt{\ell^2-M^2}(2r+M-\ell) \sin ^2 \theta}{(2 r-\ell)^2-\ell^2 \cos ^2 \theta-M^2 \sin ^2 \theta} .
\end{equation}
Similar to its electric counterpart, the magnetic field approaches $0$ for $r>\ell$ as the configuration approaches its horizon scale, $\ell\to M$, with all magnetic multipoles vanishing as the configuration. The field becomes increasingly localized around the sources, as indicated by
\begin{equation}
    dH |_{r\sim \ell} = \left[-\frac{4M(2\ell-(\ell-M)\sin^2\theta)}{(\ell-M)^\frac{3}{2} \sqrt{\ell+M}\,\sin^2\theta} \+ \cO(r-\ell) \right] \,dr \underset{\ell \to M}{\to} \infty .
\end{equation}
Consequently, the electric entrapment and the enhancement of electric degrees of freedom at the horizon scale naturally extend to magnetic charges and fields.

\section{The distorted Schwarzschild black hole}
\label{sec:DisSchw}

Drawing inspiration from the previous sections, one can anticipate the existence of ultra-compact spacetime structures supported by nontrivial electromagnetic flux, which exhibit an entrapment limit. In this scenario, these geometries are expected to closely resemble vacuum solutions, except for a small region near the sources where intense electromagnetic fluxes start to manifest. This entrapment should arise when these spacetime structures reach a significant redshift, ultimately approaching vacuum solutions characterized by a region of infinite redshift: a \emph{black geometry}. However, this  solution may not correspond to a standard black hole, as the horizon could be singular. For instance, the gravitational potential corresponding to the solution \eqref{eq:LimGravElecPot} is not the potential of a Schwarzschild black hole.

Before analyzing explicit ultra-compact structures supported by electromagnetic flux, it is crucial to investigate the vacuum limit toward which they converge. In this section, we introduce and analyze the \emph{distorted Schwarzschild black hole} corresponding to the gravitational potential \eqref{eq:LimGravElecPot}. As we will see, this solution lacks spherical symmetry, resulting in a nontrivial deformation of the two-sphere, with singular features at the horizon. Hence, this solution can be regarded as a deformed version of a Schwarzschild black hole, with its singularity shifted from $r=0$ to the horizon. However, it is worth mentioning that there is no need for concern about this singular horizon, as we will subsequently create nontrivial, regular structures supported by entrapped electromagnetic flux to replace it.

\subsection{Spacetime structure}

The solution referred to as a \emph{distorted Schwarzschild black hole} is a vacuum solution in four-dimensional general relativity. It is characterized by a single parameter, the ADM mass denoted as $M$. The metric is given by (in unit $G_4=1$)
\begin{equation}
ds_\text{dis-Schw}^2 = -\left(1-\frac{M}{r}\right)^2 dt^2+\frac{1}{1-\frac{M}{r}}\,\left[ \left(1+\dfrac{M^2 \sin^2\theta}{4r(r-M)}\right)^{-3} \left(\dfrac{dr^2}{1-\frac{M}{r}}+r^2 d\theta^2 \right) +r^2\sin^2\theta \,d\phi^2 \right].
\label{eq:DisSchw}
\end{equation} 
The solution is defined in the coordinate range $r\geq M$, and is asymptotic to flat Minkowski spacetime. For $r>M$, the spacetime is regular with a stretched S$^2$ topology. One can check that the $\phi$-angle degenerates smoothly at the poles, $\theta=0$ or $\pi$, when $\phi$ is $2\pi$ periodic. 

The locus $r=M$ corresponds to a horizon where the timelike Killing vector, $\partial_t$, shrinks. However, this locus is also a singularity due to the two-sphere deformation. More precisely, the local geometry, given by $r=M(1+\bar{r})$ with $\bar{r}\to 0$, is described by
\begin{equation}
    ds_\text{dis-Schw}^2 \sim -\bar{r}^2 dt^2 + \left(1+\frac{\sin^2 \theta}{4\bar{r}} \right)^{-3} \frac{M^2 d\bar{r}^2}{\bar{r}^2} + \frac{M^2}{\bar{r}} \left(\left(1+\frac{\sin^2 \theta}{4\bar{r}} \right)^{-3} d\theta^2 + \sin^2\theta \,d\phi^2 \right).
\end{equation}
At the poles of the S$^2$, the $(t,\bar{r})$ subspace describes an AdS$_2$ geometry while the S$^2$ blows up in size. Outside the poles, it is a singular horizon where the longitude direction $(\theta)$ shrinks and the latitude direction $(\phi)$ blows.

The topology of the two-sphere can be illustrated by deriving the geodesic distance from the North to South pole $\delta_{NS}$ and half the equator $\delta_E$, at $r=M$ and asymptotically.  We find
\begin{align}
\delta_{NS} &\underset{r\to M}{\sim} 4M\,,\qquad \delta_{NS} \underset{r\to \infty}{\sim} \pi r\,, \nn\\
\delta_{E} &\underset{r\to M}{\sim} \frac{\pi M}{\sqrt{1-\frac{M}{r}}}\,,\qquad \delta_{E} \underset{r\to \infty}{\sim} \pi r\,.
\end{align}

Despite the divergences at the horizon,  the S$^2$ area remains finite in the whole spacetime
\begin{equation}
\text{Area}(S^2) = 4\pi r^2 \left(1-\frac{M}{2r} \right)^{-2}\,.
\label{eq:AreaDisSchw}
\end{equation}
Remarkably,  the area of the horizon is identical to that of a Schwarzschild black hole with the same mass despite the nontrivial deformation: 
\begin{equation}
    \cA \= \cA_\text{Schw} \= 16 \pi M^2 \qquad \rightsquigarrow \qquad S = S_\text{Schw} \=4\pi M^2.
    \label{eq:AreaHorDis}
\end{equation}
At first glance, it seems incorrect to associate the horizon area with entropy, given the singularity at the horizon. However, in subsequent sections, we will show that the singular horizon can undergo a regular geometric transition, leading to spacetime structures consisting of regular sources with entrapped electromagnetic flux. These spacetime structures will provide a better definition of this area as the entropy of the singular black hole. In future research, it will be interesting to investigate whether other thermodynamic quantities, such as temperature, and thermodynamic laws, can also be derived for the distorted Schwarzschild black hole.

In summary, the distorted Schwarzschild black hole serves as an illustration of how the gravitational deformation of the S$^2$ component of spacetime can give rise to intriguing singular effects while still preserving essential properties of Schwarzschild black holes, including the finite area of the horizon.

\subsection{Gravitational signatures}
\label{sec:GravSig}

The differences between the metrics of Schwarzschild black holes and distorted Schwarzschild solutions may suggest that they are fundamentally different entities. However, the spacelike coordinates $(r,\theta,\phi)$ of the distorted solution cannot be truly compared to those of Schwarzschild's due to the absence of spherical symmetry.

In this section, we focus on listing various geometric and gravitational characteristics of distorted Schwarzschild black holes, comparing them with those of Schwarzschild. All computational details can be found in Appendix \ref{app:DisSchw}.

\begin{itemize}
    \item \underline{Photon ring and shadow:}

    Similar to Schwarzschild, the horizon of a distorted Schwarzschild is circumscribed by an unstable photon sphere delimiting a shadow. The photon sphere is around 95\% spherically symmetric, and its scattering characteristics, in terms of size, Lyapunov exponent, and angular velocity of photons at the orbit, are remarkably close to those of the Schwarzschild photon sphere.

    Furthermore, it is the photon ring of black holes, rather than their horizon, that dictates their gravitational signature, such as gravitational lensing \cite{Gralla:2019xty} or quasi-normal modes \cite{Cardoso:2008bp}. Consequently, one can infer that the gravitational signature of a distorted Schwarzschild black hole closely resembles that of the Schwarzschild black hole, despite differing substantially in the near-horizon region.

    \item \underline{Gravitational multipole moments:}

    Unlike the Schwarzschild black hole, the distorted Schwarzschild solution features non-zero even gravitational multipole moments, $\cM_{2n} = c_n M^{2n+1}$, where $c_n$ are dimensionless constants (see \eqref{eq:GravMultipole}).

    These multipole moments represent the large-scale effects of the S$^2$ deformation at the horizon, breaking spherical symmetry into axial symmetry that leads to a substantial deviation from the characteristics of the Schwarzschild black hole.

    \item \underline{Geometric size:}

   In Appendix \ref{app:GeoSize}, we define a background-independent radial coordinate that allows us to compare the area of the S$^2$ around a Schwarzschild black hole and its distorted version. We found that the characteristic sizes of the S$^2$ are very similar except close to their horizon. More precisely, they deviate by less than 1\% above a distance of $M/100$ from their horizon.

    However, despite sharing the same overall S$^2$ size, the distorted Schwarzschild exhibits significant axisymmetry, leading to an increasingly flattened sphere as the horizon is approached, from 95\% spherical symmetry at the photon ring to 0\% at the horizon.
\end{itemize}

Consequently, the S$^2$ deformation that led to the distorted Schwarzschild solution has preserved numerous properties of the Schwarzschild black hole: its overall size, light ring structure, and gravitational characteristics. However, it has also induced a large-scale stretching of the two-sphere, which becomes significantly noticeable between the black hole shadow and the singular horizon. This also induces nontrivial gravitational multipoles.

\section{Horizon-scale structures from electromagnetic entrapment}
\label{sec:EntrapLimGR}

In section \ref{sec:probe}, we highlighted intriguing dynamics between gravity and electromagnetism through a probe derivation. These dynamics were extended to self-gravitating charged system in section \ref{sec:GravAndElecPot} with the study of gravitational and electromagnetic potentials generated by extremal charges.

In this section, we perform a more detailed analysis of the spacetime structures that can be produced by extremal charges in their entrapment limit. Initially, our analysis will be conducted using the static Ernst formalism in four dimensions, wherein the geometries correspond to bound states of Reissner-Nordstr\"om black holes separated by struts. Subsequently, we will resolve the struts by introducing a Kaluza-Klein circle so that the bound states define regular structures of extremal black holes on a vacuum bubble. Throughout these analyses, we demonstrate that these solutions define an extensive phase space of horizon-scale structures for the distorted Schwarzschild black hole by releasing electromagnetic degrees of freedom entrapped in the near-horizon region. We conclude the section by discussing the construction of coherent microstates, referred to as topological solitons, which are smooth horizonless geometries as compact as the distorted Schwarzschild.

\subsection{Static Ernst formalism}
\label{eq:4dFrame}

The Ernst formalism has been initially derived within a four-dimensional Einstein-Maxwell theory given by the action \cite{Ernst:1967wx,Ernst:1967by}:
\begin{equation}
S_4 \= \frac{1}{16\pi G_4} \,\int \,d^4 x \sqrt{-g}\,\left( R -\frac{1}{2} |F|^2\right)\,,
\label{eq:4daction}
\end{equation}
where $G_4$ is the four-dimensional Newton constant, $R$ is the Ricci scalar, and $F$ is a two-form field strength. Its static formulation consists in restricting to static axially-symmetric solutions characterized by $\partial_t$ and $\partial_\phi$ as the timelike and spacelike isometries, and the fields depend on a two-dimensional plane in Weyl-Papapetrou coordinates, denoted as $(\rho,z)$:
\begin{equation}
\begin{split}
ds_4^2 &\= - V^{-1} \,dt^2+ V \left[e^{4\nu}\left( d\rho^2+dz^2\right) +\rho^2 d\phi^2 \right],\qquad F \=  -dA\wedge dt\,.
\end{split}
\label{eq:StaticErnstMetric4d}
\end{equation}
The functions $V$ and $A$ are the gravitational and electric potentials, respectively, and $\nu$ determines the nature of the three-dimensional base. For simplicity, we will restrict ourselves to an electric field, but one can easily take the magnetic dual formulation,
\begin{equation}
   F_m = \star F \= dH \wedge d\phi \,,\qquad \star_2 dH = \rho V dA\,,
    \label{eq:MagDualErnst}
\end{equation}
where $H$ represents the magnetic potential and $\star_2$ is the Hodge operator in the flat $(\rho,z)$ space.\footnote{We could also consider a dyonic field $F=-\cos \eta \,dA\wedge dt + \sin \eta \, dH\wedge d\phi$, where $\eta$ serves as a dyonic parameter controlling the electric charges relative to the magnetic charges.}

The Einstein-Maxwell equations lead to electrostatic Ernst equations that we review in Appendix \ref{app:ErnstEOM}. The general solutions corresponding to an arbitrary number of non-extremal Reissner-Nordstr\"om black holes on a line have been derived in \cite{NoraBreton1998}. In this section, we focus on the extremal limits of this solution, as provided in Appendix \ref{app:NExtSources}, and the binary configuration in Appendix \ref{app:TwoExtSources} \cite{Alekseev:2007re,Alekseev:2007gt,Manko:2007hi,Manko:2008gb,Bah:2022yji,Bah:2023ows}.

In Weyl-Papapetrou coordinates, the extremal black holes are point sources situated on the axis of symmetry at $\rho=0$, and $z=\alpha_n$, $n=1,\ldots,N$. We introduce the overall length of the configuration $\ell\equi \alpha_N - \alpha_1$ and fix the origin of the $z$-axis as in \eqref{eq:SizeConfAndOrigin}.

Occasionally, we prefer to express the solutions in spherical coordinates centered around the configuration:\footnote{The transformation from Weyl-Papapetrou coordinates to spherical coordinates gives
 \begin{equation}
 d\rho^2 + dz^2 = \left( 1- \frac{\ell\cos^2 \frac{\theta}{2}}{r}\right)\left( 1- \frac{\ell\sin^2 \frac{\theta}{2}}{r}\right) \, \left[\frac{dr^2}{\left( 1- \frac{\ell}{r}\right)} +r^2 \,d\theta^2\right]\,.
 \end{equation}}
\begin{equation}
\rho= \sqrt{r(r-\ell)}\,\sin \theta \,,\qquad z = \left(r-\frac{\ell}{2} \right) \cos \theta +\frac{\alpha_1+\alpha_N}{2} \,.
\label{eq:SpherCoord}
\end{equation}
In this coordinate system, $r\geq \ell$, $0\leq \theta\leq \pi$, and the centers are located at $r=\ell$ and at discrete values of $\theta=\theta_n$, where $\cos \theta_n = 1-\frac{2(\alpha_N-\alpha_n)}{\ell}$. Moreover, the metric and field \eqref{eq:StaticErnstMetric4d} reads:
\begin{equation}
\begin{split}
ds_4^2 = - V^{-1} \,dt^2+ V\left(1-\frac{\ell}{r}\right) \left[G \left( \frac{dr^2}{1-\frac{\ell}{r}}+r^2 d\theta^2\right) +r^2 \sin^2\theta \, d\phi^2 \right],\quad F \= - dA\wedge dt\,,
\end{split}
\label{eq:2RNBS}
\end{equation}
where $G\equiv e^{4\nu}\,\left( 1- \frac{\ell\cos^2 \frac{\theta}{2}}{r}\right)\left( 1- \frac{\ell\sin^2 \frac{\theta}{2}}{r}\right)\left(1-\frac{\ell}{r}\right)^{-1}$.

\subsection{Neutral bound state of two extremal Reissner-Nordstr\"om black holes}
\label{sec:BSTwoBH4d}

The solution corresponding to two extremal Reissner-Nordstr\"om black holes of opposite charges, total mass $M$ and separated by a distance $\ell\geq M$  is given by the metric and field \eqref{eq:2RNBS} with
\begin{align}
V & = \left(1+\frac{2M(2r+M-\ell)}{(2r-\ell)^2-\ell^2\cos^2\theta-M^2 \sin^2 \theta}\right)^2, \qquad A = \frac{4M \sqrt{\ell^2-M^2}\,\cos\theta}{(2r+M-\ell)^2-(\ell^2-M^2)\cos^2\theta} \,,\nn \\
 e^{\nu} & =1-\frac{M^2\sin^2\theta}{(2r-\ell)^2-\ell^2\cos^2\theta} , \label{eq:2RNBSfields}
\end{align}
and its magnetic formulation \eqref{eq:MagDualErnst} gives:
\begin{equation}
    H =\frac{2M \sqrt{\ell^2-M^2}(2r+M-\ell)\,\sin^2\theta}{(2r-\ell)^2-\ell^2\cos^2\theta-M^2 \sin^2 \theta}\,.
\end{equation}
The solution is asymptotically flat,  $\mathbb{R}^{1,3}$, and is regular for $r>\ell$, where the $\phi$-circle smoothly degenerates at the poles of the two-sphere. When viewed from a distance, the solution appears as neutral gravitational object of mass $M$ and electromagnetic dipole moment $\cJ$:
\begin{equation}
 \cJ \=  M\sqrt{\ell^2-M^2}\,.
\end{equation}

At $r=\ell$ and $\theta=0,\pi$, the function $V$ diverges, indicating the presence of two extremal black holes. These two black holes are separated by a segment where the $\phi$-circle once again degenerates, occurring at $r=\ell$ and $0<\theta<\pi$. In this segment, we expect a conical excess, suggesting the presence of a strut --- a string with negative tension --- whose purpose is to prevent the black holes from collapsing towards each other.

\begin{figure}[t]
\centering
    \begin{tikzpicture}
\def\deb{-10} 
\def\inter{0.7} 
\def\ha{2.8} 
\def\zaxisline{3.5} 
\def\rodsize{1.7} 
\def\numrod{1.7} 
\def\fin{\deb+1+2*\rodsize+\numrod*\rodsize} 


\draw (\deb+0.5+\rodsize+0.5*\numrod*\rodsize,\ha+2-\inter) node{{{\it Two Extremal Reissner-Nordstr\"om Black Holes on a Strut}}}; 

\draw[draw=black] (\deb+0.5+\rodsize-0.3+0.3,\ha) rectangle (\deb+0.5+\rodsize-0.3-0.3,\ha-\zaxisline*\inter-0.3) ;
\draw[gray] (\deb+0.5+\rodsize-0.3,\ha+0.2) node{{\scriptsize 4d BH $(+q)$}};

\draw[draw=black] (\deb+0.5+3*\rodsize-0.3+0.3,\ha) rectangle (\deb+0.5+3*\rodsize-0.3-0.3,\ha-\zaxisline*\inter-0.3) ;
\draw[gray] (\deb+0.5+3*\rodsize-0.3,\ha+0.2) node{{\scriptsize 4d BH $(-q)$}};

\draw [decorate, 
    decoration = {brace,
        raise=5pt,
        amplitude=5pt},line width=0.2mm,gray] (\deb-0.7,\ha-1.5*\inter+0.05) --  (\deb-0.7,\ha-0.5*\inter-0.05);
        \draw [decorate, 
    decoration = {brace,
        raise=5pt,
        amplitude=5pt},line width=0.2mm,gray] (\deb-0.7,\ha-2.5*\inter+0.05) --  (\deb-0.7,\ha-1.5*\inter-0.05);
        
\draw[gray] (\deb-1.4,\ha-1*\inter) node{S$^2$};
\draw[gray] (\deb-1.4,\ha-2*\inter) node{{\scriptsize time}};


\draw[black,thin] (\deb+1,\ha-\inter) -- (\fin-1,\ha-\inter);
\draw[black,thin] (\deb,\ha-2*\inter) -- (\fin,\ha-2*\inter);

\draw[black,->,line width=0.3mm] (\deb-0.4,\ha-\zaxisline*\inter) -- (\fin+0.2,\ha-\zaxisline*\inter);

\draw (\deb-0.4,\ha-\inter) node{$\phi$};
\draw (\deb-0.4,\ha-2*\inter) node{$t$};

\draw (\fin+0.2,\ha-\zaxisline*\inter-0.3) node{$z$};


\draw[black, dotted, line width=1mm] (\deb,\ha-\inter) -- (\deb+0.5,\ha-\inter);
\draw[black,line width=1mm] (\deb+0.5,\ha-\inter) -- (\deb+0.5+\rodsize-0.36,\ha-\inter);
\draw[black,line width=1mm] (\fin-0.44-\rodsize+0.2,\ha-1*\inter) -- (\fin-0.58,\ha-1*\inter);
\draw[black, dotted,line width=1mm] (\fin-0.5,\ha-1*\inter) -- (\fin,\ha-1*\inter);


\draw[gray,line width=1mm] (\deb+0.5+\rodsize-0.24,\ha-1*\inter) -- (\deb+0.44+3*\rodsize-0.3,\ha-1*\inter);
\draw[decorate,decoration={zigzag,segment length=1.3mm, amplitude=.7mm},thick]  (\deb+0.5+\rodsize-0.27,\ha-1*\inter) -- (\deb+0.47+3*\rodsize-0.3,\ha-1*\inter);

\draw[black,line width=1mm] (\deb+0.5+\rodsize-0.3,\ha-2*\inter) circle[radius=2pt];
\draw[black,line width=1mm] (\deb+0.5+3*\rodsize-0.3,\ha-2*\inter) circle[radius=2pt];


\draw[burgundy,decorate,decoration={zigzag,segment length=1.3mm, amplitude=.7mm},thick,opacity=0.5]  (\deb+0.5+\rodsize-0.3,\ha-\zaxisline*\inter) -- (\deb+0.5+3*\rodsize-0.3,\ha-\zaxisline*\inter);
\draw[black,line width=1mm,opacity=0.5] (\deb+0.5+\rodsize-0.3,\ha-\zaxisline*\inter) circle[radius=2pt];
\draw[black,line width=1mm,opacity=0.5] (\deb+0.5+3*\rodsize-0.3,\ha-\zaxisline*\inter) circle[radius=2pt];


\draw[gray,dotted,line width=0.2mm] (\deb+0.5+\rodsize-0.3,\ha-\inter) -- (\deb+0.5+\rodsize-0.3,\ha-\zaxisline*\inter);
\draw[gray,dotted,line width=0.2mm] (\deb+0.5+3*\rodsize-0.3,\ha-\inter) -- (\deb+0.5+3*\rodsize-0.3,\ha-\zaxisline*\inter);

\draw[line width=0.3mm] (\deb+0.5+\rodsize-0.3,\ha-\zaxisline*\inter+0.1) -- (\deb+0.5+\rodsize-0.3,\ha-\zaxisline*\inter-0.1);
\draw[line width=0.3mm] (\deb+0.5+3*\rodsize-0.3,\ha-\zaxisline*\inter+0.1) -- (\deb+0.5+3*\rodsize-0.3,\ha-\zaxisline*\inter-0.1);

\draw (\deb+0.5+1*\rodsize-0.3,\ha-\zaxisline*\inter-0.6) node{{\small $-\frac{\ell}{2}$}};
\draw (\deb+0.5+3*\rodsize-0.3,\ha-\zaxisline*\inter-0.6) node{{\small $\frac{\ell}{2}$}};

\draw[gray] (\deb+0.5+2*\rodsize-0.3,\ha-\zaxisline*\inter-0.3) node{{\scriptsize strut}};

\end{tikzpicture}
\caption{Spacetime structure of a bound state comprising two extremal Reissner-Nordstr\"om black holes of opposite charges and held apart by a string with negative tension. The diagram represents the behavior of the time and $\phi$ fibers on the symmetry axis, i.e. the $z$-axis at $\rho=0$. A thick line or a dot corresponds to spacetime region where the corresponding fiber degenerates. The oscillating line indicates a conical excess at the degeneracy, leading to a strut.}
\label{fig:rodsourceRNBS}
\end{figure}
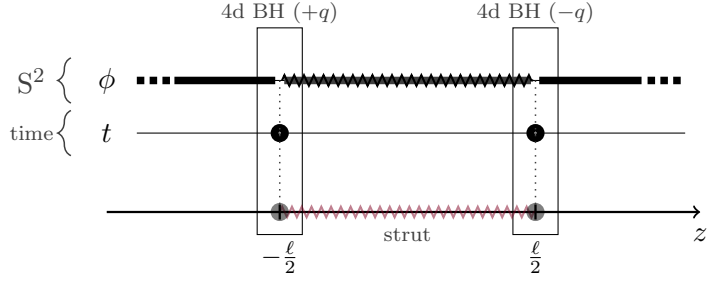  

In essence, we have constructed a bound state consisting of two Reissner-Nordstr\"om black holes that are held apart by a strut in four dimensions. The structure of these sources on the symmetry axis is depicted in Fig.\ref{fig:rodsourceRNBS}.

\subsubsection{Internal structure}

We study the structure of the bound state at the coordinate boundary, $r=\ell$, which corresponds to the segment on the symmetry axis, $\rho=0$ and $-\ell/2\leq z\leq \ell/2$ in the Weyl-Papapetrou coordinates \eqref{eq:SpherCoord}. More detailed analysis can be found in Appendix \ref{app:TwoBH4d}.

First, we focus on the strut at $r=\ell$ and $0< \theta <\pi$. The local geometry is most accurately described by considering $r = \ell + \frac{\bar{r}^2}{4}$ and taking the limit $\bar{r} \to 0$. The induced metric on the $(\bar{r},\phi)$ subspace is given by:
\begin{equation}
ds_2^2 \,\underset{\bar{r}\to 0}{\propto} \, d\bar{r}^2 +\delta^2 \,\bar{r}^2\,d\phi^2 ,\qquad \delta \equi \left(1-\frac{M^2}{\ell^2} \right)^{-2}.
\label{eq:StrutMetric}
\end{equation}
Therefore, the locus at $r=\ell$ and $\theta\neq 0,\pi$ represents a coordinate degeneracy of the $\phi$-circle. However, since $\phi$ is $2\pi$ periodic, this degeneracy results in a conical excess characterized by the parameter $\delta>1$, which becomes infinite as the inter-center distance approaches its minimal value $\ell \to M$. This excess represents a string-like structure with negative tension that counterbalances the gravitational attraction between the two black holes. Despite the divergent conical excess, the energy associated with the strut is always finite and can be expressed as \cite{Costa:2000kf,Bah:2021owp}:
\begin{equation}
E \= -\frac{(1-\delta^{-1}) \ell}{4}\,,
\end{equation}
which tends towards $-M/4$ as $\ell \to M$.  \\

Next, we analyze the nature of the black holes within the bound state, beginning with the black holes at the North pole, located at $r=\ell$ and $\theta=0$ ($\rho=0$ and $z=\ell/2$). As detailed in Appendix \ref{app:TwoBH4d}, the local geometry has an AdS$_2\times$S$^2$ topology, which corresponds to the near-horizon geometry of an extremal Reissner-Nordström black hole. This region carries a charge $-q$ given by:\footnote{The charge at the black hole is given by $\frac{1}{2\pi} \int \star F = \int dH \,,$ where the integral contour is the two-sphere around the black hole.}
\begin{equation}
q \equi \frac{M}{2} \sqrt{\frac{\ell+M}{\ell-M}} \,.
\end{equation}
As an extremal black hole, it possesses zero temperature but a finite entropy, which is given by the S$^2$ area at the horizon with the Bekenstein-Hawking formula (in units where $G_4=1$):
\begin{equation}
S_1 \= \frac{M^2 (\ell+M)^2 \pi}{4\ell^2}\,.
\end{equation}

The South pole, at $r=\ell$ and $\theta=\pi$, corresponds to the anti-extremal partner with an identical geometry but with opposite charge $q$. This leads to another extremal Reissner-Nordström black hole with the same entropy, $S_2=S_1$. Therefore, the total entropy of the bound state is given by:
\begin{equation}
S^{(2)} \= S_1 +S_2 \=   \frac{M^2 (\ell+M)^2 \pi}{2\ell^2}\,.
\label{eq:EntropBS2RN}
\end{equation}
This aligns with the Bekenstein bound \cite{PhysRevD.23.287}, where the entropy of a black hole serves as an upper limit to the energy-to-entropy ratio of any system within an equivalent area. More precisely, we have
\begin{equation}
    \frac{S^{(2)}}{M} \= \frac{S_\text{Schw}}{M}\times \frac{1}{8} \left(1+\frac{M}{\ell} \right)^2,
\end{equation}
where $S_\text{Schw}=4\pi M^2$ is the entropy of a Schwarzschild black hole with the same mass. Notably, in the ultra-compact limit where $\ell \to M$, the entropy approaches its maximal value with respect to the Schwarzschild entropy:
\begin{equation}
S^{(2)} \to 2\pi M^2 \= \frac{1}{2} S_\text{Schw},
\label{eq:EntropyRNBSscaling}
\end{equation}

\subsubsection{Entrapment limit}
\label{sec:EntrapLim4d}

In section \ref{sec:TwoExCharges}, the study of the gravitational and electric potentials revealed an entrapment limit towards a vacuum solution as the charges approach a critical distance. We reexamine this limit for our four-dimensional solution, taking into account the entire structure of spacetime.

Thus, we consider the same scenario where the separation between the black holes approaches its minimal size, $\ell = M (1+\epsilon)$, with $\epsilon \ll 1$.
The electromagnetic flux \eqref{eq:2RNBSfields} increasingly concentrates around the sources at $r=\ell$, instead of completely vanishing, while the redshift intensifies in this region. Specifically, the electromagnetic energy density and redshift scale as $\epsilon\to 0$:
\begin{equation}
\left|F\right|^2 \,\propto\,  \frac{M^4}{r^6} \left(1-\frac{M}{r} \right)^{-4} \, \epsilon\,,\qquad \left|F\right|^2  \Bigl|_{r=\ell}\,\propto \frac{\epsilon^{-3}}{M^2}\,, \qquad \left.V \right|_{r=\ell} \propto \frac{\epsilon^{-2}}{\sin^4 \theta}\,.
\end{equation}
Consequently, the bound state evolves into an ultra-compact structure with an almost infinite redshift at the locations of the black holes and the strut. Simultaneously, the flux get primarily entrapped around the sources. Thus, for $r\gtrsim M(1+\cO(\epsilon^\frac{1}{4}))$, the solution closely resembles a vacuum black-hole-like solution in four dimensions.

More precisely,  we have,  for $r\gtrsim M(1+\cO(\epsilon^\frac{1}{4}))$,  
\begin{equation}
V^{-1} \sim  \left(1-\frac{M}{r}\right)^2\,,\qquad G \sim \left(1+\dfrac{M^2 \sin^2\theta}{4r(r-M)}\right)^{-3} \,,  \qquad A \sim 0\,. 
\label{eq:EntrapLimFields4d}
\end{equation}

\begin{figure}[t]
\begin{center}
\includegraphics[width= \textwidth]{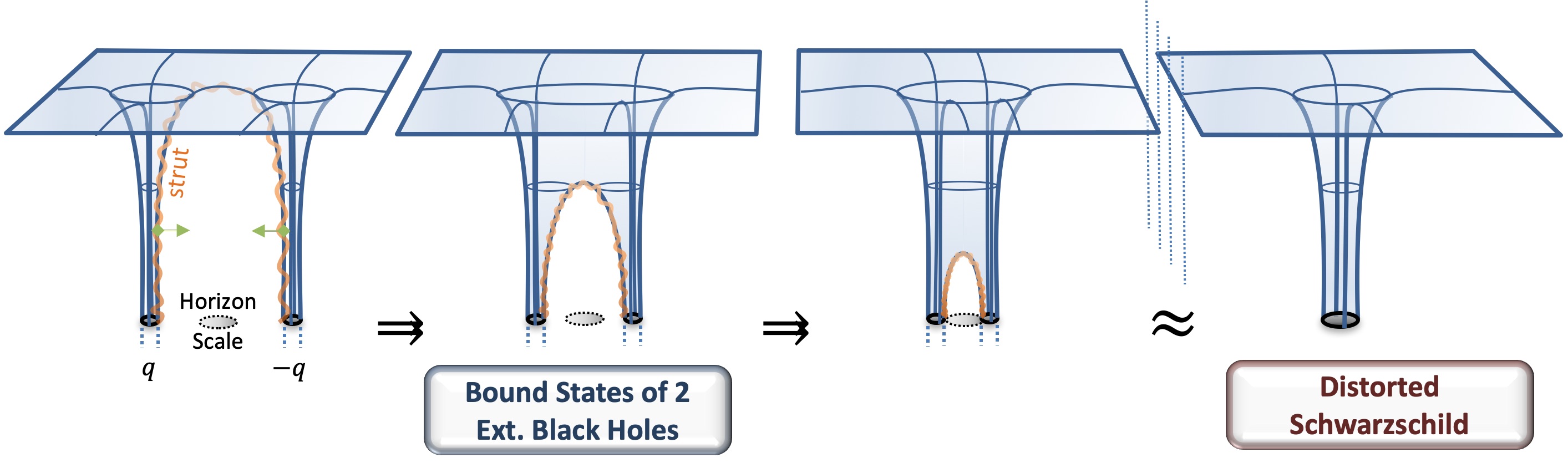}
\caption{Schematic description of the entrapment limit for a bound state of two extremal Reissner-Nordstro\"om black holes of opposite charges supported by a strut in four dimensions.}
\label{fig:EntrapLim}
\end{center}
\end{figure}

Thus, the bound state becomes indistinguishable from a distorted Schwarzschild black hole \eqref{eq:DisSchw}, except in an infinitesimal region above its horizon. In Fig.\ref{fig:EntrapLim}, we have schematically represented the effect of the entrapment limit on the spacetime as $\ell\to M$. All the characteristics of the bound state become entrapped in an infinitesimal region of size $M\epsilon$ above the sources. The bound state ``replaces" the singular horizon of the distorted Schwarzschild by releasing highly confined electromagnetic degrees of freedom at its horizon scale. 

More remarkably, the phase space spanned by these degrees of freedom, as given by the entropy of both extremal black holes \eqref{eq:EntropyRNBSscaling}, exhibits the same $M^2$ behavior as the entropy of the Schwarzschild black hole and its distorted version \eqref{eq:AreaHorDis}. Constructing a gravitational geometry that is as compact as Schwarzschild and has a phase space of microstates of comparable size is non-trivial. Moreover, this matching was not something we could fine-tune; it naturally emerged during the construction. 

Therefore, this bound state explicitly illustrates that electromagnetic entrapment might be a key mechanism for describing the microstructure within the Schwarzschild black hole and its distorted geometry, as well as the new physics that is expected to emerge at the horizon scale.

\subsection{Neutral bound state of $N$ extremal Reissner-Nordstr\"om black holes}
\label{sec:NExBH4d}

An arbitrary solution with $N$ extremal black holes on a line can be derived from the metric and electromagnetic gauge fields \eqref{eq:2RNBS} using the generic expressions \eqref{eq:NExErnstPot}, \eqref{eq:NExGravElecPot}. However, this solution cannot be further simplified and does not admit a compact form. Thus, only specific examples can be analyzed through numerical derivation.

The argument presented in section \ref{sec:NExCharges}, regarding the gravitational and electric potentials, suggests that generic neutral solutions exhibit the same entrapment limit as the two-black-hole bound state discussed in the previous section, as the configuration size $\ell$ approaches the total mass $M$. Thus, we encounter similar physics to that described in Fig.\ref{fig:EntrapLim}, but now with multiple extremal black holes arranged in a line.

Additionally, the solutions involve $2N$ independent parameters, which consist of the center positions on the symmetry axis, $\alpha_n$, and $N$ parameters generating the charges at each center, $\beta_n$. By enforcing neutrality as in \eqref{eq:neutralityCond} and setting the origin on the $z$ axis according to \eqref{eq:SizeConfAndOrigin}, we are left with $2(N-1)$ free parameters. Among these, two can be considered fixed quantities: the total mass $M$ and the configuration size $\ell$, which is taken very close to $M$ in the entrapment limit. Thus, we ultimately have $2(N-2)$ internal free parameters that determine the internal configuration of extremal black hole.

In the entrapment limit, the effects of these parameters emerge only in the immediate vicinity around the sources. As a result, these parameters define a vast phase space of internal electromagnetic degrees of freedom that generate extremal black holes, effectively replacing the horizons of the distorted Schwarzschild black holes. \\

\begin{figure}[t]
\centering
    \begin{tikzpicture}
\def\deb{-10} 
\def\inter{0.7} 
\def\ha{2.8} 
\def\zaxisline{3.5} 
\def\rodsize{1.7} 
\def\numrod{1.7} 
\def\fin{\deb+1+2*\rodsize+\numrod*\rodsize} 


\draw (\deb+0.5+\rodsize+0.5*\numrod*\rodsize,\ha+2-\inter) node{{{\it Three Extremal Reissner-Nordstr\"om Black Holes on a Strut}}}; 

\draw[draw=black] (\deb+0.5+\rodsize-0.3+0.3,\ha) rectangle (\deb+0.5+\rodsize-0.3-0.3,\ha-\zaxisline*\inter-0.3) ;
\draw[gray] (\deb+0.5+\rodsize-0.3,\ha+0.2) node{{\scriptsize 4d BH $(q)$}};

\draw[draw=black] (\deb+0.5+2*\rodsize-0.3+0.3,\ha) rectangle (\deb+0.5+2*\rodsize-0.3-0.3,\ha-\zaxisline*\inter-0.3) ;
\draw[gray] (\deb+0.5+2*\rodsize-0.3,\ha+0.2) node{{\scriptsize 4d BH $(-2q)$}};

\draw[draw=black] (\deb+0.5+3*\rodsize-0.3+0.3,\ha) rectangle (\deb+0.5+3*\rodsize-0.3-0.3,\ha-\zaxisline*\inter-0.3) ;
\draw[gray] (\deb+0.5+3*\rodsize-0.3,\ha+0.2) node{{\scriptsize 4d BH $(q)$}};

\draw [decorate, 
    decoration = {brace,
        raise=5pt,
        amplitude=5pt},line width=0.2mm,gray] (\deb-0.7,\ha-1.5*\inter+0.05) --  (\deb-0.7,\ha-0.5*\inter-0.05);
        \draw [decorate, 
    decoration = {brace,
        raise=5pt,
        amplitude=5pt},line width=0.2mm,gray] (\deb-0.7,\ha-2.5*\inter+0.05) --  (\deb-0.7,\ha-1.5*\inter-0.05);
        
\draw[gray] (\deb-1.4,\ha-1*\inter) node{S$^2$};
\draw[gray] (\deb-1.4,\ha-2*\inter) node{{\scriptsize time}};


\draw[black,thin] (\deb+1,\ha-\inter) -- (\fin-1,\ha-\inter);
\draw[black,thin] (\deb,\ha-2*\inter) -- (\fin,\ha-2*\inter);

\draw[black,->,line width=0.3mm] (\deb-0.4,\ha-\zaxisline*\inter) -- (\fin+0.2,\ha-\zaxisline*\inter);

\draw (\deb-0.4,\ha-\inter) node{$\phi$};
\draw (\deb-0.4,\ha-2*\inter) node{$t$};

\draw (\fin+0.2,\ha-\zaxisline*\inter-0.3) node{$z$};


\draw[black, dotted, line width=1mm] (\deb,\ha-\inter) -- (\deb+0.5,\ha-\inter);
\draw[black,line width=1mm] (\deb+0.5,\ha-\inter) -- (\deb+0.5+\rodsize-0.36,\ha-\inter);
\draw[black,line width=1mm] (\fin-0.44-\rodsize+0.2,\ha-1*\inter) -- (\fin-0.58,\ha-1*\inter);
\draw[black, dotted,line width=1mm] (\fin-0.5,\ha-1*\inter) -- (\fin,\ha-1*\inter);


\draw[gray,line width=1mm] (\deb+0.5+\rodsize-0.24,\ha-1*\inter) -- (\deb+0.44+2*\rodsize-0.3,\ha-1*\inter);
\draw[decorate,decoration={zigzag,segment length=1.3mm, amplitude=.7mm},thick]  (\deb+0.5+\rodsize-0.27,\ha-1*\inter) -- (\deb+0.47+2*\rodsize-0.3,\ha-1*\inter);

\draw[gray,line width=1mm] (\deb+0.5+2*\rodsize-0.24,\ha-1*\inter) -- (\deb+0.44+3*\rodsize-0.3,\ha-1*\inter);
\draw[decorate,decoration={zigzag,segment length=1.3mm, amplitude=.7mm},thick]  (\deb+0.5+2*\rodsize-0.27,\ha-1*\inter) -- (\deb+0.47+3*\rodsize-0.3,\ha-1*\inter);

\draw[black,line width=1mm] (\deb+0.5+\rodsize-0.3,\ha-2*\inter) circle[radius=2pt];
\draw[black,line width=1mm] (\deb+0.5+2*\rodsize-0.3,\ha-2*\inter) circle[radius=2pt];
\draw[black,line width=1mm] (\deb+0.5+3*\rodsize-0.3,\ha-2*\inter) circle[radius=2pt];


\draw[burgundy,decorate,decoration={zigzag,segment length=1.3mm, amplitude=.7mm},thick,opacity=0.5]  (\deb+0.5+\rodsize-0.3,\ha-\zaxisline*\inter) -- (\deb+0.5+3*\rodsize-0.3,\ha-\zaxisline*\inter);
\draw[black,line width=1mm,opacity=0.5] (\deb+0.5+\rodsize-0.3,\ha-\zaxisline*\inter) circle[radius=2pt];
\draw[black,line width=1mm,opacity=0.5] (\deb+0.5+2*\rodsize-0.3,\ha-\zaxisline*\inter) circle[radius=2pt];
\draw[black,line width=1mm,opacity=0.5] (\deb+0.5+3*\rodsize-0.3,\ha-\zaxisline*\inter) circle[radius=2pt];


\draw[gray,dotted,line width=0.2mm] (\deb+0.5+\rodsize-0.3,\ha-\inter) -- (\deb+0.5+\rodsize-0.3,\ha-\zaxisline*\inter);
\draw[gray,dotted,line width=0.2mm] (\deb+0.5+3*\rodsize-0.3,\ha-\inter) -- (\deb+0.5+3*\rodsize-0.3,\ha-\zaxisline*\inter);

\draw[line width=0.3mm] (\deb+0.5+\rodsize-0.3,\ha-\zaxisline*\inter+0.1) -- (\deb+0.5+\rodsize-0.3,\ha-\zaxisline*\inter-0.1);
\draw[line width=0.3mm] (\deb+0.5+3*\rodsize-0.3,\ha-\zaxisline*\inter+0.1) -- (\deb+0.5+3*\rodsize-0.3,\ha-\zaxisline*\inter-0.1);

\draw (\deb+0.5+1*\rodsize-0.3,\ha-\zaxisline*\inter-0.6) node{{\small $-\frac{\ell}{2}$}};
\draw (\deb+0.5+2*\rodsize-0.3,\ha-\zaxisline*\inter-0.6) node{{\small $0$}};
\draw (\deb+0.5+3*\rodsize-0.3,\ha-\zaxisline*\inter-0.6) node{{\small $\frac{\ell}{2}$}};

\draw[gray] (\deb+0.5+1.5*\rodsize-0.3,\ha-\zaxisline*\inter-0.3) node{{\scriptsize strut}};
\draw[gray] (\deb+0.5+2.5*\rodsize-0.3,\ha-\zaxisline*\inter-0.3) node{{\scriptsize strut}};

\end{tikzpicture}
\caption{Spacetime structure of three extremal Reissner-Nordstr\"om black holes with zero net charge and electromagnetic dipole, and held apart by two struts.}
\label{fig:rodsource3RNBS}
\end{figure}
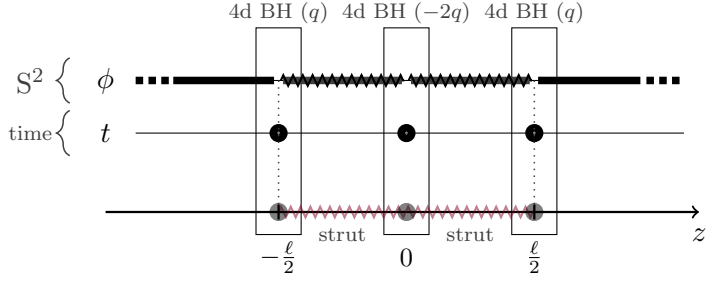  

We study the same configuration  as in section \ref{sec:NExCharges}. This configuration leads to a neutral bound state of three extremal black holes with no electric dipole. We have represented the spacetime structure on the symmetry axis in Fig.\ref{fig:rodsource3RNBS}. This configuration is completely determined by imposing, in addition to neutrality, the absence of an electromagnetic dipole moment. Consequently, the positions of the black holes are fixed at
\begin{equation}
    \alpha_1 \= -\frac{\ell}{2} \,,\qquad \alpha_2 \= 0 \,,\qquad \alpha_3 \= \frac{\ell}{2}\,,
\end{equation}
and the remaining $(\beta_1,\beta_2,\beta_3)$ are fixed by
\begin{equation}
    M= -\sum_{n=1}^3 \beta_n, \qquad Q = \sum_{n=1}^3 f_n =0 \,,\qquad \cJ =  \sum_{n=1}^3 f_n \beta_n =0\,,
\end{equation}
where the norm of $f_n$ are given in terms of $(\alpha_n,\beta_n)$ in \eqref{eq:AxisDataN} and their sign must be carefully chosen so that solutions exist (for the present configuration, $f_2$ is taken negative).

By numerically studying several solutions, we found that the charges at the black holes are
\begin{equation}
    q_1 \= q_3 \= -2 q_2 \=  \frac{M}{2} \times \frac{\ell+M}{\ell-M}\,.
\end{equation}
and the black holes are separated by two struts described by the same type of conical singularity as in \eqref{eq:StrutMetric}, but with a more intense excess:
\begin{equation}
    \delta \= \left(1-\frac{M^2}{\ell^2} \right)^{-4}\,.
\end{equation}
The charge structure confirms that the configuration remains neutral and with zero dipole moment for arbitrary values of $\ell$.

By carefully studying the limit where the extremal black holes become entrapped near the horizon scale of the configuration, we have confirmed the conclusion of section \ref{sec:NExCharges}: the spacetime converges toward the distorted Schwarzschild black hole and resolves its singular horizon through three extremal black holes held apart by struts.

Moreover, we were able to numerically find the entropy of each extremal black hole, denoted as $(S_1, S_2, S_3)$, as functions of $\ell$ and $M$ (in unit $G_4=1$)
\begin{equation}
S_1 =S_3 = \frac{\pi M^2 (\ell+M)^4}{16 \ell^4}\,,\qquad S_2 \= \frac{\pi M^2(\ell-M)^2 (\ell+M)^6}{4\ell^8}\,.
\end{equation}
Thus,  in the entrapment limit,  as $\ell$ approaches $M$, the following relationships hold:
\begin{equation}
    S_1 = S_3 \to \pi M^2 \,,\qquad S_2 \to 0 \quad \Rightarrow \quad S^{(3)} \to 2\pi M^2 = \frac{1}{2} S_\text{Schw},
    \label{eq:EntropLim3BH}
\end{equation}
where $S^{(3)}$ represents the total entropy of the system. Thus, the black hole in the middle becomes microscopic despite possessing substantial charges.

Furthermore, the total entropy is remarkably identical to that of the two-black-hole bound state and is half of the Schwarzschild entropy. Consequently, it is reasonable to assume that any bound states of $N$ extremal black holes will exhibit the same entrapment limit towards the distorted Schwarzschild geometry and will have similar entropy (at least of order $M^2$). 
Since the solutions are given by $2(N-2)$ continuous parameters, it might seem that the total phase space is infinite. However, this apparent infinite dimensionality is a consequence of the effective description in four dimensions.

It is important to note that our construction in four dimensions focuses primarily on replacing the horizon of vacuum black holes using emergent electromagnetic degrees of freedom that materialize in terms of extremal black holes and struts. While this approach provides a clear demonstration of electromagnetic entrapment, we cannot expect to reach a comprehensive definition of microstructure. Nevertheless, our results indicate that the physics at the horizon of vacuum black geometries allows for a vast phase space of electromagnetic degrees of freedom.

\subsection{Physical geometries with a compact dimension}

One of the main drawbacks of previous constructions is the presence of struts, which are unphysical strings with negative tension. In \cite{Elvang:2002br}, it has been shown that struts can be smoothly replaced by topological spacetime bubbles when an extra compact dimension is added and deformed on top of the geometries. While this result was initially demonstrated for static and axially-symmetric spacetime in vacuum, it has been generalized to spacetime with electromagnetic flux in \cite{Bah:2021owp}. Topological bubbles provide the necessary repulsion force and resistance to contraction without being singular \cite{Elvang:2002br}. Moreover, while the strut could have arbitrary tension, and therefore arbitrary length, the bubble will have a fixed size so that the topological pressure will counterbalance the gravitational attraction of the geometries. Thus, it provides a scale to the configurations and contains physical information that was missing in four dimensions.

In this section, we apply this mechanism to our constructions and obtain a more comprehensive definition of the horizon-scale structure that emerges from electromagnetic entrapment. For that purpose, we uplift the four-dimensional framework detailed in section \ref{eq:4dFrame} to five dimensions by adding one extra compact dimension $y$, of radius $R_y$ so that its periodicity is
\begin{equation}
    y \= y + 2\pi R_y\,.
\end{equation}
Additional details on this five-dimensional framework and its generalization of the static Ernst formalism can be found in Appendix \ref{app:5dErnst}.

Then, we fix the deformation along the extra dimension so that it corresponds to a vacuum bubble of size $\ell$ where this extra dimension degenerates. As such, the generic ansatz for the five-dimensional metric and field is given by
\begin{equation}
\begin{split}
ds_5^2 = - \frac{dt^2}{V}+ \sqrt{V}\left[ \left(1-\frac{\ell}{r}\right)dy^2 + e^{3\nu} \left( \frac{dr^2}{1-\frac{\ell}{r}}+r^2 d\theta^2\right) +r^2 \sin^2\theta \, d\phi^2 \right],\quad F \= - \frac{\sqrt{3}}{2}\,dA\wedge dt\,, \label{eq:5dmetric}
\end{split}
\end{equation}
where we have used the same spherical coordinates as in \eqref{eq:SpherCoord}. The fields $(V,A,\nu)$ are governed by the same static Ernst equations as in four dimensions, allowing us to use the same solutions corresponding to extremal point charges. Note that if one takes $V=1$ and $\nu=A=0$, the geometry corresponds to a vacuum bubble or Euclidean Schwarzschild geometry, indicating that, compared to the four-dimensional solution, we have generated nontrivial topology in the spacetime.

The dimensional reduction along the S$^1$ yields solutions of a four-dimensional Einstein-Maxwell-dilaton theory:
\begin{align}
d s_4^2 & =-\frac{\sqrt{1-\frac{\ell}{r}}\, d t^2}{V^\frac{3}{4}}+\frac{V^\frac{3}{4}}{\sqrt{1-\frac{\ell}{r}}}\,\left[e^{3 \nu} \left(\frac{dr^2}{1-\frac{\ell}{r}} + r^2 d\theta^2 \right) +r^2 \sin^2 \theta d\phi^2\right]\,, \\
e^{\frac{2}{\sqrt{3}}\Phi} & =\frac{1}{\left(1-\frac{\ell}{r} \right)\,\sqrt{V}}\,, \qquad F=-\frac{\sqrt{3}}{2}\,d A \wedge d t .\nonumber
\end{align}
In four dimensions, the coordinate boundary $r=\ell$ is singular, as some metric components and the scalar diverge. This singularity highlights a breakdown of the four-dimensional description and suggests a resolution in higher dimensions as a coordinate degeneracy of a KK circle.

\subsubsection{Two extremal black holes on a KK bubble}

We take $(V,A,\nu)$ to be the fields given by \eqref{eq:2RNBSfields} so that it sources two extremal black holes at the extremity of the bubble, $r=\ell$ and $\theta=0$ and $\pi$. For $r>\ell$, the solution is similar to the four-dimensional bound state of two Reissner-Nordström black holes. It is asymptotic to $\mathbb{R}^{1,3}\times\text{S}^1$ and has a mass and an electric dipole given by (in units $G_4=1$):\footnote{Note that we have now defined the ADM mass as $\cM$, as it no longer matches the $M$ parameter in the $(V, A, \nu)$ fields.}
\begin{equation}
    \cM \= \frac{\ell+3M}{4}\,,\qquad \cJ \= \frac{\sqrt{3}\,M}{2} \sqrt{\ell^2-M^2}.
\end{equation}
However, the internal structure has undergone significant modifications. At the place of the strut, $r=\ell$ and $0<\theta<\pi$, the degeneracy of the $\phi$-circle has shifted into a degeneracy of the $y$-circle \eqref{eq:5dmetric}. Therefore, the $(\bar{r},y)$ subspace is described by
\begin{equation}
    ds_2^2 \propto d\bar{r}^2 + \frac{\ell^4 \bar{r}^2}{4(\ell^2-M^2)^3}\,dy^2\,,
    \label{eq:R2localSpace}
\end{equation}
where $\bar{r}^2=4(r-\ell)\to 0$ is the local radial coordinate at the bubble. This defines a regular origin of a $\mathbb{R}^2$ space if one fixes the bubble size as 
\begin{equation}
    R_{y} \= \frac{2(\ell^2-M^2)^\frac{3}{2}}{\ell^2}\,.
    \label{eq:RegCond2BMPV}
\end{equation}
The loci $r=\ell$ and $\theta=0$ and $\pi$ still host two extremal black holes with opposite charges. The distinction from the four-dimensional bound state is that the near-horizon regions have an AdS$_2\times$S$^3$ topology, characterizing them as static five-dimensional BMPV black holes \cite{Breckenridge:1996is}. In four dimensions, the $\phi$-circle was the only angle degenerating at both edges of the black holes, but in the five-dimensional case, $\phi$ and $y$ are both degenerating at one edge, defining an S$^3$. The entropy of the bound state is slightly modified and is given by
\begin{equation}
   S^{(2)} \= S_1 + S_2 \= 2S_1 \=\frac{\pi(M(\ell+M))^{\frac{3}{2}}}{\sqrt{2} \ell}
\end{equation}
converging similarly towards $\frac{1}{2} S_\text{Schw}$ as $\ell \sim M\sim \cM$.

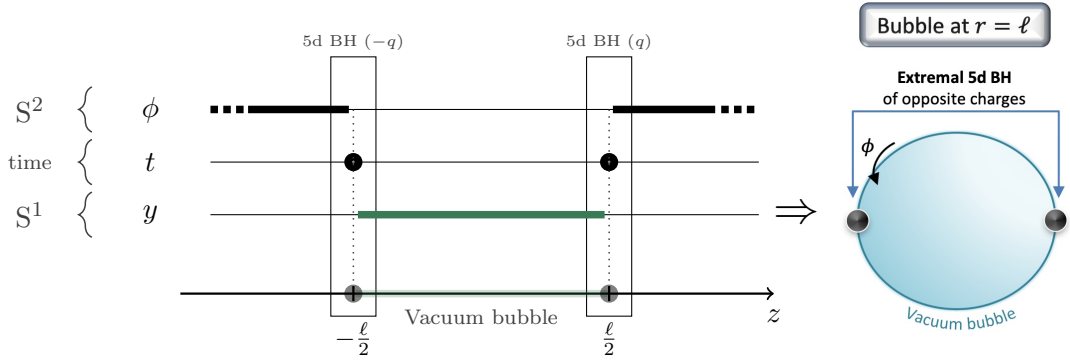
\begin{figure}[t]
\centering
    \begin{tikzpicture}
\def\deb{-10} 
\def\inter{0.7} 
\def\ha{2.8} 
\def\zaxisline{4.5} 
\def\rodsize{1.7} 
\def\numrod{1.7} 
\def\fin{\deb+1+2*\rodsize+\numrod*\rodsize} 


\draw (\deb+0.5+\rodsize+0.5*\numrod*\rodsize,\ha+2-\inter) node{{{\it Two static BMPV Black Holes on a vacuum bubble}}}; 

\draw[draw=black] (\deb+0.5+\rodsize-0.3+0.3,\ha) rectangle (\deb+0.5+\rodsize-0.3-0.3,\ha-\zaxisline*\inter-0.3) ;
\draw[gray] (\deb+0.5+\rodsize-0.3,\ha+0.2) node{{\tiny 5d BH $(-q)$}};

\draw[draw=black] (\deb+0.5+3*\rodsize-0.3+0.3,\ha) rectangle (\deb+0.5+3*\rodsize-0.3-0.3,\ha-\zaxisline*\inter-0.3) ;
\draw[gray] (\deb+0.5+3*\rodsize-0.3,\ha+0.2) node{{\tiny 5d BH $(q)$}};

\draw [decorate, 
    decoration = {brace,
        raise=5pt,
        amplitude=5pt},line width=0.2mm,gray] (\deb-1.4,\ha-1.5*\inter+0.05) --  (\deb-1.4,\ha-0.5*\inter-0.05);
        \draw [decorate, 
    decoration = {brace,
        raise=5pt,
        amplitude=5pt},line width=0.2mm,gray] (\deb-1.4,\ha-2.5*\inter+0.05) --  (\deb-1.4,\ha-1.5*\inter-0.05);
                \draw [decorate, 
    decoration = {brace,
        raise=5pt,
        amplitude=5pt},line width=0.2mm,gray] (\deb-1.4,\ha-3.5*\inter+0.05) --  (\deb-1.4,\ha-2.5*\inter-0.05);
        
\draw[gray] (\deb-2.4,\ha-1*\inter) node{S$^2$};
\draw[gray] (\deb-2.4,\ha-2*\inter) node{{\scriptsize time}};
\draw[gray] (\deb-2.4,\ha-3*\inter) node{S$^1$};


\draw[black,thin] (\deb+1,\ha-\inter) -- (\fin-1,\ha-\inter);
\draw[black,thin] (\deb,\ha-2*\inter) -- (\fin,\ha-2*\inter);
\draw[black,thin] (\deb,\ha-3*\inter) -- (\fin,\ha-3*\inter);

\draw[black,->,line width=0.3mm] (\deb-0.4,\ha-\zaxisline*\inter) -- (\fin+0.2,\ha-\zaxisline*\inter);

\draw (\deb-0.8,\ha-\inter) node{$\phi$};
\draw (\deb-0.8,\ha-2*\inter) node{$t$};
\draw (\deb-0.8,\ha-3*\inter) node{$y$};

\draw (\fin+0.2,\ha-\zaxisline*\inter-0.3) node{$z$};


\draw[black, dotted, line width=1mm] (\deb,\ha-\inter) -- (\deb+0.5,\ha-\inter);
\draw[black,line width=1mm] (\deb+0.5,\ha-\inter) -- (\deb+0.5+\rodsize-0.36,\ha-\inter);
\draw[black,line width=1mm] (\fin-0.44-\rodsize+0.2,\ha-1*\inter) -- (\fin-0.58,\ha-1*\inter);
\draw[black, dotted,line width=1mm] (\fin-0.5,\ha-1*\inter) -- (\fin,\ha-1*\inter);


\draw[amazon,line width=1mm] (\deb+0.5+\rodsize-0.24,\ha-3*\inter) -- (\deb+0.44+3*\rodsize-0.3,\ha-3*\inter);

\draw[black,line width=1mm] (\deb+0.5+\rodsize-0.3,\ha-2*\inter) circle[radius=2pt];
\draw[black,line width=1mm] (\deb+0.5+3*\rodsize-0.3,\ha-2*\inter) circle[radius=2pt];


\draw[amazon,line width=1mm,opacity=0.25] (\deb+0.5+\rodsize-0.27,\ha-\zaxisline*\inter) -- (\deb+0.5+3*\rodsize-0.32,\ha-\zaxisline*\inter);
\draw[black,line width=1mm,opacity=0.5] (\deb+0.5+\rodsize-0.3,\ha-\zaxisline*\inter) circle[radius=2pt];
\draw[black,line width=1mm,opacity=0.5] (\deb+0.5+3*\rodsize-0.3,\ha-\zaxisline*\inter) circle[radius=2pt];


\draw[gray,dotted,line width=0.2mm] (\deb+0.5+\rodsize-0.3,\ha-\inter) -- (\deb+0.5+\rodsize-0.3,\ha-\zaxisline*\inter);
\draw[gray,dotted,line width=0.2mm] (\deb+0.5+3*\rodsize-0.3,\ha-\inter) -- (\deb+0.5+3*\rodsize-0.3,\ha-\zaxisline*\inter);

\draw[line width=0.3mm] (\deb+0.5+\rodsize-0.3,\ha-\zaxisline*\inter+0.1) -- (\deb+0.5+\rodsize-0.3,\ha-\zaxisline*\inter-0.1);
\draw[line width=0.3mm] (\deb+0.5+3*\rodsize-0.3,\ha-\zaxisline*\inter+0.1) -- (\deb+0.5+3*\rodsize-0.3,\ha-\zaxisline*\inter-0.1);

\draw (\deb+0.5+1*\rodsize-0.3,\ha-\zaxisline*\inter-0.6) node{{\small $-\frac{\ell}{2}$}};
\draw (\deb+0.5+3*\rodsize-0.3,\ha-\zaxisline*\inter-0.6) node{{\small $\frac{\ell}{2}$}};

\draw[gray] (\deb+0.5+2*\rodsize-0.3,\ha-\zaxisline*\inter-0.3) node{{\scriptsize Vacuum bubble}};

\node[anchor=south,inner sep=-1cm] at (\fin+2.6,0) {\includegraphics[width=.2\textwidth]{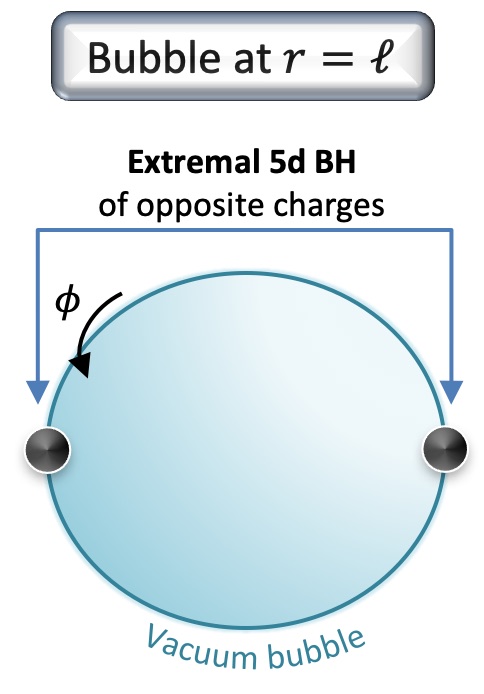}};

\draw (\fin+0.5,\ha-3*\inter) node{{\LARGE $\Rightarrow$}};

\end{tikzpicture}
\caption{Spacetime structure of two BMPV black holes of opposite charges.}
\label{fig:rodsourceBMPVBS2}
\end{figure}  

We have successfully constructed a regular bound state of two BMPV black holes with opposite charges, separated by a vacuum bubble. The spacetime structure is illustrated in Fig. \ref{fig:rodsourceBMPVBS2}. Unlike the four-dimensional construction with struts, the spacetime is generated only by physical sources, representing well-defined gravitational entities.

Moreover, the regularity condition \eqref{eq:RegCond2BMPV} fixes the size of the bubble, determining the overall size of the configuration in terms of the ADM mass, $\cM$, and the radius of the extra dimension, $R_{y}$. Consequently, $\ell-M$ cannot be freely modified anymore, freezing the geometry and leading to significant physical consequences:
\begin{itemize}
    \item The regularity constraint at fixed $R_{y}$ imposes:
    \begin{equation}
        \cM \geq \frac{R_{y}}{8}\,,\qquad \ell \geq \frac{R_{y}}{2}\,.
    \end{equation}
    This sets a lower bound on the validity range of the bound state, ensuring it cannot exist for a mass lower than the Kaluza-Klein scale.
    \item Macroscopic solutions, with $\cM \gg R_y$, inherently approach their entrapment limit:
    \begin{equation}
    \ell \sim \cM \left(1+\frac{3\epsilon}{8} \right)\,,\qquad M \sim \cM  \left(1-\frac{\epsilon}{8} \right)\,,\qquad \epsilon \= \left( \frac{R_{y}}{2 \cM} \right)^\frac{2}{3} \ll 1\,.
    \label{eq:MacroscopicConstraint}
    \end{equation}
    The geometry becomes indistinguishable from a rigid S$^1$ over a distorted Schwarzschild black hole:
    \begin{equation}
    ds_{5}^2 \sim ds_\text{dis-Schw}^2 +  dy^2\,,\qquad F \sim 0\,.
    \label{eq:FiberDisSch}
    \end{equation}
    for $r\gtrsim \cM(1+\cO(\epsilon^\frac{1}{4}))$. It replaces the horizon with a regular spacetime structure formed by a smooth topological bubble and extremal black holes. The structure is confined at the horizon scale in an infinitesimal region set by $\epsilon$, the size of the extra dimension. Conversely, the constraint \eqref{eq:MacroscopicConstraint} forces the geometries to be in their black-hole-like regime, describing a novel form of gravitational matter that only exists at the horizon of black holes.

    Additionally, the high-redshift region has entrapped not only the electromagnetic flux but also the topological deformation required to form the geometry. The metric \eqref{eq:FiberDisSch} indicates a sudden transformation from a rigid S$^1$ to a degenerating S$^1$, forming the bubble and the five-dimensional black holes.

    \item The regularity constraint implies that:
    \begin{equation}
         \left. \frac{\partial \ell}{\partial \cM} \right|_{R_y=\text{cst}} \,>\, 0\,,\quad \text{and}\qquad  \left. \frac{\partial \ell}{\partial \cM} \right|_{R_y=\text{cst}} \underset{\cM \gg R_y}{\sim} 1\,.\label{eq:GrowthGN}
    \end{equation}
    Therefore, The structure necessarily grows with the Newton constant and approaches a linear growth for macroscopic objects.\footnote{We remind that we work in unit of $G_4=1$. So the relation \eqref{eq:GrowthGN} implies derivatives in terms of $G_4 \cM$.} This characteristic is uncommon for any form of matter in gravity, as ordinary matter usually contracts as gravity intensifies. In contrast, black holes, as entities that expand in size with growing gravitational energy, display a linear growth when static. This criterion stands as a crucial factor for constructing pertinent horizon-scale microstructures, a criterion automatically satisfied by our construction.
\end{itemize}

In summary, the embedding in five dimensions has provided a more comprehensive definition of the geometries as novel horizon-scale structures. It has restricted their existence to their entrapment limit, making macroscopic solutions indistinguishable from distorted Schwarzschild black holes. The structure grows as gravity intensifies, similar to black holes. Importantly, a simple bound state of two extremal black holes spans half of the entropy of the Schwarzschild black hole and its distorted version. This phase space has a clear description in terms of electromagnetic degrees of freedom emerging at the horizon scale from electromagnetic entrapment, shedding light on the microstructure of black holes in vacuum.

\subsubsection{$N$ extremal black holes on a KK bubble}

Similarly to the four-dimensional constructions, one can solve the fields $(V,A,\nu)$ using the generic solution for $N$ extremal sources given in Appendix \ref{app:NExtSources}. The solutions will closely resemble the four-dimensional ones analyzed in section \ref{sec:NExBH4d}, exhibiting the same entrapment limit toward the distorted Schwarzschild geometry for macroscopic solutions, with black holes arranged along a line. The key distinction is that the black holes are no longer separated by struts but are distributed along a vacuum bubble.

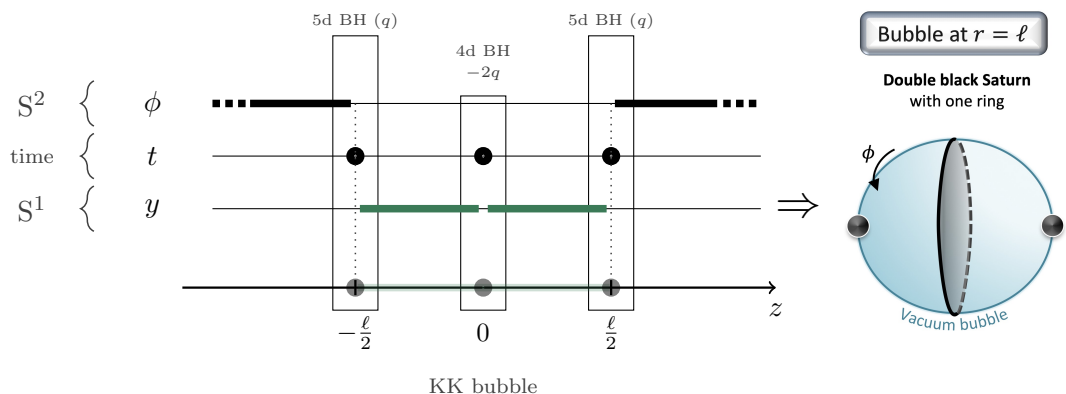
\begin{figure}[t]
\centering
    \begin{tikzpicture}
\def\deb{-10} 
\def\inter{0.7} 
\def\ha{2.8} 
\def\zaxisline{4.5} 
\def\rodsize{1.7} 
\def\numrod{1.7} 
\def\fin{\deb+1+2*\rodsize+\numrod*\rodsize} 


\draw (\deb+0.5+\rodsize+0.5*\numrod*\rodsize,\ha+2-\inter) node{{{\it Three Extremal Black Holes on a KK bubble}}}; 

\draw[draw=black] (\deb+0.5+\rodsize-0.3+0.3,\ha+0.2) rectangle (\deb+0.5+\rodsize-0.3-0.3,\ha-\zaxisline*\inter-0.3) ;
\draw[gray] (\deb+0.5+\rodsize-0.3,\ha+0.4) node{{\tiny 5d BH $(q)$}};

\draw[draw=black] (\deb+0.5+2*\rodsize-0.3+0.3,\ha-0.6) rectangle (\deb+0.5+2*\rodsize-0.3-0.3,\ha-\zaxisline*\inter-0.3) ;
\draw[gray] (\deb+0.5+2*\rodsize-0.3,\ha) node{{\tiny 4d BH}};
\draw[gray] (\deb+0.5+2*\rodsize-0.3,\ha-0.3) node{{\tiny $-2q$}};

\draw[draw=black] (\deb+0.5+3*\rodsize-0.3+0.3,\ha+0.2) rectangle (\deb+0.5+3*\rodsize-0.3-0.3,\ha-\zaxisline*\inter-0.3) ;
\draw[gray] (\deb+0.5+3*\rodsize-0.3,\ha+0.4) node{{\tiny 5d BH $(q)$}};

\draw [decorate, 
    decoration = {brace,
        raise=5pt,
        amplitude=5pt},line width=0.2mm,gray] (\deb-1.4,\ha-1.5*\inter+0.05) --  (\deb-1.4,\ha-0.5*\inter-0.05);
        \draw [decorate, 
    decoration = {brace,
        raise=5pt,
        amplitude=5pt},line width=0.2mm,gray] (\deb-1.4,\ha-2.5*\inter+0.05) --  (\deb-1.4,\ha-1.5*\inter-0.05);
                \draw [decorate, 
    decoration = {brace,
        raise=5pt,
        amplitude=5pt},line width=0.2mm,gray] (\deb-1.4,\ha-3.5*\inter+0.05) --  (\deb-1.4,\ha-2.5*\inter-0.05);
        
\draw[gray] (\deb-2.4,\ha-1*\inter) node{S$^2$};
\draw[gray] (\deb-2.4,\ha-2*\inter) node{{\scriptsize time}};
\draw[gray] (\deb-2.4,\ha-3*\inter) node{S$^1$};


\draw[black,thin] (\deb+1,\ha-\inter) -- (\fin-1,\ha-\inter);
\draw[black,thin] (\deb,\ha-2*\inter) -- (\fin,\ha-2*\inter);
\draw[black,thin] (\deb,\ha-3*\inter) -- (\fin,\ha-3*\inter);

\draw[black,->,line width=0.3mm] (\deb-0.4,\ha-\zaxisline*\inter) -- (\fin+0.2,\ha-\zaxisline*\inter);

\draw (\deb-0.8,\ha-\inter) node{$\phi$};
\draw (\deb-0.8,\ha-2*\inter) node{$t$};
\draw (\deb-0.8,\ha-3*\inter) node{$y$};

\draw (\fin+0.2,\ha-\zaxisline*\inter-0.3) node{$z$};


\draw[black, dotted, line width=1mm] (\deb,\ha-\inter) -- (\deb+0.5,\ha-\inter);
\draw[black,line width=1mm] (\deb+0.5,\ha-\inter) -- (\deb+0.5+\rodsize-0.36,\ha-\inter);
\draw[black,line width=1mm] (\fin-0.44-\rodsize+0.2,\ha-1*\inter) -- (\fin-0.58,\ha-1*\inter);
\draw[black, dotted,line width=1mm] (\fin-0.5,\ha-1*\inter) -- (\fin,\ha-1*\inter);


\draw[amazon,line width=1mm] (\deb+0.5+\rodsize-0.24,\ha-3*\inter) -- (\deb+0.44+2*\rodsize-0.3,\ha-3*\inter);
\draw[amazon,line width=1mm] (\deb+0.5+2*\rodsize-0.24,\ha-3*\inter) -- (\deb+0.44+3*\rodsize-0.3,\ha-3*\inter);

\draw[black,line width=1mm] (\deb+0.5+\rodsize-0.3,\ha-2*\inter) circle[radius=2pt];
\draw[black,line width=1mm] (\deb+0.5+2*\rodsize-0.3,\ha-2*\inter) circle[radius=2pt];
\draw[black,line width=1mm] (\deb+0.5+3*\rodsize-0.3,\ha-2*\inter) circle[radius=2pt];


\draw[amazon,line width=1mm,opacity=0.25] (\deb+0.5+\rodsize-0.27,\ha-\zaxisline*\inter) -- (\deb+0.5+3*\rodsize-0.32,\ha-\zaxisline*\inter);
\draw[black,line width=1mm,opacity=0.5] (\deb+0.5+\rodsize-0.3,\ha-\zaxisline*\inter) circle[radius=2pt];
\draw[black,line width=1mm,opacity=0.5] (\deb+0.5+2*\rodsize-0.3,\ha-\zaxisline*\inter) circle[radius=2pt];
\draw[black,line width=1mm,opacity=0.5] (\deb+0.5+3*\rodsize-0.3,\ha-\zaxisline*\inter) circle[radius=2pt];


\draw[gray,dotted,line width=0.2mm] (\deb+0.5+\rodsize-0.3,\ha-\inter) -- (\deb+0.5+\rodsize-0.3,\ha-\zaxisline*\inter);
\draw[gray,dotted,line width=0.2mm] (\deb+0.5+3*\rodsize-0.3,\ha-\inter) -- (\deb+0.5+3*\rodsize-0.3,\ha-\zaxisline*\inter);

\draw[line width=0.3mm] (\deb+0.5+\rodsize-0.3,\ha-\zaxisline*\inter+0.1) -- (\deb+0.5+\rodsize-0.3,\ha-\zaxisline*\inter-0.1);
\draw[line width=0.3mm] (\deb+0.5+3*\rodsize-0.3,\ha-\zaxisline*\inter+0.1) -- (\deb+0.5+3*\rodsize-0.3,\ha-\zaxisline*\inter-0.1);

\draw (\deb+0.5+1*\rodsize-0.3,\ha-\zaxisline*\inter-0.6) node{{\small $-\frac{\ell}{2}$}};
\draw (\deb+0.5+2*\rodsize-0.3,\ha-\zaxisline*\inter-0.6) node{{\small $0$}};
\draw (\deb+0.5+3*\rodsize-0.3,\ha-\zaxisline*\inter-0.6) node{{\small $\frac{\ell}{2}$}};

\draw[gray] (\deb+0.5+2*\rodsize-0.3,\ha-\zaxisline*\inter-1.3) node{{\scriptsize KK bubble}};

\node[anchor=south,inner sep=-1cm] at (\fin+2.6,0) {\includegraphics[width=.2\textwidth]{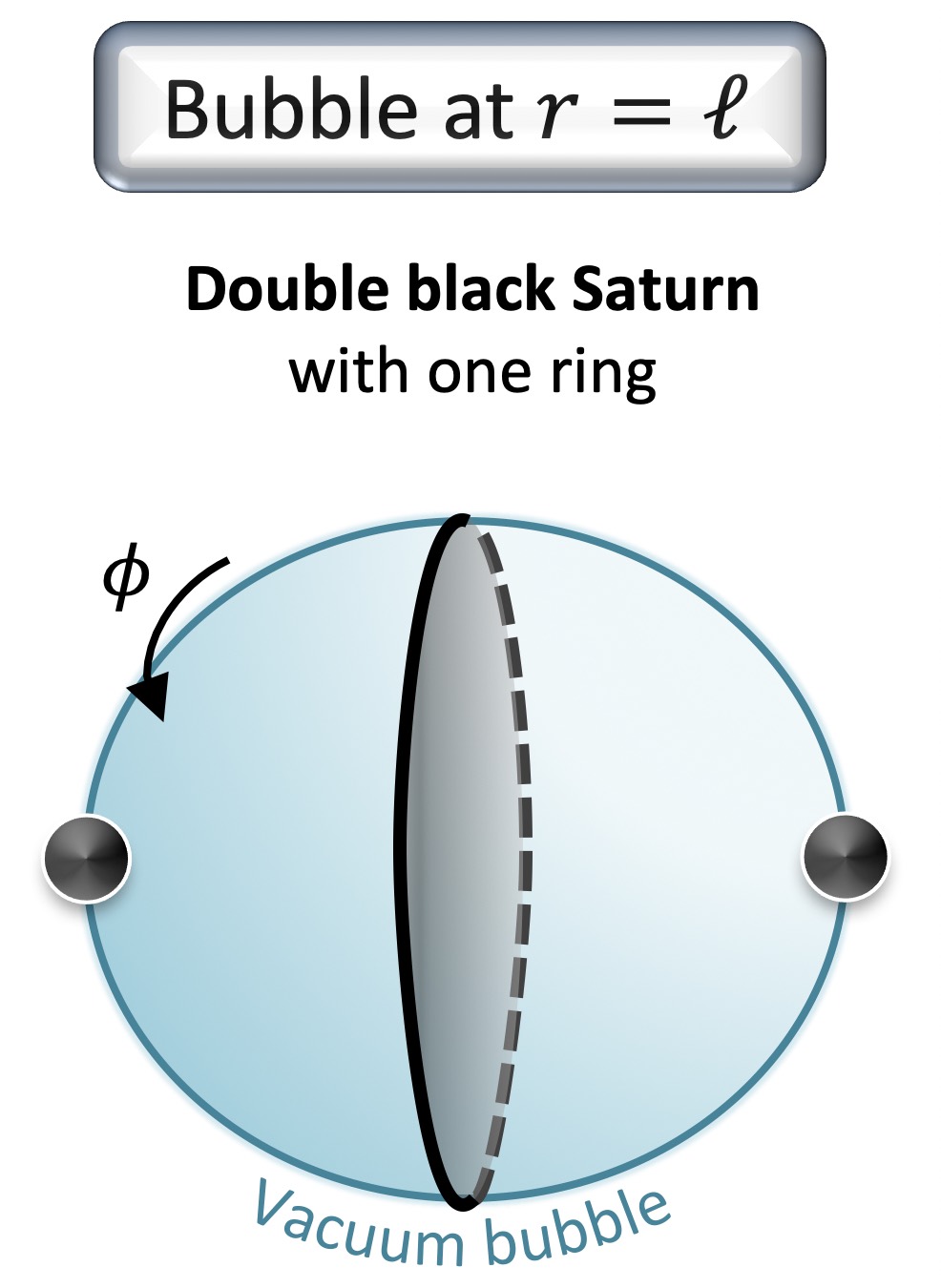}};

\draw (\fin+0.5,\ha-3*\inter) node{{\LARGE $\Rightarrow$}};

\end{tikzpicture}
\caption{Spacetime structure of three extremal black holes on a KK bubble.}
\label{fig:rodsourceBMPVBS35d}
\end{figure}  

In Fig.\ref{fig:rodsourceBMPVBS35d}, we have illustrated a generic spacetime structure at the coordinate boundary $r=\ell$ using the same three-black hole bound state as in section \ref{sec:NExBH4d}. As described in the figure, the extremal black holes in the middle extend along the $\phi$-circle, resulting in a horizon topology of S$^2\times$S$^1$, unlike the S$^3$ horizons of the black holes at the edge. Therefore, the configurations resemble extremal and static double black Saturn geometries \cite{Elvang:2007rd}, with as many rings as there are extremal sources in the middle of the bubble.

Moreover, the regularity of the bubble requires imposing $N-1$ constraints, so that each segment between black holes corresponds to a smooth $\IR^2$ space, $d\bar{r}^2 + \bar{r}^2 \frac{dy^2}{R_y^2}$, as in \eqref{eq:R2localSpace}. Therefore, from $2(N-1)$ free parameters characterizing the parameter space of the geometries in four dimensions, we are now reduced to $N-2$ parameters by considering the ADM mass fixed as well. These parameters can be associated with some freedom in the local black-hole charges. These local charges are necessarily bounded by a function of the ADM mass and quantized in terms of an elementary charge, so that the parameter space is now discrete. 

Therefore, the total entropy accounting for all bound state is
\begin{equation}
    S^{(T)} \= \log \left(\sum_{\text{config.}} e^{S^{(n)}}\right) \approx \log \left(\sum_{\text{config.}} e^{\frac{1}{2} S_\text{Schw}} \right) \= \frac{1}{2} S_\text{Schw} + \log \left( \#_\text{config.}\right).
    \label{eq:TotalEntropyConf}
\end{equation}
Having a clear estimate of the number of configurations, $\#_\text{config.}$, is highly involved and could be the subject of future study. Nevertheless, the phase space features an entropy that necessarily ranges in between $\frac{1}{2}S_\text{Schw}$ and $S_\text{Schw}$, and having exactly $S_\text{Schw}$ is not unrealistic.

\vspace{0.2cm}

In conclusion, our analysis has shown that electromagnetic entrapment allows for the construction of explicit microstructures of black geometries in a vacuum. We have constructed extensive spacetime structures with a phase space as vast as the Schwarzschild entropy that are indistinguishable from a distorted Schwarzschild black hole but replace its near horizon with characterizable structures generated by entrapped electromagnetic fields and topological deformations. As such, these structures offer a comprehensive description of the degrees of freedom within neutral black holes. This characteristic has been illustrated in Fig.\ref{fig:IntroFig} in the introduction.

\subsection{Toward coherent microstates}

The replacement of a vacuum black hole with bound states of extremal black holes has enabled a comparison of their phase space of microstates. However, this process still involves substituting black holes with other black holes. Consequently, the constructed bound states should not be regarded as complete classical resolutions or specific microstate descriptions.

To achieve a more thorough resolution, further steps are needed, where extremal black holes are replaced by smooth and horizonless topological microstructures supported by an equivalent amount of electromagnetic flux. Such geometries, referred to as \emph{topological solitons}, correspond to specific (likely atypical) microstates of the phase space. While not expected to cover the entire phase space, constructing coherent microstates represents a culmination point where electromagnetic entrapment at the horizon scale allows the explicit construction of black-hole microstates.

In this section, while we do not construct topological solitons explicitly, we discuss how this can be achieved using existing work. For simplicity, we only discuss the resolution of the bound state of two BMPV black holes on a vacuum bubble.

To construct topological solitons corresponding to explicit microstates of the distorted Schwarzschild black hole, one needs to ``dilute'' the electromagnetic flux generating the extremal black holes into smooth topological microstructures. This classical resolution can be approached through two potential scenarios. First, one can use the extensive knowledge on the microstructure of BMPV black holes developed in the microstate geometry program \cite{Warner:2019jll,Bena:2022ldq,Bena:2022rna}. Second, one can replace the extremal BMPV black holes with near-extremal topological solitons using the integrability of the Ernst formalism and the procedure developed in \cite{Bah:2022yji,Bah:2023ows}.

In the first approach, BMPV black holes, being supersymmetric black holes in string theory, possess smooth horizonless microstate geometries for a fraction of their microstates \cite{Shigemori:2019orj,Mayerson:2020acj}. Two known categories of such microstructures are superstrata \cite{Bena:2015bea,Bena:2017xbt,Ceplak:2018pws,Heidmann:2019zws,Heidmann:2019xrd,Shigemori:2020yuo} and multicenter bubbling geometries \cite{Bena:2007kg,Heidmann:2017cxt,Bena:2017fvm,Bena:2018bbd,Heidmann:2018vky}.\footnote{Although additional generic microstate geometries with a larger phase space are anticipated, their supergravity descriptions remain incomplete \cite{Bena:2022wpl,Bena:2022fzf}.} To resolve the bound state of interest, composed of two BMPV black holes on opposite sides of a vacuum bubble, each individual black hole can be locally resolved deep within their AdS$_2$ near-horizon throats. The backreaction can then be handled by using scaling microstates. This approach resolves the BMPV black hole horizons while maintaining the sources locally extremal. This uses  known smooth microstructures of supersymmetric black holes in string theory, thereby making a direct connection between the microstructures of neutral black holes and extremal black holes.

The second approach maintains the integrable structure of the equations, replacing extremal black holes with near-extremal bubbles  \cite{Bah:2022yji,Bah:2023ows}. In this context, each BMPV black hole is substituted by a chain of topological stars \cite{Bah:2020ogh}. This resolution scheme differs from the first approach by not preserving the local extremality of the sources. Despite this, the advantage lies in achieving a fully analytical construction by using the Ernst formalism in higher dimensions, as demonstrated in other similar contexts \cite{Bah:2022yji,Bah:2023ows}. 

In summary, electromagnetic entrapment provides a concrete description of the emerging physics at the horizon scale, uncovering an extensive phase space of electromagnetic and topological systems spreading within the near-horizon region.  In this context, topological solitons emerge as coherent, smooth, horizonless manifestations of these novel degrees of freedom at the horizon scale, revealing a classical mechanism in gravity that aptly describes the highly-entropic microstructure of black holes.

\section{Conclusion and outlook}
\label{sec:Conclusion}

In this paper, we have highlighted a significant interplay between electromagnetism and gravity through the principle of electromagnetic entrapment. Electromagnetic entrapment introduces novel dynamics to electromagnetic degrees of freedom in highly-redshifted spacetimes, shedding light on new physics at the horizon scale.

We first demonstrated this principle through probe derivations, where electromagnetic fields were generated by test particles in fixed gravitational backgrounds. Subsequently, we constructed explicit geometries produced by massive and charged sources in four dimensions and in five dimensions with one dimension being compact. We showed that electromagnetic entrapment affects their dynamics, leading to a critical size where they become indistinguishable from a vacuum black geometry. Notably, these geometries differ only in the near-horizon region by replacing the vacuum horizon with unique gravitational microstructures that remain entrapped.

The five-dimensional solutions, in particular, provide a comprehensive description of these microstructures in terms of smooth topological configurations with extremal black holes distributed along the topology. We showed that these structures meet the criteria to describe the novel gravitational phase of matter that is expected to emerge at the horizon scale. They are confined to exist at this scale, they are as compact as the black hole horizon, they grow in size as the Newton constant increases, and they have an extensive phase space that scales like $M^2$. Finally, we argue that electromagnetic entrapment also allows the existence of coherent microstates described in terms of smooth horizonless geometries produced by topological and electromagnetic degrees of freedom --- topological solitons. 
\vspace{0.1cm}

Electromagnetic entrapment raises numerous questions that we aim to explore in future investigations. First, our probe derivation demonstrates that this is a highly generic property not restricted to black holes, potentially yielding significant consequences in other domains of physics. An intriguing avenue is the impact of electromagnetic entrapment on an infalling observer. As electromagnetic interactions diminish at the horizon, this could result in an ``electromagnetic spaghettization'', affecting molecular interactions and atom quantization. For example, examining the quantization of energy for a hydrogen atom near a Schwarzschild black hole and the extension of electromagnetic entrapment to non-abelian gauge fields could be interesting in that matter. Additionally, application of these properties to other highly-redshifted spacetimes, such as neutron stars or the early universe, could provide interesting insights.

Generalizing the entrapment effect to other types of fields is another avenue of exploration. The potential extension to non-abelian gauge fields and the evaluation of its impact on weak and strong interactions could unravel new phenomena. Moreover, we have indirectly demonstrated in this paper that highly-redshifted regions of spacetime can entrap topological deformations, which, from a lower-dimensional perspective, correspond to scalar fields. Investigating whether scalar gauge theory in highly-redshifted curved spacetime develops an entrapment feature is an intriguing prospect that could explain physics of non-supersymmetric topological solitons \cite{Bah:2021owp,Bah:2021rki,Heidmann:2021cms,Bah:2022yji,Bah:2023ows} or even boson stars \cite{JETZER1992163}.

Regarding our current constructions, a significant aspect requiring further study is the fact that we resolved a distorted Schwarzschild spacetime rather than the Schwarzschild black hole itself. It is crucial to address whether our discussion can be extended to Schwarzschild or is specific to this deformed spacetime. We believe the former scenario is plausible and that the convergence of all constructed geometries to a distorted Schwarzschild geometry is a consequence of restrictive assumptions. Our restriction to  extremal sources played a role, and we anticipate that resolving the Schwarzschild black hole can be achieved with more ``extended'' charge configurations in their entrapment limit — an area we plan to explore in the future. 

Moreover, it will be interesting to quantify more rigorously the phase space described by the horizon-scale structures and compare to the Schwarzschild entropy. While we have successfully shown that most of the five-dimensional constructions have an entropy of $\frac{1}{2} S_\text{Schw}$, one still needs to look for a more microscopic description of this entropy. This can be done by generalizing the construction to string theory where the electromagnetic degrees of freedom are associated to brane/anti-brane dynamics. This will be the subject of an upcoming study \cite{PRLHeidmann}.

A second pivotal area for study concerns the stability of the bound states. In flat space, massive particles and anti-particles experience strong gravitational and electromagnetic attraction. Even if electromagnetic entrapment and topological pressure mitigate these attractions to some extent, the stability of bound states of extremal and anti-extremal black holes on a KK bubble is uncertain. Notably, neutral black holes are thermodynamically unstable under Hawking evaporation. Evaluating the instability of the bound states as different centers exchange particles and anti-particles and understanding if this radiation of the microstructure connects to Hawking radiation, as in previous studies \cite{Chowdhury:2007jx,Bena:2015dpt}, is an intriguing avenue for further investigation.

\vspace{-0.2cm}
\section*{Acknowledgements}
We  would like to thank  Ibrahima Bah, Iosif Bena, Bogdan Ganchev, Marcel Hughes, Enoch Leung, Brandon Manley, Samir D. Mathur and Thomas Waddleton for useful discussions. The work of PH is supported by the Department of Physics at The Ohio State University. The work of MM is supported by DOE grant DE-SC0011726.

\appendix

\section{Static Ernst formalism}
\label{app:Ernst}

In this section, we review the static Ernst formalism derived from  Einstein-Maxwell theory in four dimensions, along with the solutions associated with $N$ extremal point charges positioned along a line.  We end the section by reviewing the generalization of the static Ernst formalism in M-theory as initially derived in \cite{Heidmann:2021cms}.

\subsection{Equations of motion}
\label{app:ErnstEOM}

We consider the four-dimensional Einstein-Maxwell theory given by the action \eqref{eq:4daction}.The static Ernst formalism involves constraining the solution to axially symmetric conditions, as provided by the ansatz:
\begin{equation}
\begin{split}
ds_4^2 &\= - V^{-1} \,dt^2+ V \left[e^{4\nu}\left( d\rho^2+dz^2\right) +\rho^2 d\phi^2 \right],\\
 F &\= - \cos (\eta) \, dA\wedge dt +  \sin (\eta) \, dH \wedge d\phi\,.
\end{split}
\label{eq:4dErnstAnsApp}
\end{equation}
where $V,A,H$ and $\nu$ depends on $(\rho,z)$, $\eta$ is a constant dyonic parameter and the electric and magnetic potentials satisfy the duality relation:
\begin{equation}
    \star_2 dH = \rho V dA\,,
    \label{eq:MagDual}
\end{equation}
where $\star_2$ is the Hodge dual in the flat $(\rho,z)$ subspace.

The Einstein-Maxwell equations lead to
\begin{align}
& \Delta \log V + \frac{V}{2} \,  \nabla A . \nabla A \,=\, 0\,, \qquad  \nabla .  \left( \rho V \nabla A\right) \,=\, 0 \,,\nonumber \\
&\frac{4}{\rho}\, \partial_z \nu \,=\, \partial_\rho \log V \,\partial_z \log V - V \,\partial_\rho  A\partial_z  A\,, \label{eq:EOMErnst}\\
& \frac{8}{\rho}\, \partial_\rho \nu \,=\,  \left( \partial_\rho \log V\right)^2 - \left(\partial_z \log V\right)^2 - V \, \left((\partial_\rho A)^2-(\partial_z A)^2 \right)\,, \nonumber
\end{align}
where we have defined the gradient and Laplacian:
\begin{equation}
\Delta \equi \frac{1}{\rho} \,\partial_\rho \left( \rho \,\partial_\rho\right) + \partial_z^2\,,\qquad \nabla \equi (\partial_\rho,\partial_z)\,.
\end{equation}
Those equations correspond to the electrostatic limit of the Ernst equations with the following Ernst potentials:
\begin{equation}
    \Psi= \tfrac{1}{2}\,A,\qquad \mathcal{E}= V^{-1} - \tfrac{1}{4}\,A^2.
    \label{eq:RelationErnstGravElecPot}
\end{equation}
More precisely,  the equations \eqref{eq:EOMErnst} can be written in the generic Ernst form \begin{align}
&(\text{Re}(\mathcal{E}) + \Psi^* \Psi ) \,\Delta \mathcal{E} = (\nabla \mathcal{E}  +2 \Psi^* \nabla \Psi) \nabla \mathcal{E}\,,\quad (\text{Re}(\mathcal{E})+ \Psi^* \Psi ) \,\Delta \Psi = (\nabla \mathcal{E} +2 \Psi^* \nabla \Psi) \nabla \Psi\,, \nn\\
&(\text{Re}(\mathcal{E}) + \Psi^* \Psi ) \,R^{(3)}_{ij} = \frac{1}{2} \partial_{(i}\mathcal{E} \, \partial_{j)}\mathcal{E}^* + \Psi \,\partial_{(i} \mathcal{E} \,\partial_{j)} \Psi^* + \Psi^* \,\partial_{(i} \mathcal{E}^* \,\partial_{j)} \Psi - (\mathcal{E}+\mathcal{E}^*)\, \partial_{(i} \Psi \,\partial_{j)} \Psi^*\,,
\end{align}
where $R^{(3)}$ is the Ricci tensor of the three-dimensional base $ds_3^2= e^{4\nu}\left( d\rho^2+dz^2\right) +\rho^2 d\phi^2$.

Note that in the static case, $\mathcal{E}$  and $\Psi$ are real functions, which simplifies the equations to the static Ernst equations:
\begin{equation}
    \Delta \mathcal{E}=\nabla\left[\log \left(\mathcal{E}+\Psi^2\right)\right]  \nabla \mathcal{E}, \quad \Delta \Psi=\nabla\left[\log \left(\mathcal{E}+\Psi^2\right)\right] \nabla \Psi
\end{equation}

\subsection{Solutions for $N$ extremal sources}
\label{app:NExtSources}

In \cite{NoraBreton1998}, a solution sourced by $N$ rods, which are segments on the symmetry, has been derived. The solution is characterized by the $2N$ rod coordinates, $[\alpha_{2k-1},\alpha_{2k}]$, and $N$ additional parameters $\beta_k$. In four dimensions, this solution corresponds to a chain of Reissner-Nordstr\"om black holes on a line.

In this section, our focus lies on the extremal limit where all rods degenerate into point sources, specifically $\alpha_{2k-1} \to \alpha_{2k} \equiv \alpha_k$. Consequently, the solution is described in terms of $2N$ parameters, denoted as $(\alpha_k, \beta_k)$ for $k=1,\ldots,N$. As suggested in \cite{VladimirManko2000}, it appears that the extremal limit is not a trivial limit of the general solution presented in \cite{NoraBreton1998}. Therefore, we extend the results of \cite{VladimirManko2000} to a solution of the static Ernst equations generated by $N$ extremal sources.

These solutions are obtained using Sibgatullin's method \cite{Manko_1993} from two generating functions $(e(z),f(z))$:
\begin{equation}
    e(z) \= 1 + \sum_{n=1}^N \frac{e_n}{z-\beta_n}\,,\qquad f(z) = \sum_{n=1}^N \frac{f_n}{z-\beta_n}.
\end{equation}
These generating functions are named ``axis data" as they correspond to the Ernst potentials above the sources on the symmetry axis:
\begin{equation}
 \mathcal{E}(\rho=0,z\geq \alpha_N) = e(z)\,,\qquad  \Psi(\rho=0,z\geq \alpha_N) = f(z).
 \label{eq:AxisDataPotApp}
\end{equation}
The axis data determine the positions of the extremal centers as the zeroes of their gravitational part:
\begin{equation}
    e(z)+f(z)^2 =  \left(\prod_{n=1}^N \frac{z-\alpha_n}{z-\beta_n}\right)^2.
\end{equation}
This fixes the $(e_n,f_n)$ parameters in terms of $(\alpha_n,\beta_n)$ as
\begin{equation}
\begin{split}
    f_k^2 &\= (\beta_k-\alpha_k)^2 \,\left(\prod_{n\neq k} \frac{\beta_k-\alpha_n}{\beta_k-\beta_n}\right)^2\,,\\
    e_k &\=\left.\frac{\mathrm{d}}{\mathrm{d} z}\left((z-\alpha_k)^2 \,\left(\prod_{n\neq k} \frac{z-\alpha_n}{z-\beta_n}\right)^2\right)\right|_{z=\beta_k}-2 f_k \sum_{n \neq k} \frac{f_n}{\beta_k-\beta_n}.
    \label{eq:AxisDataNApp}
\end{split}
\end{equation}
The Ernst potentials and the base factor in the whole spacetime are determined by determinants of $(2N+1)\times(2N+1)$ and $2N\times 2N$ matrices as follows
\begin{equation}
    \mathcal{E} \= \frac{E_+}{E_-}\,,\qquad \Psi \= \frac{F}{E_-}\,,\qquad e^{4\nu} \= \frac{E_{+} E_{-}+F^2}{K^2 \prod_{n=1}^{N} r_n^4}.
    \label{eq:NExErnstPot}
\end{equation}
The determinants are defined by
\begin{align}
    E_\pm &\= \begin{vmatrix}
 1 & \begin{matrix} 1 &\ldots & 1 & \frac{z-\alpha_1}{r_1} & \dots & \frac{z-\alpha_N}{r_N} \end{matrix} \\
 \begin{matrix} \pm 1 \\ \vdots \\ \pm 1 \\ 0 \\ \vdots \\ 0 \end{matrix} & \text{$D$}
\end{vmatrix},   \quad   
 F \= \begin{vmatrix}
 0 & \begin{matrix} f(\alpha_1) &\ldots & f(\alpha_N) & P_1 & \dots & P_N \end{matrix} \\
 \begin{matrix} - 1 \\ \vdots \\ - 1 \\ 0 \\ \vdots \\ 0 \end{matrix} & \text{$D$}
\end{vmatrix}, \label{eq:NExMatrixDet}
\end{align}
\begin{align}
K & \= \left|\begin{array}{cccccc}
\frac{1}{\alpha_1-\beta_1} & \cdots & \frac{1}{\alpha_{N}-\beta_1} & -\frac{1}{\left(\alpha_1-\beta_1\right)^2} & \cdots & -\frac{1}{\left(\alpha_N-\beta_1\right)^2} \\
\vdots & \ddots & \vdots & \vdots & \ddots & \vdots \\
\frac{1}{\alpha_1-\beta_N} & \cdots & \frac{1}{\alpha_{N}-\beta_N} & -\frac{1}{\left(\alpha_1-\beta_N\right)^2} & \cdots & -\frac{1}{\left(\alpha_N-\beta_N\right)^2} \\
\frac{h_1\left(\alpha_1\right)}{\alpha_1-\beta_1} & \cdots & \frac{h_1\left(\alpha_{N}\right)}{\alpha_{N}-\beta_1} & \partial_{\alpha_1}\left(\frac{h_1\left(\alpha_1\right)}{\alpha_1-\beta_1}\right) & \cdots & \partial_{\alpha_N}\left(\frac{h_1\left(\alpha_N\right)}{\alpha_N-\beta_1}\right)\\
\vdots & \ddots & \vdots & \vdots & \ddots & \vdots \\
\frac{h_N\left(\alpha_1\right)}{\alpha_1-\beta_N} & \cdots & \frac{h_N\left(\alpha_{N}\right)}{\alpha_{N}-\beta_N} & \partial_{\alpha_1}\left(\frac{h_N\left(\alpha_1\right)}{\alpha_1-\beta_N}\right) & \cdots & \partial_{\alpha_N}\left(\frac{h_N\left(\alpha_N\right)}{\alpha_N-\beta_N}\right)
\end{array}\right|, \nn \hspace{3.5cm}\quad
\end{align}
where $P_j= r_j^2 \partial_{\alpha_j}\left(\frac{f(\alpha_j)}{r_j}\right)$ and the $2N\times 2N$ matrix $D$ is defined by
\begin{equation}
   D= \left(\begin{array}{cccccc}
\frac{r_1}{\alpha_1-\beta_1} & \cdots & \frac{r_{N}}{\alpha_{N}-\beta_1} & -\frac{r_1^2}{\left(\alpha_1-\beta_1\right)^2} & \cdots & -\frac{r_N^2}{\left(\alpha_N-\beta_1\right)^2} \\
\vdots & \ddots & \vdots & \vdots & \ddots & \vdots \\
\frac{r_1}{\alpha_1-\beta_N} & \cdots & \frac{r_{N}}{\alpha_{N}-\beta_N} & -\frac{r_1^2}{\left(\alpha_1-\beta_N\right)^2} & \cdots & -\frac{r_N^2}{\left(\alpha_N-\beta_N\right)^2} \\
\frac{h_1\left(\alpha_1\right)}{\alpha_1-\beta_1} & \cdots & \frac{h_1\left(\alpha_{N}\right)}{\alpha_{N}-\beta_1} & r_1^2 \partial_{\alpha_1}\left(\frac{h_1\left(\alpha_1\right)}{\left(\alpha_1-\beta_1\right) r_1}\right) & \cdots & r_N^2 \partial_{\alpha_N}\left(\frac{h_1\left(\alpha_N\right)}{\left(\alpha_N-\beta_1\right) r_N}\right)\\
\vdots & \ddots & \vdots & \vdots & \ddots & \vdots \\
\frac{h_N\left(\alpha_1\right)}{\alpha_1-\beta_N} & \cdots & \frac{h_N\left(\alpha_{N}\right)}{\alpha_{N}-\beta_N} & r_1^2 \partial_{\alpha_1}\left(\frac{h_N\left(\alpha_1\right)}{\left(\alpha_1-\beta_N\right) r_1}\right) & \cdots & r_N^2 \partial_{\alpha_N}\left(\frac{h_N\left(\alpha_N\right)}{\left(\alpha_N-\beta_N\right) r_N}\right)
\end{array}\right),
\end{equation}
and we have introduced
\begin{equation}
    r_n=\sqrt{\rho^2+\left(z-\alpha_n\right)^2}, \qquad h_l\left(\alpha_n\right)=e_l+2 f_l f\left(\alpha_n\right).
    \label{eq:DistanceAndhDef}
\end{equation}
Furthermore, we assumed that $(e_n,f_n)$ are independent of $\alpha_n$ when employing partial derivatives with respect to $\alpha_n$. For instance, we considered $\partial_{\alpha_n}(h_l(\alpha_n))= 2 f_l \,f'(\alpha_n)$.

The gravitational and electric potentials can be deduced from \eqref{eq:RelationErnstGravElecPot} as follows:
\begin{equation}
    V \= \frac{E_-^2}{E_+ E_- +F^2}\,,\qquad A \= \frac{2F}{E_-}\,.
    \label{eq:NExGravElecPot}
\end{equation}

Finally, the derivation of the magnetic dual potential, $H$, directly from \eqref{eq:MagDual} appears to be a challenging task. This obstacle can be overcome by utilizing the fact that $H$ represents the real part of Kinnersley’s potential $\Psi_2$, whose construction within Sibgatullin’s method is similar to that of the Ernst potentials $\mathcal{E}$ and $\Psi$. In \cite{PhysRevD.51.4192}, the Kinnersley’s potential was derived in the non-extremal case, from which we extracted its extremal limit:
\begin{equation}
    H \=  -\frac{2I}{E_-}\,,\qquad  I \= \begin{vmatrix}
 \sum_{n=1}^N f_n & \begin{matrix} 0 & f(\alpha_1) &\ldots & f(\alpha_N) & P_1 & \dots & P_N \end{matrix} \\
 \begin{matrix} z \\ - \beta_1 \\ \vdots \\ - \beta_N \\ e_1 \\ \vdots \\ e_N \end{matrix} & \text{$E_-$}
\end{vmatrix}.
\label{eq:NExMagnDual}
\end{equation}
This concludes the construction of the solution of Ernst equations corresponding to $N$ extremal point charges on a line.

\subsection{Solutions for two extremal sources of opposite charges}
\label{app:TwoExtSources}

The reduction of solutions sourced by $N$ charged rods to $N=2$ has received significant attention in the analysis of black hole bound states in equilibrium in four dimensions \cite{Alekseev:2007re, Alekseev:2007gt, Manko:2007hi, Manko:2008gb}. A further simplification of this solution was achieved in \cite{Bah:2022yji}. In this section, we revisit the extremal limit discussed in \cite{Bah:2022yji, Bah:2023ows}, where both point sources carry opposite charges, resulting in an overall neutral system. This solution can also be directly derived from the solution presented in the previous section for $N=2$. The $N=2$ solution is characterized by four parameters, but by enforcing opposite charges and the origin of the symmetry axis as in \eqref{eq:SizeConfAndOrigin}, only two independent parameters remain. We select these parameters to be the configuration size, $\ell=\alpha_2-\alpha_1$, and the total mass in four dimensions, $M$, expressed in terms of the $\beta_n$ in \eqref{eq:ConsChargeNex}.

The fields in \eqref{eq:4dErnstAnsApp} are given by
\begin{align}
V & = \left(1+\frac{2M(2r+M-\ell)}{(2r-\ell)^2-\ell^2\cos^2\theta-M^2 \sin^2 \theta}\right)^2, \quad e^{\nu}  =1-\frac{M^2\sin^2\theta}{(2r-\ell)^2-\ell^2\cos^2\theta} ,  \label{eq:BPSantiBPSSol}\\
A & = \frac{4M \sqrt{\ell^2-M^2}\,\cos\theta}{(2r+M-\ell)^2-(\ell^2-M^2)\cos^2\theta} \,,\quad H =\frac{2M \sqrt{\ell^2-M^2}(2r+M-\ell)\,\sin^2\theta}{(2r-\ell)^2-\ell^2\cos^2\theta-M^2 \sin^2 \theta}\,.\nonumber 
\end{align}
where $(r,\theta)$ are the spherical coordinates centered around the configuration and given in terms of $(\rho,z)$ as
\begin{equation}
    r \equi \frac{r_1 +r_2 +\ell}{2}\,,\qquad \cos \theta \equi \frac{r_1 -r_2}{\ell}\,,
\end{equation}
and $(r_1,r_2)$ are given in \eqref{eq:DistanceAndhDef} with $\alpha_2= -\alpha_1 = \ell/2$.

\subsection{Entrapment limit of $N$ extremal charges}
\label{app:NExtEntrap}

\begin{figure}[t]
\begin{center}
\includegraphics[width= 0.7\textwidth]{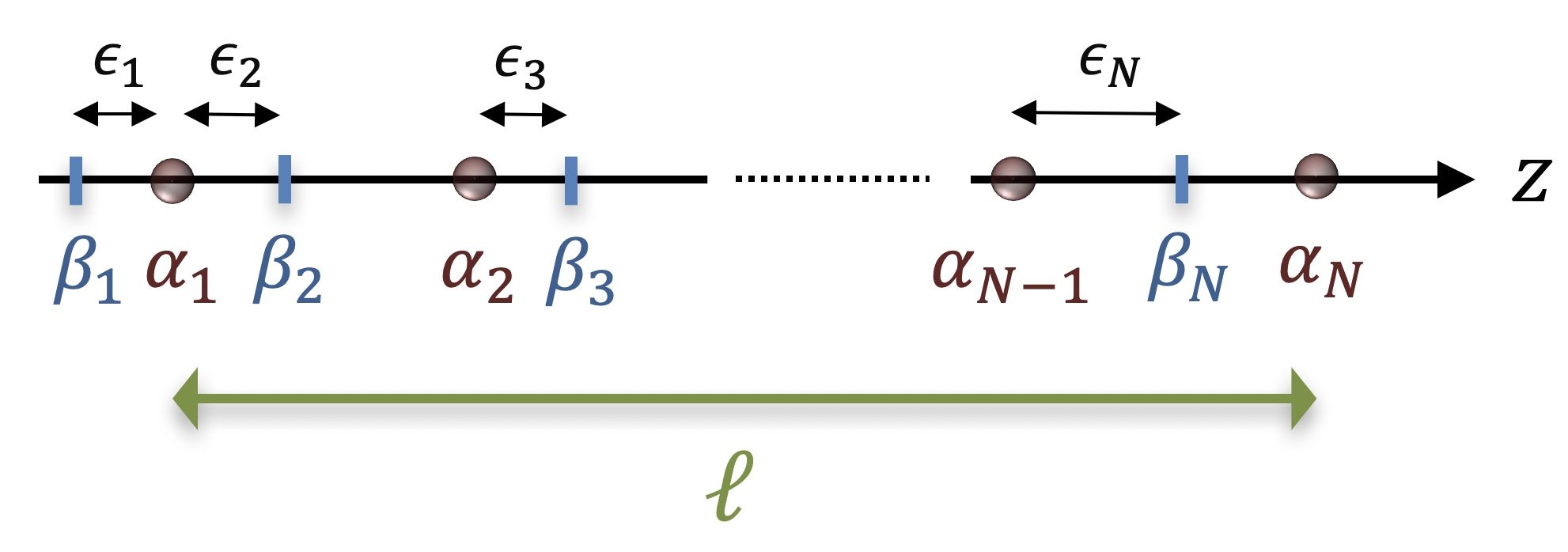}
\caption{Schematic description of a configuration of $N$ extremal charges along a line, characterized by $2N$ parameters: the $\alpha_n$ denote their positions, and the $\beta_n$ are non-physical parameters.}
\label{fig:NChargeConf}
\end{center}
\end{figure}

We consider a configuration of $N$ extremal and electric charges on a line. The solution is described using $2N$ parameters, $\alpha_n$, corresponding to the charge positions on the symmetry axis, and $\beta_n$ giving rise to the charges, where $n=1,\ldots,N$. Without loss of generality, we can assume that the parameters are ordered such that $x_1 \leq x_2 \leq \ldots \leq x_N$, for $x=\alpha$ or $\beta$. A depiction of a generic configuration, along with some conventions, is illustrated in Fig.\ref{fig:NChargeConf}. Additionally, we introduce the overall size of the charge distribution and position the origin of the symmetry axis as follows:
\begin{equation}
    \ell \equi \alpha_N - \alpha_1\,, \qquad \sum_{n=1}^N \alpha_n =0\,.
    \label{eq:SizeConfAndOriginApp}
\end{equation}

To prove that the charge configuration admits an entrapment limit, we first derive analytically this limit on the symmetry axis. The Ernst potentials of the configuration express on the symmetry axis above the last source, $z\geq \alpha_N$, in terms of two functions, $(e(z),f(z))$, named axis data \eqref{eq:AxisDataPotApp}. As such, the gravitational and electric potentials are given by\footnote{We remind that we define the gravitational potential such as $V=-g^{tt}=(\epsilon+\Psi^2)^{-1}$.}
\begin{equation}
    V(\rho=0, z>\alpha_N) = \left(\prod_{n=1}^N \frac{z-\beta_n}{z-\alpha_n}\right)^2,\qquad A(\rho=0,z>\alpha_N) =  2\sum_{n=1}^N \frac{f_n}{z-\beta_n},
    \label{eq:AxisDataGEPotApp}
\end{equation}
where the $f_k$ are given in terms of $(\alpha_n,\beta_n)$ in \eqref{eq:AxisDataNApp}. It is worth noting that the sign of $f_k$ is entirely arbitrary.

The axis data reveal much more about the entire solution than one might initially anticipate. In fact, it has been demonstrated in \cite{Simon:1983kz,Sotiriou:2004ud} that all gravitational and electric multipole moments of an axisymmetric electrovacuum solution can be obtained by expanding these functions at large $z$. For instance, the total mass, the total charge, and the electric dipole moments are respectively:
\begin{equation}
    M \= -\sum_{n=1}^N \beta_n\,,\qquad Q = \sum_{n=1}^N f_n\,,\qquad \cJ = M Q + \sum_{n=1}^N f_n \beta_n\,.
    \label{eq:ConsChargeNexApp}
\end{equation}
The constraint of having a neutral configuration simply requires that:
\begin{equation}
    \sum_{n=1}^N f_n \=0.
    \label{eq:neutralityCondApp}
\end{equation}
It is important to note that the $f_n$ are not directly related to the individual masses and charges within the configuration. Deriving these quantities is much more involved, as it necessitates knowledge of the gravitational and electric potentials within the configuration $z\leq \alpha_N$.

We parametrize the $\beta_n$ parameters around the $\alpha_n$ values as follows (see Fig.\ref{fig:NChargeConf}):
\begin{equation}
     \beta_1 = \alpha_1 +\epsilon_1 , \qquad \beta_2 = \alpha_1 +\epsilon_2\,,\qquad \beta_k \= \alpha_{k-1}+\epsilon_k \quad k\geq 3\,.
     \label{eq:EntrapLimNEx}
\end{equation}
We consider the limit where the $\epsilon_n$ are small and $\sum_{n=1}^N \epsilon_n >0$. This involves moving all charges ($\alpha_k$) toward the $\beta_{k+1}$ values, $k\geq 2$, and also considering $\beta_{1}\sim \beta_2 \sim \alpha_1$. In other words, it makes each group of successive charges approach their ``Schwarzschild radii''. For instance, one has:
\begin{equation}
    \ell \= M + \sum_{n=1}^N \epsilon_n\,,
\end{equation}
so that $\ell\to M$ and the whole configuration is approaching its critical size. Moreover, under this limit, one has\footnote{We have absorbed the freedom in the sign of each $f_n$ into the sign of $\epsilon_n$.}
\begin{equation}
\begin{split}
    f_{k} &\sim \frac{\epsilon_k \,\ell}{\epsilon_1-\epsilon_{2}} \left[1-\sum_{n=3}^{N}\frac{\epsilon_n}{\alpha_{n-1}-\alpha_{1}}-\frac{\epsilon_k}{\ell} \right]  \,,\qquad k=1,2\,. \\  
     f_n &\sim \epsilon_n\, \frac{\alpha_{N}-\alpha_{n-1}}{\alpha_{n-1}-\alpha_{1}}, \qquad n\geq 3.
\end{split}
\end{equation}
Imposing neutrality \eqref{eq:neutralityCondApp} requires
\begin{equation}
    \epsilon_1 = -\epsilon_{2} \left(1-\epsilon_{2}-\sum_{n=3}^{N}\frac{\epsilon_n\,(\alpha_{N}-\alpha_{n-1})}{\alpha_{n-1}-\alpha_{1}} \right)\,.
\end{equation}
As such, only two $f_n$ are non-negligible, namely $f_1 \sim - f_{2} \sim  \frac{\ell}{2}$, while all others are of the order of $\epsilon_n$. This might suggest that our charge configuration is converging towards the two-charge system discussed in the preceding section. However, as previously argued, the $f_n$ do not correspond to the individual charges in the system. Therefore, even though most of the $f_n$ are negligible, all $N$ individual charges can still be of the same order. 

To demonstrate this, one can observe that by fixing $\epsilon_{2}$ in terms of the other $\epsilon_n$ for $n\geq 3$, it is possible to ensure that the configuration has stricly zero electric dipole moment throughout the limit, $\cJ=0$ \eqref{eq:ConsChargeNexApp}. Clearly, the two-charge system cannot impose the dipole to be strictly zero, highlighting a different charge configuration. Figures in Fig.\ref{fig:ElecFlux3charge} for a three-charge system without dipole moment demonstrate this point.

Continuing the analysis, we proceed by considering the small $\epsilon_n$ limit in the axis data \eqref{eq:AxisDataGEPotApp}. We find that for $z$ not too close to $\alpha_N$, particularly when $z> \alpha_N(1+\cO(\epsilon_N))$, the expressions are approximately given by:
\begin{equation}
    V(\rho=0,z) \sim \left( 1-\frac{\ell}{z-\alpha_1}\right)^{-2}  \,,\qquad A(\rho=0,z) \sim 0\,. \label{eq:AxisDataApproxApp}
\end{equation}
Therefore, similar to the two-charge system, the configuration of $N$ extremal charges converges toward a vacuum solution, at least on the symmetry axis, as it approaches an infinite redshift limit, signifying its horizon scale.

Furthermore, the axis data of the vacuum solutions \eqref{eq:AxisDataApproxApp} correspond to the vacuum solution obtained from the entrapment of the two-charge system \eqref{eq:LimGravElecPot}. Thus, the configuration of $N$ extremal charges converges toward the exact same solution and is similarly indistinguishable (at least on the symmetry axis) from this solution, except in an infinitesimal region around the sources where the electric degrees of freedom used to produce the nontrivial charge distribution start to manifest.

We now need to extend the entrapment from the symmetry axis to the entire space. A rigorous treatment would involve revisiting Sibgatullin's method and demonstrating that when two sets of axis data match just outside the source locations, the final solutions should also match except in a small region around the sources. Sibgatullin's method relies on integral forms of the axis data, indicating that an analytic proof should be achievable. However, due to the potential complexity of such an analysis, it falls beyond the scope of this paper and is deferred to future projects. For indirect arguments demonstrating the extension of the entrapment derived here on the symmetry axis to the entire $(\rho,z)$ space, we direct the reader to Section \ref{sec:NExCharges}.

\subsection{Five-dimensional framework}
\label{app:5dErnst}

The Ernst formalism can be generalized in higher dimensions for suitable choices of gauge fields \cite{Bah:2020pdz,Bah:2021owp,Heidmann:2021cms}. We introduce a five-dimensional framework that can be obtained from the truncation of a more intricate supergravity frame in \cite{Heidmann:2021cms}. It consists in an Einstein-Maxwell theory with a one-form electric field governed by the following action:
\begin{equation}
S_5 \= \frac{1}{16\pi G_5} \,\int \,d^5 x \sqrt{-g}\,\left( R -\frac{1}{2} |F|^2\right)\,,
\label{eq:5daction}
\end{equation}
where $G_5$ is the five-dimensional Newton constant, $R$ is the Ricci scalar, and $F$ is an electric two-form field strength. We consider static and axially symmetric solutions with a $\partial_y$ isometry, where $y$ parametrizes the fifth dimension:
\begin{equation}
\begin{split}
ds_5^2 &\= - V^{-1} \,dt^2+ \sqrt{V} \left[\frac{dy^2}{\sqrt{W}}+\sqrt{W}\left(e^{3\nu+\nu_W}\left( d\rho^2+dz^2\right) +\rho^2 d\phi^2 \right)\right],\quad F \=  -\frac{\sqrt{3}}{2}\,dA\wedge dt\,.
\end{split}
\label{eq:StaticErnstMetric5d}
\end{equation}
One can also consider the magnetic dual formulation with
\begin{equation}
   F_m = \star F \= \frac{\sqrt{3}}{2}\,dH \wedge d\phi \wedge dy\,,\qquad \star_2 dH = \rho V dA\,.
\end{equation}
The Einstein-Maxwell equations decompose into two electrostatic Ernst equations. The sector $(V,A,\nu)$ is governed by the same equations as in four dimensions \eqref{eq:EOMErnst}, while the sector $(W,\nu_W)$ corresponds to the vacuum limit of \eqref{eq:EOMErnst}, i.e.  $(V,A,\nu)\to (W,0,\nu_W)$.

As a result, the five-dimensional system inherits the integrable structure of the Ernst formalism derived in four dimensions in section \ref{app:ErnstEOM}. One can therefore solve the $(V,A,\nu)$ using the solution corresponding to $N$ extremal point sources introduced before, and the $(W,\nu_W)$ sector using vacuum Weyl solution. In this paper, we solved $(W,\nu_W)$ with one single rod source in between the first and last extremal sources, $\alpha_1$ and $\alpha_N$ respectively. This corresponds to a vacuum bubble along $y$ given by:
\begin{equation}
    W \= \left( 1-\frac{\ell}{r}\right)^{-2}\,,\qquad e^{\nu_W} \= \frac{1-\frac{\ell}{r}}{\left( 1- \frac{\ell\sin^2 \frac{\theta}{2}}{r}\right)\left( 1- \frac{\ell\cos^2 \frac{\theta}{2}}{r}\right)},
\end{equation}
where $(r,\theta)$ is given in terms of $(\rho,z)$ in \eqref{eq:SpherCoord}.

Upon dimensional reduction along the $y$ direction, the solution leads to: 
\begin{align}
d s_4^2 & =-\frac{dt^2}{(V^3 W)^\frac{1}{4}}+(V^3 W)^\frac{1}{4}\,\left[e^{3 \nu+\nu_W} \left(d\rho^2+ dz^2 \right) +\rho^2\, d\phi^2\right]\,, \\
e^{\frac{2}{\sqrt{3}}\Phi} & =\sqrt{\frac{W}{V}}\,, \qquad F=-\frac{\sqrt{3}}{2}\,d A \wedge d t .\nonumber
\end{align}


\section{Gravitational signatures of distorted Schwarzschild black holes}
\label{app:DisSchw}

In this section, we derive the gravitational characteristics of the distorted Schwarzschild solution that have been listed in section \ref{sec:GravSig}.

\subsection{Photon ring and shadow}

We aim to describe the light ring structure of the distorted Schwarzschild black hole and compare it with that of the Schwarzschild light ring/shadow. Given that the distorted solution lacks spherical symmetry but possesses a $\mathbb{Z}_2$ symmetry under $\theta\to \pi-\theta$, our analysis will be restricted to the equatorial plane $\theta=\pi/2$.

Consequently, we study null geodesics in the equatorial plane governed by the following geodesic equations:\footnote{The equation is derived by imposing $\cH=\frac{1}{2} g^{\mu \nu} p_\mu p_\nu =0$, where $p_\mu$ represents the conjugate momenta $p_\mu=g_{\mu\nu} \dot{x}^\nu$, $x^\nu = (t,r,\theta,\phi)$, and the dot signifies the derivative with respect to the affine parameter along the geodesic. Due to the $\phi$ and time independence, we further set $p_t=-1$ and $p_\phi= \text{ constant}$.}
\begin{equation}
    \dot{r}^2 - \cV =0 \,,\qquad \cV \equi \left(1+\frac{M^2 \sin^2\theta}{4r(r-M)}\right)^{-3} \left( 1-\frac{\left(1-\frac{M}{r}\right)^3}{r^2 \sin^2 \theta}\,p_\phi^2\right)\,,
\end{equation}
where $p_\phi$ is the impact parameter and $\dot{r}$ represents the derivative of $r$ with respect to the affine parameter along the geodesic.

 The solution has a single photon orbit on the equatorial plane, determined by $\cV = \partial_r \cV = 0$. It can be shown that $\partial_r^2 \cV >0$, indicating the instability of the photon orbit, akin to the Schwarzschild black hole. This orbit is located at $r=\hat{r}$ with an impact parameter $p_\phi$, exhibiting the following scattering properties:
 \begin{equation}
 \begin{split}
 \hat{r} =  \frac{5M}{2}\,,\quad p_\phi =  \left(\frac{5}{3} \right)^{\frac{3}{2}}\hat{r} \approx 5.4 M\,,\quad \Omega =\frac{18 }{25\sqrt{15}\,M} \approx \frac{0.19}{M}\,, \quad  \lambda  = \frac{128}{625\, M} \approx \frac{0.20}{M}\,,
 \end{split}
 \end{equation}
where $\Omega= \frac{\dot{\phi}}{\dot{t}}$ is the angular velocity of photons at the orbit, and $\lambda= \sqrt{\frac{1}{2\dot{t}^2} \frac{d^2 \cV}{dr^2}}$ is the Lyapunov exponent, measuring the instability timescale at the orbit.

For the photon orbit of a Schwarzschild black hole with the same mass, the corresponding values are:
 \begin{equation}
 \hat{r}_\text{S}=3 M \,,\quad p_{\phi \text{S}} = 3\sqrt{3} M\approx 5.2 M\,,\quad \Omega_\text{S}=\lambda_\text{S}=\frac{1}{3\sqrt{3}M} \approx \frac{0.2}{M}\,.
 \end{equation}
Thus, the values closely resemble those of the Schwarzschild black hole, except for the radius $\hat{r}$. However, as $\hat{r}$ is coordinate and background dependent, the pertinent length of the orbit for comparison is the arc length:
 \begin{equation}
  L \= 2\pi \sqrt{g_{\phi \phi}} |_{r=\hat{r}} \= 10\pi \sqrt{\frac{5}{3}} M \,\approx\, 1.08 \,L_\text{S},
 \end{equation}
 where $L_\text{S}=6\pi M$ denotes the Schwarzschild value. At first, this striking correspondence is restricted to the equatorial plane, necessitating a comprehensive derivation of the entire photon sphere for a complete analysis. However, as the spacetime is more than 95\% spherically symmetric at $r=\hat{r}$,\footnote{By stating that the spacetime is 95\% spherically symmetric, we imply that the ratio of the longitudinal and latitudinal lengths of the sphere is above 0.95.} it is reasonable to assume that the complete photon sphere defining the shadow of the distorted Schwarzschild will largely share the same characteristics as its reduction on the equatorial plane. This assumption can be supported by imaging simulations on analogous geometries \cite{Heidmann:2022ehn,Bah:2023ows}.
 
 Consequently, the properties of the unstable light ring circumscribing the distorted Schwarzschild black hole closely resemble those of the Schwarzschild light ring, despite significant deformations that considerably stretch the sphere at and above the horizon.

\subsection{Gravitational multipole moments}

In the absence of spherical symmetry, it is expected that the distorted Schwarzschild black hole will manifest nontrivial gravitational multipole moments, in contrast to the Schwarzschild black hole. We will derive these multipole moments in this section using Thorne's approach \cite{Thorne:1980ru,Bena:2020uup,Bah:2021jno}, which involves expressing the solution in Asymptotically Cartesian and Mass Centered coordinates (ACMC) and expanding the $g_{tt}$ component.

The ACMC coordinates are obtained by asymptotically expanding the metric components in a series of Legendre Polynomials and ensuring that the $g_{rr}$ components have no $r^{-\ell} P_\ell(\cos \theta)$ coefficients (see \cite{Bena:2020uup} for more detailed information). The ACMC coordinates for distorted Schwarzschild black holes, denoted as $(r_c,\theta_c)$, are defined by:
\begin{equation}
r\sqrt{1-\frac{M}{r}}\,\sin \theta = \sqrt{r_c^2 -M^2}\,\sin \theta_c \,,\qquad \left( r-\frac{M}{2}\right)\cos \theta = r_c \cos \theta_c\,.
\label{eq:ACMCcoor}
\end{equation}

We further expand the $g_{tt}$ component using the same approach and determine that distorted Schwarzschild black holes possess non-zero even gravitational multipole moments, which can be expressed as follows:
\begin{equation}
\mathcal{M}_{2n} \= c_n M^{2n+1} \,,\qquad c_n = \left(1,\frac{1}{4},-\frac{3}{112},-\frac{1}{704},\frac{631}{549120},\ldots\right).
\label{eq:GravMultipole}
\end{equation}

\subsection{Geometric size}
\label{app:GeoSize}

To conduct a meaningful geometric comparison between the distorted Schwarzschild and the Schwarzschild black hole, a relevant quantity defining the geometric size of these gravitational objects needs to be introduced. The absence of spherical symmetry adds a layer of complexity to this analysis. Firstly, one requires a background-independent radial coordinate that can be defined for both solutions.

To address this, we consider the ACMC coordinates \eqref{eq:ACMCcoor}, which is simply the radial coordinate for the Schwarzschild black hole. We take the radial part above the horizon, defining $r_\text{iso}=r_c-M$ for the distorted solution and $r_\text{iso}=r_c-2M$ for the Schwarzschild case.

\begin{figure}[t]
\begin{center}
\includegraphics[width= 0.95\textwidth]{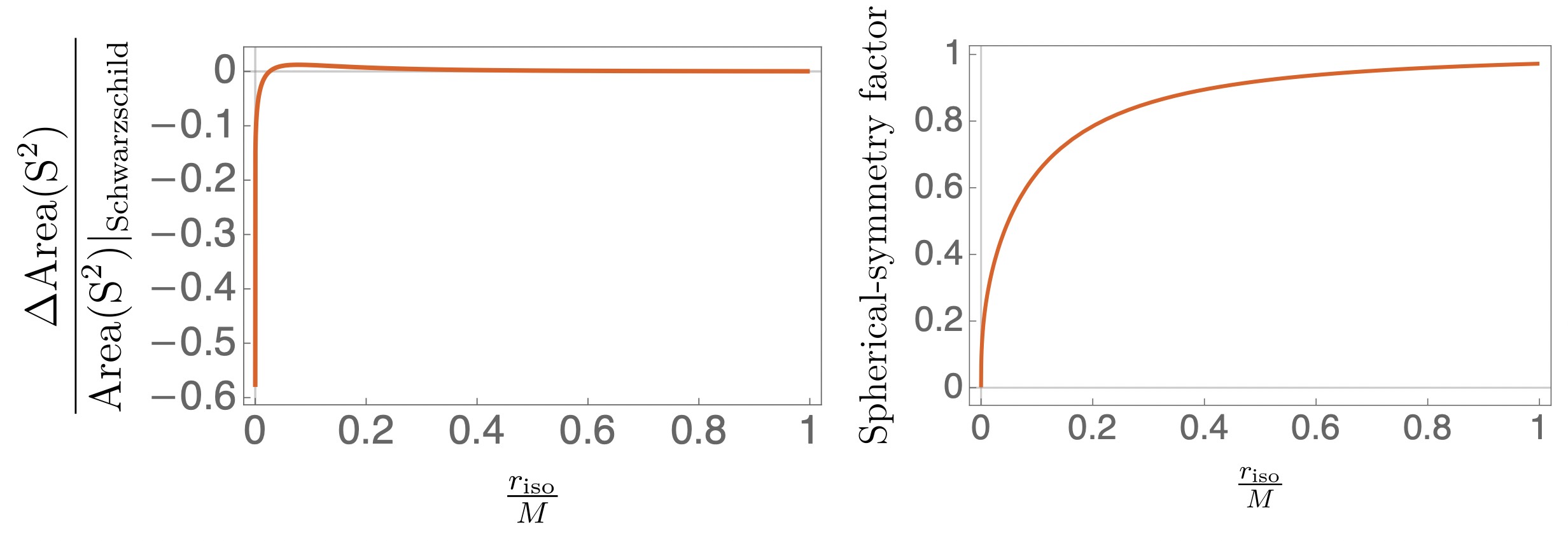}
\caption{Geometric properties of the two-sphere for the distorted Schwarzschild black hole. On the left: relative difference in the S$^2$ Area between the distorted Schwarzschild and the Schwarzschild black holes. On the right: the spherical-symmetry factor corresponding to the ratio between the longitudinal and latitudinal lengths of the two-sphere.}
\label{fig:S2size}
\end{center}
\end{figure}

The left plot in Fig.\ref{fig:S2size} compares the areas of the two-sphere at a constant $r_\text{iso}$ for the distorted Schwarzschild and the Schwarzschild black holes. The areas begin to deviate by less than 1\% for $r_\text{iso} \geq M/100$. Consequently, despite the metric differences, the sizes of the S$^2$ are remarkably similar up to the vicinity of the horizons. It is worth noting that the areas differ significantly at $r_\text{iso}=0$, while we have shown that the areas at the horizons are equal. This discrepancy arises because the computation of the area at constant $r$ or $r_\text{iso}$ does not yield the same surface.

The right plot in Fig.\ref{fig:S2size} illustrates the spherical-symmetry factor of the distorted Schwarzschild black hole. This factor is defined as the ratio of the longitudinal and latitudinal lengths of the two-sphere, with the latitude being determined at the equator. The plot demonstrates that despite sharing the same overall S$^2$ size, the distorted Schwarzschild exhibits significant axisymmetry, leading to an increasingly flattened sphere as the horizon is approached.


\section{Neutral bound state of two extremal black holes}
\label{app:TwoBH4d}

In this section, we analyze the spacetime structure produced by two extremal Reissner-Nordstr\"om black holes of opposite charges separated by a strut. The metric and gauge fields are given in \eqref{eq:2RNBS}. \\

We start with the structure of the bound state at the locus $r=\ell$. We begin by focusing on the strut at $r=\ell$ and $\theta\neq 0,\pi$. We introduce a new radial coordinate $\bar{r}$ defined as $r = \ell + \frac{\bar{r}^2}{4}$ and take the limit $\bar{r} \to 0$.  In this limit, the local geometry is given by:
\begin{align}
ds_4^2 \underset{\bar{r}\to 0}{\sim} &- \frac{(\ell-M)^2}{\left(\ell-M+\frac{2M}{\sin^2\theta} \right)^2} dt^2 \\
&+ \frac{(\ell-M)^2(\ell+M)^4\left(\ell-M+\frac{2M}{\sin^2\theta} \right)^2\sin^2\theta}{4 \ell^7} \left(d\bar{r}^2 +\frac{\bar{r}^2}{\left(1-\frac{M^2}{\ell^2} \right)^{4} } d\phi^2 +\ell d\theta^2\right) ,\nn\\
F  \underset{\bar{r}\to 0}{\sim} & -d\left[\frac{4M\sqrt{\ell^2-M^2}\,\cos \theta}{(\ell+M)^2-(\ell^2-M^2)\cos^2\theta}\right] \wedge dt\,.\nn
\end{align}
Hence, the locus at $r=\ell$ and $0<\theta<\pi$ represents a coordinate degeneracy of the $\phi$-circle. However, since $\phi$ must be $2\pi$ periodic to satisfy the asymptotic flatness of the solution, this degeneracy results in a conical excess characterized by the parameter $\delta$:
\begin{equation}
\delta= \left(1-\frac{M^2}{\ell^2} \right)^{-2} >1\,,
\end{equation}
and it becomes infinite as the inter-center distance approaches its minimal value $\ell \to M$. This conical excess represents a string-like structure with negative tension that is essential to counterbalance the gravitational attraction between the two black holes. \\

Next, we examine the nature of the black holes within the bound state, beginning with the black holes at the North pole, located at $r=\ell$ and $\theta=0$. In this region, we introduce the following local coordinates:
\begin{equation}
\sqrt{r(r-\ell)}\,\sin \theta \= \frac{M^2(\ell+M)^2}{4\ell^2}\,\bar{r} \sin \tau\,,\quad \left( r-\frac{\ell}{2}\right) \cos \theta \= \frac{M^2(\ell+M)^2}{4\ell^2}\,\bar{r} \cos \tau +\frac{\ell}{2}.  \label{eq:BHBSCoor4d}
\end{equation}
The local geometry at $\bar{r}\to0$ gives
\begin{align}
ds_4^2 \underset{\bar{r}\to 0}{\sim} &\frac{M^2(\ell+M)^2\left(\ell^2-M^2 \sin^2 \frac{\tau}{2} \right)^2}{4\ell^6}\left[\frac{d\bar{r}^2}{\bar{r}^2}-\bar{r}^2 dt^2 +d\tau^2 + \frac{\sin^2 \tau}{\left(1-\frac{M^2}{\ell^2} \sin^2 \frac{\tau}{2}  \right)^4}\,d\phi^2 \right] ,\nn\\
F  \underset{\bar{r}\to 0}{\sim} & -2q \left(1-\frac{M^2}{\ell^2} \right) \,d\bar{r} \wedge dt\,,\qquad \star F\underset{\bar{r}\to 0}{\sim}-q \frac{2\ell^2(\ell^2-M^2) \,\sin\tau}{\left( \ell^2-M^2 \sin^2 \frac{\tau}{2}\right)^2} d\tau \wedge d\phi\,.\nn
\end{align}
We recognize a S$^2$ vibration over an AdS$_2$ geometry, which corresponds to the near-horizon geometry of an extremal Reissner-Nordstr\"om black hole (with a stretched sphere). This region carries a charge $-q$ given by:
\begin{equation}
q \equi \frac{M}{2} \sqrt{\frac{\ell+M}{\ell-M}} \,.
\end{equation}
As an extremal black hole, it possesses zero temperature but a finite entropy, which is described by the Bekenstein-Hawking formula (in units where $G_4=1$):
\begin{equation}
S_1 \= \frac{M^2 (\ell+M)^2 \pi}{4\ell^2}\,.
\end{equation}

The South pole, located at $r=\ell$ and $\theta=\pi$, corresponds to the anti-extremal partner with an identical geometry but with opposite charge $q$. This creates another extremal Reissner-Nordstr\"om black hole with the same entropy, $S_2=S_1$. Therefore, the total entropy of the bound state is given by:
\begin{equation}
S^{(2)} \= S_1 +S_2 \=   \frac{M^2 (\ell+M)^2 \pi}{2\ell^2}\,.
\end{equation}

\bibliographystyle{utphys}      

\bibliography{microstates}       

\providecommand{\href}[2]{#2}\begingroup\raggedright\begin{thebibliography}{10}

\bibitem{Bekenstein:1973ur}
J.~D. Bekenstein, ``{Black holes and entropy},''
\href{http://dx.doi.org/10.1103/PhysRevD.7.2333}{{\em Phys. Rev.} {\bfseries D7} (1973) 2333--2346}.

\bibitem{Hawking:1975vcx}
S.~W. Hawking, ``{Particle Creation by Black Holes},'' \href{http://dx.doi.org/10.1007/BF02345020}{{\em Commun. Math. Phys.} {\bfseries 43} (1975) 199--220}. [Erratum: Commun.Math.Phys. 46, 206 (1976)].

\bibitem{Mathur:2009hf}
S.~D. Mathur, ``{The information paradox: A pedagogical introduction},'' \href{http://dx.doi.org/10.1088/0264-9381/26/22/224001}{{\em Class. Quant. Grav.} {\bfseries 26} (2009) 224001},
\href{http://arxiv.org/abs/0909.1038}{{\ttfamily arXiv:0909.1038 [hep-th]}}.

\bibitem{Almheiri:2012rt}
A.~Almheiri, D.~Marolf, J.~Polchinski, and J.~Sully, ``{Black Holes: Complementarity or Firewalls?},'' \href{http://dx.doi.org/10.1007/JHEP02(2013)062}{{\em JHEP} {\bfseries 1302} (2013) 062},
\href{http://arxiv.org/abs/1207.3123}{{\ttfamily arXiv:1207.3123 [hep-th]}}.

\bibitem{Mathur:2023uoe}
S.~D. Mathur and M.~Mehta, ``{The universality of black hole thermodynamics},'' \href{http://arxiv.org/abs/2305.12003}{{\ttfamily arXiv:2305.12003 [hep-th]}}.

\bibitem{Strominger:1996sh}
A.~Strominger and C.~Vafa, ``{Microscopic Origin of the Bekenstein-Hawking Entropy},'' \href{http://dx.doi.org/10.1016/0370-2693(96)00345-0}{{\em Phys. Lett.} {\bfseries B379} (1996) 99--104},
\href{http://arxiv.org/abs/hep-th/9601029}{{\ttfamily arXiv:hep-th/9601029}}.

\bibitem{Gibbons:2013tqa}
G.~Gibbons and N.~Warner, ``{Global structure of five-dimensional fuzzballs},'' \href{http://dx.doi.org/10.1088/0264-9381/31/2/025016}{{\em Class.Quant.Grav.} {\bfseries 31} (2014) 025016},
\href{http://arxiv.org/abs/1305.0957}{{\ttfamily arXiv:1305.0957 [hep-th]}}.

\bibitem{Bena:2013dka}
I.~Bena and N.~P. Warner, ``{Resolving the Structure of Black Holes: Philosophizing with a Hammer},''
\href{http://arxiv.org/abs/1311.4538}{{\ttfamily arXiv:1311.4538 [hep-th]}}.

\bibitem{Bena:2022rna}
I.~Bena, E.~J. Martinec, S.~D. Mathur, and N.~P. Warner, ``{Fuzzballs and Microstate Geometries: Black-Hole Structure in String Theory},'' \href{http://arxiv.org/abs/2204.13113}{{\ttfamily arXiv:2204.13113 [hep-th]}}.

\bibitem{Copson1928ElectrostaticsIA}
E.~T. Copson, ``Electrostatics in a gravitational field,'' {\em Proceedings of the Royal Society of Edinburgh: Section A Mathematics} {\bfseries 80} (1928) 201 -- 211. \url{https://api.semanticscholar.org/CorpusID:121932619}.

\bibitem{1968CMaPh...8..245I}
W.~{Israel}, ``{Event horizons in static electrovac space-times},'' \href{http://dx.doi.org/10.1007/BF01645859}{{\em Communications in Mathematical Physics} {\bfseries 8} no.~3, (Sept., 1968) 245--260}.

\bibitem{1971JMP....12.1845C}
J.~M. {Cohen} and R.~M. {Wald}, ``{Point Charge in the Vicinity of a Schwarzschild Black Hole},'' \href{http://dx.doi.org/10.1063/1.1665812}{{\em Journal of Mathematical Physics} {\bfseries 12} no.~9, (Sept., 1971) 1845--1849}.

\bibitem{PhysRevD.8.3259}
R.~S. Hanni and R.~Ruffini, ``Lines of force of a point charge near a schwarzschild black hole,'' \href{http://dx.doi.org/10.1103/PhysRevD.8.3259}{{\em Phys. Rev. D} {\bfseries 8} (Nov, 1973) 3259--3265}. \url{https://link.aps.org/doi/10.1103/PhysRevD.8.3259}.

\bibitem{Linet:1976sq}
B.~Linet, ``{Electrostatics and magnetostatics in the Schwarzschild metric},'' \href{http://dx.doi.org/10.1088/0305-4470/9/7/010}{{\em J. Phys. A} {\bfseries 9} (1976) 1081--1087}.

\bibitem{Garfinkle_2021}
D.~Garfinkle, ``Electric field of a charge in the vicinity of a higher dimensional black hole,'' \href{http://dx.doi.org/10.1103/physrevd.103.024056}{{\em Physical Review D} {\bfseries 103} no.~2, (Jan., 2021) }. \url{http://dx.doi.org/10.1103/PhysRevD.103.024056}.

\bibitem{EntrapmentProbe}
I.~Bah, P.~Heidmann, and M.~Mehta, ``{to appear},''.

\bibitem{Ernst:1967wx}
F.~J. Ernst, ``{New formulation of the axially symmetric gravitational field problem},'' \href{http://dx.doi.org/10.1103/PhysRev.167.1175}{{\em Phys. Rev.} {\bfseries 167} (1968) 1175--1179}.

\bibitem{Ernst:1967by}
F.~J. Ernst, ``{New Formulation of the Axially Symmetric Gravitational Field Problem. II},'' \href{http://dx.doi.org/10.1103/PhysRev.168.1415}{{\em Phys. Rev.} {\bfseries 168} (1968) 1415--1417}.

\bibitem{Bah:2020pdz}
I.~Bah and P.~Heidmann, ``{Topological stars, black holes and generalized charged Weyl solutions},'' \href{http://dx.doi.org/10.1007/JHEP09(2021)147}{{\em JHEP} {\bfseries 09} (2021) 147}, \href{http://arxiv.org/abs/2012.13407}{{\ttfamily arXiv:2012.13407 [hep-th]}}.

\bibitem{Bah:2021owp}
I.~Bah and P.~Heidmann, ``{Smooth bubbling geometries without supersymmetry},'' \href{http://dx.doi.org/10.1007/JHEP09(2021)128}{{\em JHEP} {\bfseries 09} (2021) 128}, \href{http://arxiv.org/abs/2106.05118}{{\ttfamily arXiv:2106.05118 [hep-th]}}.

\bibitem{Heidmann:2021cms}
P.~Heidmann, ``{Non-BPS floating branes and bubbling geometries},'' \href{http://dx.doi.org/10.1007/JHEP02(2022)162}{{\em JHEP} {\bfseries 02} (2022) 162}, \href{http://arxiv.org/abs/2112.03279}{{\ttfamily arXiv:2112.03279 [hep-th]}}.

\bibitem{NoraBreton1998}
N.~Bretón, V.~S. Manko, and J.~A. Sánchez, ``On the equilibrium of charged masses in general relativity: the electrostatic case,'' \href{http://dx.doi.org/10.1088/0264-9381/15/10/013}{{\em Classical and Quantum Gravity} {\bfseries 15} no.~10, (Oct, 1998) 3071}. \url{https://dx.doi.org/10.1088/0264-9381/15/10/013}.

\bibitem{Alekseev:2007re}
G.~A. Alekseev and V.~A. Belinski, \href{http://dx.doi.org/10.1142/9789812834300_0022}{``{Superposition of fields of two Reissner - Nordstrom sources},''} in {\em {11th Marcel Grossmann Meeting on General Relativity}}, pp.~2490--2492.
\newblock 10, 2007.
\newblock \href{http://arxiv.org/abs/0710.2515}{{\ttfamily arXiv:0710.2515 [gr-qc]}}.

\bibitem{Alekseev:2007gt}
G.~A. Alekseev and V.~A. Belinski, ``{Equilibrium configurations of two charged masses in General Relativity},'' \href{http://dx.doi.org/10.1103/PhysRevD.76.021501}{{\em Phys. Rev. D} {\bfseries 76} (2007) 021501}, \href{http://arxiv.org/abs/0706.1981}{{\ttfamily arXiv:0706.1981 [gr-qc]}}.

\bibitem{Manko:2007hi}
V.~S. Manko, ``{The Double-Reissner-Nordstrom solution and the interaction force between two spherically symmetric charged particles},'' \href{http://dx.doi.org/10.1103/PhysRevD.76.124032}{{\em Phys. Rev. D} {\bfseries 76} (2007) 124032}, \href{http://arxiv.org/abs/0710.2158}{{\ttfamily arXiv:0710.2158 [gr-qc]}}.

\bibitem{Manko:2008gb}
V.~S. Manko, E.~Ruiz, and J.~Sanchez-Mondragon, ``{Analogs of the double-Reissner-Nordstrom solution in magnetostatics and dilaton gravity: mathematical description and some physical properties},'' \href{http://dx.doi.org/10.1103/PhysRevD.79.084024}{{\em Phys. Rev. D} {\bfseries 79} (2009) 084024}, \href{http://arxiv.org/abs/0811.2029}{{\ttfamily arXiv:0811.2029 [gr-qc]}}.

\bibitem{Bah:2021rki}
I.~Bah and P.~Heidmann, ``{Bubble bag end: a bubbly resolution of curvature singularity},'' \href{http://dx.doi.org/10.1007/JHEP10(2021)165}{{\em JHEP} {\bfseries 10} (2021) 165}, \href{http://arxiv.org/abs/2107.13551}{{\ttfamily arXiv:2107.13551 [hep-th]}}.

\bibitem{Bah:2022yji}
I.~Bah, P.~Heidmann, and P.~Weck, ``{Schwarzschild-like Topological Solitons},'' \href{http://arxiv.org/abs/2203.12625}{{\ttfamily arXiv:2203.12625 [hep-th]}}.

\bibitem{Bah:2022pdn}
I.~Bah and P.~Heidmann, ``{Non-BPS bubbling geometries in AdS$_{3}$},'' \href{http://dx.doi.org/10.1007/JHEP02(2023)133}{{\em JHEP} {\bfseries 02} (2023) 133}, \href{http://arxiv.org/abs/2210.06483}{{\ttfamily arXiv:2210.06483 [hep-th]}}.

\bibitem{Heidmann:2022zyd}
P.~Heidmann and A.~Houppe, ``{Solitonic excitations in AdS$_{2}$},'' \href{http://dx.doi.org/10.1007/JHEP07(2023)186}{{\em JHEP} {\bfseries 07} (2023) 186}, \href{http://arxiv.org/abs/2212.05065}{{\ttfamily arXiv:2212.05065 [hep-th]}}.

\bibitem{Bah:2023ows}
I.~Bah and P.~Heidmann, ``{Geometric Resolution of Schwarzschild Horizon},'' \href{http://arxiv.org/abs/2303.10186}{{\ttfamily arXiv:2303.10186 [hep-th]}}.

\bibitem{PhysRevMajumdar}
S.~D. Majumdar, ``A class of exact solutions of einstein's field equations,'' \href{http://dx.doi.org/10.1103/PhysRev.72.390}{{\em Phys. Rev.} {\bfseries 72} (Sep, 1947) 390--398}. \url{https://link.aps.org/doi/10.1103/PhysRev.72.390}.

\bibitem{Papaetrou:1947ib}
A.~Papaetrou, ``{A Static solution of the equations of the gravitational field for an arbitrary charge distribution},'' {\em Proc. Roy. Irish Acad. A} {\bfseries 51} (1947) 191--204.

\bibitem{PhysRevD.23.287}
J.~D. Bekenstein, ``Universal upper bound on the entropy-to-energy ratio for bounded systems,'' \href{http://dx.doi.org/10.1103/PhysRevD.23.287}{{\em Phys. Rev. D} {\bfseries 23} (Jan, 1981) 287--298}. \url{https://link.aps.org/doi/10.1103/PhysRevD.23.287}.

\bibitem{Elvang:2002br}
H.~Elvang and G.~T. Horowitz, ``{When black holes meet Kaluza-Klein bubbles},'' \href{http://dx.doi.org/10.1103/PhysRevD.67.044015}{{\em Phys. Rev. D} {\bfseries 67} (2003) 044015}, \href{http://arxiv.org/abs/hep-th/0210303}{{\ttfamily arXiv:hep-th/0210303}}.

\bibitem{Manko_1993}
V.~S. Manko and N.~R. Sibgatullin, ``Construction of exact solutions of the einstein-maxwell equations corresponding to a given behaviour of the ernst potentials on the symmetry axis,'' \href{http://dx.doi.org/10.1088/0264-9381/10/7/014}{{\em Classical and Quantum Gravity} {\bfseries 10} no.~7, (Jul, 1993) 1383}. \url{https://dx.doi.org/10.1088/0264-9381/10/7/014}.

\bibitem{Simon:1983kz}
W.~Simon, ``{The Multipole Expansion of Stationary Einstein-maxwell Fields},'' \href{http://dx.doi.org/10.1063/1.526271}{{\em J. Math. Phys.} {\bfseries 25} (1984) 1035}.

\bibitem{Sotiriou:2004ud}
T.~P. Sotiriou and T.~A. Apostolatos, ``{Corrected multipole moments of axisymmetric electrovacuum spacetimes},'' \href{http://dx.doi.org/10.1088/0264-9381/21/24/003}{{\em Class. Quant. Grav.} {\bfseries 21} (2004) 5727--5733}, \href{http://arxiv.org/abs/gr-qc/0407064}{{\ttfamily arXiv:gr-qc/0407064}}.

\bibitem{Gralla:2019xty}
S.~E. Gralla, D.~E. Holz, and R.~M. Wald, ``{Black Hole Shadows, Photon Rings, and Lensing Rings},'' \href{http://dx.doi.org/10.1103/PhysRevD.100.024018}{{\em Phys. Rev. D} {\bfseries 100} no.~2, (2019) 024018}, \href{http://arxiv.org/abs/1906.00873}{{\ttfamily arXiv:1906.00873 [astro-ph.HE]}}.

\bibitem{Cardoso:2008bp}
V.~Cardoso, A.~S. Miranda, E.~Berti, H.~Witek, and V.~T. Zanchin, ``{Geodesic stability, Lyapunov exponents and quasinormal modes},'' \href{http://dx.doi.org/10.1103/PhysRevD.79.064016}{{\em Phys. Rev. D} {\bfseries 79} no.~6, (2009) 064016}, \href{http://arxiv.org/abs/0812.1806}{{\ttfamily arXiv:0812.1806 [hep-th]}}.

\bibitem{Costa:2000kf}
M.~S. Costa and M.~J. Perry, ``{Interacting black holes},'' \href{http://dx.doi.org/10.1016/S0550-3213(00)00577-0}{{\em Nucl. Phys. B} {\bfseries 591} (2000) 469--487}, \href{http://arxiv.org/abs/hep-th/0008106}{{\ttfamily arXiv:hep-th/0008106}}.

\bibitem{Breckenridge:1996is}
J.~Breckenridge, R.~C. Myers, A.~Peet, and C.~Vafa, ``{D-branes and spinning black holes},'' \href{http://dx.doi.org/10.1016/S0370-2693(96)01460-8}{{\em Phys.Lett.} {\bfseries B391} (1997) 93--98},
\href{http://arxiv.org/abs/hep-th/9602065}{{\ttfamily arXiv:hep-th/9602065 [hep-th]}}.

\bibitem{Elvang:2007rd}
H.~Elvang and P.~Figueras, ``{Black Saturn},'' \href{http://dx.doi.org/10.1088/1126-6708/2007/05/050}{{\em JHEP} {\bfseries 05} (2007) 050}, \href{http://arxiv.org/abs/hep-th/0701035}{{\ttfamily arXiv:hep-th/0701035}}.

\bibitem{Warner:2019jll}
N.~P. Warner, ``{Lectures on Microstate Geometries},'' \href{http://arxiv.org/abs/1912.13108}{{\ttfamily arXiv:1912.13108 [hep-th]}}.

\bibitem{Bena:2022ldq}
I.~Bena, E.~J. Martinec, S.~D. Mathur, and N.~P. Warner, ``{Snowmass White Paper: Micro- and Macro-Structure of Black Holes},'' \href{http://arxiv.org/abs/2203.04981}{{\ttfamily arXiv:2203.04981 [hep-th]}}.

\bibitem{Shigemori:2019orj}
M.~Shigemori, ``{Counting Superstrata},'' \href{http://dx.doi.org/10.1007/JHEP10(2019)017}{{\em JHEP} {\bfseries 10} (2019) 017}, \href{http://arxiv.org/abs/1907.03878}{{\ttfamily arXiv:1907.03878 [hep-th]}}.

\bibitem{Mayerson:2020acj}
D.~R. Mayerson and M.~Shigemori, ``{Counting D1-D5-P microstates in supergravity},'' \href{http://dx.doi.org/10.21468/SciPostPhys.10.1.018}{{\em SciPost Phys.} {\bfseries 10} no.~1, (2021) 018}, \href{http://arxiv.org/abs/2010.04172}{{\ttfamily arXiv:2010.04172 [hep-th]}}.

\bibitem{Bena:2015bea}
I.~Bena, S.~Giusto, R.~Russo, M.~Shigemori, and N.~P. Warner, ``{Habemus Superstratum! A constructive proof of the existence of superstrata},'' \href{http://dx.doi.org/10.1007/JHEP05(2015)110}{{\em JHEP} {\bfseries 05} (2015) 110},
\href{http://arxiv.org/abs/1503.01463}{{\ttfamily arXiv:1503.01463 [hep-th]}}.

\bibitem{Bena:2017xbt}
I.~Bena, S.~Giusto, E.~J. Martinec, R.~Russo, M.~Shigemori, D.~Turton, and N.~P. Warner, ``{Asymptotically-flat supergravity solutions deep inside the black-hole regime},'' \href{http://dx.doi.org/10.1007/JHEP02(2018)014}{{\em JHEP} {\bfseries 02} (2018) 014},
\href{http://arxiv.org/abs/1711.10474}{{\ttfamily arXiv:1711.10474 [hep-th]}}.

\bibitem{Ceplak:2018pws}
N.~Ceplak, R.~Russo, and M.~Shigemori, ``{Supercharging Superstrata},''
\href{http://arxiv.org/abs/1812.08761}{{\ttfamily arXiv:1812.08761 [hep-th]}}.

\bibitem{Heidmann:2019zws}
P.~Heidmann and N.~P. Warner, ``{Superstratum Symbiosis},''
\href{http://arxiv.org/abs/1903.07631}{{\ttfamily arXiv:1903.07631 [hep-th]}}.

\bibitem{Heidmann:2019xrd}
P.~Heidmann, D.~R. Mayerson, R.~Walker, and N.~P. Warner, ``{Holomorphic Waves of Black Hole Microstructure},'' \href{http://dx.doi.org/10.1007/JHEP02(2020)192}{{\em JHEP} {\bfseries 02} (2020) 192}, \href{http://arxiv.org/abs/1910.10714}{{\ttfamily arXiv:1910.10714 [hep-th]}}.

\bibitem{Shigemori:2020yuo}
M.~Shigemori, ``{Superstrata},'' \href{http://dx.doi.org/10.1007/s10714-020-02698-8}{{\em Gen. Rel. Grav.} {\bfseries 52} no.~5, (2020) 51}, \href{http://arxiv.org/abs/2002.01592}{{\ttfamily arXiv:2002.01592 [hep-th]}}.

\bibitem{Bena:2007kg}
I.~Bena and N.~P. Warner, ``{Black holes, black rings and their microstates},'' \href{http://dx.doi.org/10.1007/978-3-540-79523-0}{{\em Lect. Notes Phys.} {\bfseries 755} (2008) 1--92},
\href{http://arxiv.org/abs/hep-th/0701216}{{\ttfamily arXiv:hep-th/0701216}}.

\bibitem{Heidmann:2017cxt}
P.~Heidmann, ``{Four-center bubbled BPS solutions with a Gibbons-Hawking base},'' \href{http://dx.doi.org/10.1007/JHEP10(2017)009}{{\em JHEP} {\bfseries 10} (2017) 009},
\href{http://arxiv.org/abs/1703.10095}{{\ttfamily arXiv:1703.10095 [hep-th]}}.

\bibitem{Bena:2017fvm}
I.~Bena, P.~Heidmann, and P.~F. Ramirez, ``{A systematic construction of microstate geometries with low angular momentum},'' \href{http://dx.doi.org/10.1007/JHEP10(2017)217}{{\em JHEP} {\bfseries 10} (2017) 217},
\href{http://arxiv.org/abs/1709.02812}{{\ttfamily arXiv:1709.02812 [hep-th]}}.

\bibitem{Bena:2018bbd}
I.~Bena, P.~Heidmann, and D.~Turton, ``{AdS$_{2}$ holography: mind the cap},'' \href{http://dx.doi.org/10.1007/JHEP12(2018)028}{{\em JHEP} {\bfseries 12} (2018) 028},
\href{http://arxiv.org/abs/1806.02834}{{\ttfamily arXiv:1806.02834 [hep-th]}}.

\bibitem{Heidmann:2018vky}
P.~Heidmann and S.~Mondal, ``{The full space of BPS multicenter states with pure D-brane charges},'' \href{http://dx.doi.org/10.1007/JHEP06(2019)011}{{\em JHEP} {\bfseries 06} (2019) 011}, \href{http://arxiv.org/abs/1810.10019}{{\ttfamily arXiv:1810.10019 [hep-th]}}.

\bibitem{Bena:2022wpl}
I.~Bena, S.~D. Hampton, A.~Houppe, Y.~Li, and D.~Toulikas, ``{The (amazing) super-maze},'' \href{http://dx.doi.org/10.1007/JHEP03(2023)237}{{\em JHEP} {\bfseries 03} (2023) 237}, \href{http://arxiv.org/abs/2211.14326}{{\ttfamily arXiv:2211.14326 [hep-th]}}.

\bibitem{Bena:2022fzf}
I.~Bena, N.~\v{C}eplak, S.~D. Hampton, A.~Houppe, D.~Toulikas, and N.~P. Warner, ``{Themelia: the irreducible microstructure of black holes},'' \href{http://arxiv.org/abs/2212.06158}{{\ttfamily arXiv:2212.06158 [hep-th]}}.

\bibitem{Bah:2020ogh}
I.~Bah and P.~Heidmann, ``{Topological Stars and Black Holes},'' \href{http://arxiv.org/abs/2011.08851}{{\ttfamily arXiv:2011.08851 [hep-th]}}.

\bibitem{JETZER1992163}
P.~Jetzer, ``Boson stars,'' \href{http://dx.doi.org/https://doi.org/10.1016/0370-1573(92)90123-H}{{\em Physics Reports} {\bfseries 220} no.~4, (1992) 163--227}. \url{https://www.sciencedirect.com/science/article/pii/037015739290123H}.

\bibitem{PRLHeidmann}
P.~Heidmann, ``{Half the Schwarzschild Entropy from Strominger and Vafa},'' \href{http://arxiv.org/abs/to appear}{{\ttfamily to appear}}.

\bibitem{Chowdhury:2007jx}
B.~D. Chowdhury and S.~D. Mathur, ``{Radiation from the non-extremal fuzzball},'' \href{http://dx.doi.org/10.1088/0264-9381/25/13/135005}{{\em Class. Quant. Grav.} {\bfseries 25} (2008) 135005},
\href{http://arxiv.org/abs/0711.4817}{{\ttfamily arXiv:0711.4817 [hep-th]}}.

\bibitem{Bena:2015dpt}
I.~Bena, D.~R. Mayerson, A.~Puhm, and B.~Vercnocke, ``{Tunneling into Microstate Geometries: Quantum Effects Stop Gravitational Collapse},'' \href{http://dx.doi.org/10.1007/JHEP07(2016)031}{{\em JHEP} {\bfseries 07} (2016) 031}, \href{http://arxiv.org/abs/1512.05376}{{\ttfamily arXiv:1512.05376 [hep-th]}}.

\bibitem{VladimirManko2000}
V.~S. Manko, E.~Ruiz, and J.~D. Sanabria-Gómez, ``Extended multi-soliton solutions of the einstein field equations: Ii. two comments on the existence of equilibrium states,'' \href{http://dx.doi.org/10.1088/0264-9381/17/18/320}{{\em Classical and Quantum Gravity} {\bfseries 17} no.~18, (Sep, 2000) 3881}. \url{https://dx.doi.org/10.1088/0264-9381/17/18/320}.

\bibitem{PhysRevD.51.4192}
E.~Ruiz, V.~S. Manko, and J.~Mart\'{\i}n, ``Extended n-soliton solution of the einstein-maxwell equations,'' \href{http://dx.doi.org/10.1103/PhysRevD.51.4192}{{\em Phys. Rev. D} {\bfseries 51} (Apr, 1995) 4192--4197}. \url{https://link.aps.org/doi/10.1103/PhysRevD.51.4192}.

\bibitem{Heidmann:2022ehn}
P.~Heidmann, I.~Bah, and E.~Berti, ``{Imaging Topological Solitons: the Microstructure Behind the Shadow},'' \href{http://arxiv.org/abs/2212.06837}{{\ttfamily arXiv:2212.06837 [gr-qc]}}.

\bibitem{Thorne:1980ru}
K.~S. Thorne, ``{Multipole Expansions of Gravitational Radiation},'' \href{http://dx.doi.org/10.1103/RevModPhys.52.299}{{\em Rev. Mod. Phys.} {\bfseries 52} (1980) 299--339}.

\bibitem{Bena:2020uup}
I.~Bena and D.~R. Mayerson, ``{Black Holes Lessons from Multipole Ratios},'' \href{http://arxiv.org/abs/2007.09152}{{\ttfamily arXiv:2007.09152 [hep-th]}}.

\bibitem{Bah:2021jno}
I.~Bah, I.~Bena, P.~Heidmann, Y.~Li, and D.~R. Mayerson, ``{Gravitational Footprints of Black Holes and Their Microstate Geometries},'' \href{http://arxiv.org/abs/2104.10686}{{\ttfamily arXiv:2104.10686 [hep-th]}}.

\end{thebibliography}\endgroup


\end{document}